\documentclass[peerreview,onecolumn,12pt]{IEEEtran}
\usepackage{setspace}
\usepackage{amsfonts,epsfig,amsmath,latexsym,amssymb,amscd,multirow,graphicx,geometry,lscape,amsmath,amssymb,bm,pifont,graphicx,amssymb,amsmath}
\usepackage{cancel,algorithm,algorithmic,algorithm,verbatim,array,multicol,palatino,amsfonts}
\usepackage{lineno}
\usepackage[normalem]{ulem}
\usepackage[dvips]{color}
\newcommand{\y}{\mathbf{y}}
\newcommand{\x}{\mathbf{x}}

\renewcommand{\u}{\mathbf{u}}
\newcommand{\A}{\mathbf{A}}

\newcommand{\Y}{\mathbf{Y}}
\newcommand{\R}{\mathbf{R}}

\newcommand{\exE}{\mathbb{E}}
\newcommand{\Ga}{\mathcal{G}\text{a}}

\def\L{{\cal L}}
\newfont{\fsc}{eusm10}                         % frenchscript letters

\newtheorem{theorem}{\vspace{0.5cm} \noindent \textbf{Theorem}}
\newtheorem{lemma}{\vspace{0.5cm} \noindent \textbf{Lemma}}
\modulolinenumbers[1]
\newtheorem{definition}{\vspace{0.5cm} \noindent \textbf{Definition}}
\newtheorem{remark}{\vspace{0.5cm} \noindent \textbf{Remark}}
\newtheorem{corollary}{\vspace{0.5cm} \noindent \textbf{Corollary}}
\newtheorem{example}{\vspace{0.5cm} \noindent \textbf{Example}}

\renewcommand{\baselinestretch}{1.4}
\begin{document}

\title{Generalized Interference Models in Doubly
Stochastic Poisson Random Fields for Wideband
Communications: the PNSC($\alpha$) model}
\author{Gareth W.~Peters$^{1,2,3}$, Ido Nevat$^{4}$, Francois Septier$^{5}$, Laurent Clavier $^{5}$
\begin{center}
{\footnotesize {\ \textit{$^{1}$ Department of Statistical Science, University College London,
London, UK; \\[0pt]
email: garethpeters@unsw.edu.au \\[0pt]
$^{2}$ CSIRO  Sydney, Locked Bag 17, North Ryde, New South Wales, 1670, Australia \\[0pt]
$^{3}$ School of Mathematics and Statistics, University of New South Wales,
Sydney, 2052, Australia; \\[0pt]
$^{4}$ Wireless \& Networking Tech. Lab, CSIRO, Sydney, Australia.\\[0pt]
$^{5}$ Telecom-Lille, Lille, France.
} }}
\end{center}
}
%
%\date{{\footnotesize {Transactions on Signal Processing, version from \today }}}
\maketitle

\begin{abstract}
\noindent 
A general stochastic model is developed for the total interference in wideband systems, denoted as the PNSC$(\alpha)$ Interference Model. It allows one to obtain, analytic representations in situations where (\textit{a}) interferers are distributed according to either a homogeneous or an inhomogeneous in time or space Cox point process and (\textit{b}) when the frequency bands occupied by each of the unknown number of interferers is also a random variable in the allowable bandwidth. 
The analytic representations obtained are generalizations of Cox processes to the family of sub-exponential models characterized by distributions from the $\alpha$-stable family. We develop general parameteric density representations for the interference models via doubly stochastic Poisson mixture representations of Scaled Mixture of Normal's via the Normal-Stable variance mixture. To illustrate members of this class of interference model we also develop two special cases for a moderately impulsive interference ($\alpha=3/2$) and a highly impulsive interference ($\alpha=2/3$) where closed form representations can be obtained either by the SMiN representation or via function expansions based on the Holtsmark distribution or Whittaker functions.
To illustrate the paper we  
%In addition we develop the Geometric SNR representation of the PNSC$(\alpha)$ model as well as 
propose expressions for the Capacity of a BPSK system under a PNSC$(\alpha)$ interference, via analytic expressions for the Likelihood Ratio Test staistic. 
%Then we finish this study of capacity with derivation of bounds on the capacity under PNSC$(\alpha)$ Interference in a BPSK transmission, that we prove can be expressed as a function the Fractional Order Moments of the PNSC$(\alpha)$ model we derived. We illustrate the properties of the model and derived GSNR and Capacity bounds in several examples.

\noindent \textbf{Keywords: } Interference models; Cox Process; Doubly Stochastic Poisson Stable Process; Isotropic $\alpha$-stable; Complex $\alpha$-stable.
\end{abstract}
%\newpage

\section{Introduction}
Modeling interference in \textit{ad hoc} or cognitive networks \cite{Sou92,Win09,Gha10,Pin10,Pin10b,Rab11} is an active research area in wireless communications. A key attribute often present is that interference in modern wireless networks exhibit an impulsive nature, evidenced by the growing literature on the understanding and study of such features in wireless communications.

Several modeling approaches for impulsive noises have been proposed, especially %in the framework of 
for impulse radio ultra wide band communications (Laplace, Generalized Gaussian, Cauchy, $\alpha$-stable, Middleton class A, an overview is, for instance, given in \cite{Bea09} and the references therein). Each of these models shares something in common: they are members of what one can define as heavy tailed or impulsive noise processes; each is a member of the sub-exponential family of interference distributions. In this sub-class of interference models there is one particular sub-family, the $\alpha$-stable distribution, which enjoys a very rich and extensive literature on its characteristics and attributes both theoretically as a generalization of the Central Limit Theorem and in applied settings. %as motivated from first principles as the limiting distribution of several interference processes in applied signal processing and information theory. 
Several papers have dealt with such models, see for example in the signal processing literature the discussion in \cite{Pin10,Pin10b}.

\subsection{Brief summary of the paper.}

In this paper we extend the family of $\alpha$-stable interference models, %from first principles, 
proving that one can obtain and characterize a total interference model in the class of doubly stochastic Poisson-Gamma-complex Isotropic Stable Cox processes, which we refer to as the PNSC$(\alpha)$ class of models. This involves three main contributions:
\begin{itemize}
	\item First we extend the model, for instance presented in \cite{Win09}, to a doubly stochastic Poisson random field. Practically, this means that not only the number of interferers is a random variable but also the number of sub-carriers occupied by each interferer from the total set of available carriers is random. There are several applied settings in which such an approach is required, for instance in cognitive networks where secondary users would use several sub-carriers to adjust their desired bit rate but not all of the possible sub-carriers available. In addition, in this context, a strong contribution of the paper is to also allow the Poisson field to be inhomogeneous in time or space. For example, we include illustrations of the cases where users are sparser when further from the access point or when directive antennas are used so that interferers are only located in a sector of the plane.
	\item Having derived and interpreted the resulting class of impulsive total interference models, labeled the PNSC$(\alpha)$ family of interference, we need to develop tools for practitioners to utilize these models. Therefore, the second contribution involves developing analytic and closed form representations. We note that a challenge of stable distributions is that, in general, they do not admit analytic expressions for the density and distributions. This makes, for instance, receiver processing difficult to adapt to such models. We overcome this important challenge utilizing several key features of the family of complex isotropic stable distributions, which we extend to the doubly-stochastic Poisson-Gamma-complex Isotropic Stable Cox process setting derived. Namely, the key attributes we utilize are that stable random vectors are closed under convolution, they admit scaled mixture of Gaussian representations under projection and they admit infinite series expansion representations under projection. We apply these in a general context and demonstrate their attributes for interference modeling under truncation approximations of such series.

Then for two special cases we develop some closed form solutions resulting from special function representations, which do not require infinite series or scale mixture representations. The first one considers situations with moderately impulsive noise ($\alpha=3/2$, which corresponds to a channel attenuation coefficient $\sigma=2.66$ if we refer to \cite{Win09,Gha10} for instance) and the second to a highly impulsive noise ($\alpha=2/3$ corresponding to $\sigma=6$). We demonstrate and discuss how such analytical results can be very useful in the understanding and design of nodes or network in impulsive interference.
	\item Thirdly we derive novel expressions for the Likelihood Ratio Test statistic and study the capacity of a binary input soft output symmetric memoryless channel for the class of PNSC$(\alpha)$ interference models. %Illustrating each of these in detailed numerical studies.
\end{itemize}

\subsection{Organization and Contribution}
The paper is organized as follows: in section \ref{Background} we provide a brief technical review of theoretical properties of $\alpha$-stable random variables and vectors that will be of direct consequence for the proofs and derivations  of results in this paper. In section \ref{SystemDescription} we describe the system and the different assumptions that we make to obtain the theoretical results. Finally in section \ref{sec:analytic} we derive the interference distributions in the different cases and provide examples to illustrate the proposed model.

Below, we highlight some specific contributions involving two generic frameworks which are developed based on analytic representations of distributions corresponding to the "total interference" across transmission bandwidth in a wide-band wireless communication system in which the number of users transmitting is treated stochastically and the bandwidth they occupy is also treated as stochastic. This involves the following key contributions:
\begin{itemize}
 \item In Theorem \ref{thm_1}, we extend the representation of \cite{Pin06,Win09,Gha10,Pin10} to derive the $\log$ Characteristic Function (CF) for the total interference at a given frequency, for a random number of potential interferers in a spatial region of transmission. We prove that the resulting $\log$ CF can be represented by the family of isotropic bivariate $\alpha$-stable distributions. In doing so, we generalize existing results in the following ways: (\textit{a)} we allow the number of users to be distributed according to three possible scenarios involving homogeneous Poisson (Model 1), temporally inhomogeneous Poisson (Model 2) and spatially inhomogeneous Poisson (Model 3) point processes; (\textit{b}) in Lemma \ref{alpha_representation} we detail analytic distributional representations for the bivariate isotropic stable distribution based on a Scaled Mixture of Normals (representation 1) and an exact Projection based univariate representation via a Cramer-Wold decomposition (representation 2).
 %%%%%%%%%%%%%%%%%%%
\item In Theorem \ref{thm_pois_smin} we utilize the isotropic $\alpha$-stable representations of the total interference at a given frequency (Theorem \ref{thm_1}), to extend these results to derive analytic solutions for both the distribution function and density function of the resulting inhomogeneous (spatially or temporally) stochastic Poisson compound processes that models the practically important total interference across the entire transmission bandwidth. This involves development of analytic solutions for the distribution of a compound process (Poisson mixture) of bivariate stable components and utilization of the closure under convolution of the stable components, based on work of \cite{peters2011analytic}.
%%%%%%%%%%%%%%%%%%%
\item In Theorem \ref{thm_pois_gamma_smin} we generalize the results derived in Theorem \ref{thm_pois_smin} to analytic representations of a Cox process (doubly stochastic compound Poisson-Stable mixture) for the total interference in the context in which the mean occupied bandwidth of each potential interferer is generalized to a stochastic model. This allows for potential scenarios such as different occupancy average requirements per user; or time varying occupancy requirements. 
%%%%%%%%%%%%%%%%%%%	
%\item In Theorem \ref{ThmMoments} we derive analytic expresions for the Fractional Lower Order Moments (FLOM's) in the PNSC$(\alpha)$ models characterised by doubly stochastic Poisson-Gamma-complex Isotropic Stable Cox processes. Then in Lemma \ref{PropGNSR} we derive a Generalised Geometric SNR (GSNR) expressions for the class of PNSC$(\alpha)$ interference models.
%%%%%%%%%%%%%%%%%%%	
\item In Theorem \ref{ThmCapacityBasic} we derive expressions for the Capacity of the PNSC$(\alpha)$ interference models expressed analytically according to a likelihood ratio test statistic. We derive several analytic expressions for the LRT in the case of PNSC$(\alpha)$ interference models, making evaluation of the capacity highly efficient for generally intractable stable models. We illustrate their accuracy as a function of the stable tail index. 
%%%%%%%%%%%%%%%%%%%	
%\item In Theorem \ref{ThmCapacityBound} we derive expressions for bounds on the capacity of a channel in the presence of PNSC$(\alpha)$ interference, as a function of the derived FLOM's and the interference model characteristics. We then illustrate this bound in examples, to show its performance.
%%%%%%%%%%%%%%%%%%%	
\end{itemize}

%%%%%%%%%%%%%%%%%%%%%%%%%%%%%%%%%%%%%%%%%%%%%%%%%%%%%%%%%%%%%%%%%%%%%%%%%%%%%%%%%%%%%%%%%%%%%%
\subsection{Notation}
The following notation is used throughout: random variables are denoted by upper case letters and their realizations by lower case letters. In addition, bold will be used to denote a vector or matrix quantity, upper subscripts will refer to a specific interferer and lower subscripts to the element of a vector or matrix.

\section{Background on Statistical Interference Modeling}
\label{Background}

\subsection{Interference modeling}
Interference from undesired active users in a network will be a strong limitation in future networks performance.
The interference model has been studied widely in information theory \cite{Car78,Han81,Sat81,Bre10}. If the exact capacity is not known some close approximations have been derived. The question on how to deal with interference is however still an open problem. In this regard a lot of work on multiuser detectors for instance have been proposed \cite{Ver98} but also, more recently, some new schemes for interference alignment \cite{Bre10} or amplifying interference \cite{Dab08} have been considered. However, those works aim at avoiding the interference and generally require some costly channel learning mechanisms or synchronization techniques. 

An alternative perspective is to consider that a certain amount of interference will be unavoidable. Under such an assumption, a robust interference model can allow an effective design of receivers and networks to limit the resulting impact of such interference. This is a powerful tool to study for example outage probability or connectivity in networks. For instance several works on stochastic geometry are based on similar interference models as we present in this paper \cite{Gan12,Kou12}. If we consider $\kappa_R$ to be a random variable representing the number of active interferers. In a rather general framework, the total interference is a random variable expressed according to	$Y=\sum_{k=1}^{\kappa_R}A_k \psi_k$ where $(\psi_k)_{k=1,\cdots,N}$ are independent, identically distributed and bounded random variables with even probability density function that depends on the physical layer design (see \cite{Sou92,Win09,Gha10} for different examples). The $(A_k)_{k=1,\cdots,N}$ are positive, independent, identically distributed random variables that depend on the channel characteristics and determine the statistical properties of the total interference $Y$. 

To proceed, the most intuitive statistical approach would be to consider the asymptotic behavior of the distribution of the total interference $Y \sim F(y)$ and to determine under what conditions such an interference would belong to the domain of attraction of a Gaussian family of distributions, denoted $\mathcal{D}_G(F)$. Such an approach involves considering an asymptotic regime where the number of interferers grows to infinity while the contribution of each interferer to $Y$ becomes infinitesimal. In non-impulsive, non-sub exponential distributional settings, this would typically result in application of a form of the celebrated Central Limit Theorem: $Y$ converges in law to a Normal distribution, such that $F(y) \in \mathcal{D}_G(F)$. However, in the general case in which impulsive noise is present it is well known that this asymptotic regime is not easily reached (see for example an in-depth study in \cite{Fio06b} for impulse radio ultra wide band signals). Instead, the domain of attraction of impulsive noise models from the sub-exponential family, which are convolved to create the total interference $Y \sim F(y)$ can belong to the domain of attraction of a stable family of distributions, denoted $\mathcal{D}_S(F)$, for which $\mathcal{D}_G(F) \subset \mathcal{D}_S(F)$.

A common requirement for convergence of such a sequence of i.i.d. interferences to converge to the Gaussian domain of attraction, involves a restriction on the variance of such summands in the sequence. This is not present in impulsive noise processes. One could argue that this feature may seem natural since it represents a channel attenuation, which by its very nature must be finite. However, the interference which is being modeled is actually compared to the desired link attenuation and can, in comparison, be "very large" and impulsive in nature. Such large impulsive realizations of the interference happen infrequently in practice but are sufficient to give an impulsive nature to interference. To capture these situations, heavy tailed distributions with infinite variance can be well suited while models with finite second order moments will fail to adequately capture such impulsive attributes observed. The generalized central limit theorem has then to be used (see \cite[p. 22]{Nik95} or \cite[p. 9]{samorodnitsky1994stable}) and states that interference (for large $\kappa_R$) falls in the domain of attraction of a random variable with a stable distribution, $F(y) \in \mathcal{D}_S(F)$.

A general framework is proposed in \cite{Pin10} and application to cognitive radio with a modified law (truncated $\alpha$-stable although the term truncated is slightly misleading) is presented in \cite{Rab11}. Here the truncation refers to a form of soft ``tempering'' of the stable distribution tails, as opposed to a hard thresholding. To prove the validity of the $\alpha$-stable assumption, the usual solution is to write its $\log$ CF as $\varphi_Y\left(\omega\right)=-\sigma |\omega|^{\alpha}$.
This can be done in many situations (users' repartition, channel conditions, physical layer etc., refer to \cite{Win09} for more details).  One strong advantage of this model over other proposed solutions is its theoretical foundations which we derive in the context of interference modeling from first principles for our domain of modelling, based purely on simple statistical assumptions on the system.

\subsection{Background on Univariate and multivariate $\alpha$-Stable distributions}
\label{backgroundstable}
In this section, we provide a brief technical survey of relevant results from the probability and statistics literature relating to sub-exponential family models of distributions, in particular the family of $\alpha$-Stable models. These results will provide sufficient coverage to understand the derivations and results we develop in this paper related to interference modeling in wireless communications.

Considered as generalizations of the Gaussian distribution, $\alpha$-Stable models are defined as the class of location-scale distributions which are closed under convolutions. In an interference modeling context, $\alpha$-stable distributions possess several useful properties, including the possibility of incorporating infinite mean and infinite variance, skewness and heavy tails, see \cite{zolotarev1983univariate} and \cite{samorodnitsky1994stable}. It is due to this inherent flexibility that they have found extensive use in practical modeling settings, both in wireless communications such as interference modelling that we consider and in many other domains of application, see a comprehensive list of such literature in the stable bibliography\begin{footnote}{\texttt{http://academic2.american.edu/~jpnolan/stable/stable.html} }\end{footnote}. 

\bigskip

\subsubsection{Univariate $\alpha$-Stable Models}
We consider a random variable $X$ with $\alpha$-stable distribution, denoted by $X \sim \mathcal{S}_{\alpha}\left(x; \beta, \gamma, \delta, 0\right)$. Where, $\mathcal{S}_{\alpha}\left(x; \beta, \gamma, \delta, 0\right)$ denotes the univariate four parameter stable distribution family under parameterization $S(0)$ as defined in \cite{nolan:2012}.  

The univariate $\alpha$-stable distribution we consider is specified by four parameters: $\alpha \in (0, 2]$ determining the rate of tail decay; $\beta \in [-1, 1]$ determining the degree and sign of asymmetry (skewness); $\gamma > 0$ the scale (under some parameterizations); and $\delta \in \mathbb{R}$ the location. The parameter $\alpha$ is termed the characteristic exponent, with small and large $\alpha$ implying heavy and light tails respectively. In general $\alpha$-stable models admit no closed-form expression for the density which can be analytically evaluated point-wise, except Gaussian $(\alpha = 2, \beta = 0)$, Cauchy $(\alpha = 1, \beta = 0)$ and Levy $(\alpha = 0.5, \beta = 1)$ distribution cases. Therefore, statistical inference typically proceeds via the characteristic function, see discussions in \cite{nolan1998parameterizations}, \cite{peters2010likelihood} and \cite{press1972estimation}. However, intractable to evaluate point-wise, importantly for wireless communication applications, simulation of random variates is very efficient (see \cite{chambers1976method}). % and the algorithm provided in Appendix \ref{Append1}. 

\bigskip

\begin{definition}
\label{defn_0}
\textit{A random variable $X$ is stable if and only if $X \stackrel{d}{=}
aZ +b$, where $0 < \alpha \leq 2$, $-1 \leq \beta \leq 1$, $a > 0$, $b \in \mathbb{R}$ and Z is a random variable with characteristic function
\begin{equation}
\exE \left[\exp(i \theta Z)\right] = \begin{cases} \exp\left(-|\theta|^{\alpha}\left[1-i\beta \tan\frac{\pi \alpha}{2}\left(\text{sign}(\theta)\right)\right]\right) & \alpha \neq 1, \\ 
\exp\left(-|\theta|^{\alpha}\left[1+i\beta \frac{2}{\pi}\left(\text{sign}(\theta)\right)\log|\theta|\right]\right) & \alpha = 1. \end{cases} 
\end{equation}
where $\text{sign}(u) = -1$ if $u < 0$, $\text{sign}(u) = 0$ if $u = 0$ and $\text{sign}(u) = 1$ if $u > 0$.
}
\end{definition}

\bigskip

From this definition one may define several practically useful reparameterizations, in this paper we consider the following parameterization denoted in \cite{nolan:2012} as the $S(0)$ parameterization. A random variable $X$ is said to have a stable distribution, $\mathcal{S}_{\alpha}(\beta,\gamma,\delta;0)$, if its CF has the following form:
\[ \exE[\text{exp}(i\theta X)] = 	\left\{              
			\begin{array}{ll}
				\text{exp}\{ -\gamma^\alpha|\theta|^\alpha(1+i\beta(\text{sign}(\theta))\tan({\pi \alpha \over 2})(|\gamma \theta|^{1-\alpha}-1))+i\delta \theta\} & \text{if   } \alpha \neq 1\\ 
				\text{exp}\{ -\gamma|\theta|(1+i\beta({2 \over \pi})(\text{sign}(\theta))\text{ln}(\gamma|\theta|))+i\delta \theta\} & \text{if   } \alpha = 1.
				
			\end{array}
			\right.
	\]

\bigskip
In the following lemmas we present some fundamental basic facts about univariate $\alpha$-Stable random variables that will be required to establish the results we develop in this paper. In particular these results will be used to construct analytic exact Poisson and doubly stochastic Poisson mixture representations of wireless communications interference processes, arising in the context in which an unknown number of interferers are present. This will be achieved by considering an important sub-family of $\alpha$-stable models, those that are symmetric and isotropic. 

\bigskip

%%%%%%%%%%%%%%%%%%%%%%%%%%%%%%%%%%%%%%%%%%%%%%%%%%%%%%%%%%%%%%%%%%%%%%%%%%%%%%%%%%%%%%%%%%%%%%%%
\begin{lemma}
\label{lemma_1}
\textit{
If $Y \sim \mathcal{S}_{\alpha}(\beta,\gamma,\delta;0)$, then for any $a \neq 0, b \in \mathbb{R}$, the transformation $Z=aY+b$ is a scaled version of the $\alpha$-stable distribution. That is $Z \sim \mathcal{S}_{\alpha}(\text{sign}(a)\beta,|a|\gamma,a\delta+b;0)$. In addition, the CF, densities and distribution functions are jointly continuous in all four parameters $(\alpha,\beta,\gamma,\delta)$ and in $x$. 
}
\end{lemma}
%%%%%%%%%%%%%%%%%%%%%%%%%%%%%%%%%%%%%%%%%%%%%%%%%%%%%%%%%%%%%%%%%%%%%%%%%%%%%%%%%%%%%%%%%%%%%%%%

These results follow from \cite{samorodnitsky1994stable} and \cite[Proposition 1.16]{nolan:2012}.
\bigskip

\begin{lemma}
\label{lemma_2}
\textit{
If for all $i \in \left\{1,\ldots,N\right\}$ one has random variables $X_i \sim \mathcal{S}_{\alpha}(\beta_i,\gamma_i,\delta_i;0)$ then the distribution of the linear combination, given N, is 
\begin{equation}
%\small
\begin{split}
Z = \sum_{i=1}^N X_i & \sim S(\alpha,\widetilde{\beta},\widetilde{\gamma},\widetilde{\delta};0)\\
\widetilde{\gamma}^{\alpha} &= \sum_{i=1}^N \gamma_i^{\alpha}, \; \; \; \; \widetilde{\beta} =  \frac{\sum_{i=1}^N \beta_i\gamma_i^{\alpha}}{\sum_{i=1}^N \gamma_i^{\alpha}}, \\
\widetilde{\delta} &= 	\left\{              
			\begin{array}{ll}
			\sum_{i=1}^N \delta_i + \tan \frac{\pi \alpha}{2}\left(\widetilde{\beta}\widetilde{\gamma} - \sum_{i=1}^N \beta_j\gamma_j\right) & \text{if   } \alpha \neq 1,\\  
			\sum_{i=1}^N \delta_i + \frac{2}{\pi}\left(\widetilde{\beta}\widetilde{\gamma}\log\widetilde{\gamma} - \sum_{i=1}^N \beta_j\gamma_j\log\gamma_i\right) & \text{if   } \alpha = 1.
			\end{array}
			\right. 
\end{split}
%\normalsize
\end{equation}
}
\end{lemma}
%%%%%%%%%%%%%%%%%%%%%%%%%%%%%%%%%%%%%%%%%%%%%%%%%%%%%%%%%%%%%%%%%%%%%%%%%%%%%%%%%%%%%%%%%%%%%%%%

\bigskip
This result follows from \cite[Section 1.2, Property 1.2.1]{samorodnitsky1994stable} and \cite[Proposition 1.17]{nolan:2012}.
\bigskip

A practically relevant sub-family of $\alpha$-stable distributions is obtained when one considers the symmetric case. A random variable $X$ is said to be distributed from a symmetric $\alpha$-Stable distribution, $X \sim \mathcal{S}_{\alpha}\left(0,\gamma,\delta\right)$, when the skewness parameter $\beta = 0$. In this case, the model still captures a spectrum of distributions ranging from Gaussian $\alpha = 2$ through to infinite mean and infinite variance models. This particular sub-class is interesting as it can be represented uniquely by a Scaled Mixture of Normals (SMiN) representation as shown in Lemma \ref{lemma_3}. 
%%%%%%%%%%%%%%%%%%%%%%%%%%%%%%%%%%%%%%%%%%%%%%%%%%%%%%%%%%%%%%%%%%%%%%%%%%%%%%%%%%%%%%%%%%%%%%%%
\bigskip

\begin{lemma}
\label{lemma_3}
\textit{
In \cite[Equations (3) and (4), pp.2]{godsill2000inference} it is shown that when $X \sim \mathcal{S}_{\alpha}\left(0,\gamma,\delta\right)$ it may be represented exactly by the following SMiN representation through the introduction of an auxiliary random variable $\lambda$,
\begin{equation}
%\begin{split}
X|\lambda \sim N\left(\delta, \gamma\lambda\right),
%\end{split}
%\label{errorMixture}
\label{EqnSMiN}
\end{equation}
with auxiliary scale variable $\lambda \sim \mathcal{S}_{\alpha/2}\left(0,1,1\right)$.
}
\end{lemma}
%%%%%%%%%%%%%%%%%%%%%%%%%%%%%%%%%%%%%%%%%%%%%%%%%%%%%%%%%%%%%%%%%%%%%%%%%%%%%%%%%%%%%%%%%%%%%%%%
\bigskip

This result will be used in our models as the basis for working with a large family of symmetric stable models that we derive. Clearly, this is advantageous as conditionally on $\lambda$, one has a Gaussian distributed random variable $X|\lambda$. However, general closed-form expansions of the probability distribution functions in terms of well-understood functions do not exist. However, since the work of \cite{bergstrom1952some} who showed that all continuous stable distributions can be written in terms of infinite series expansions of elementary functions, there has been specific examples created. There is the series expansion representations developed in \cite{metzler2000random} they transform the parameters and provide the stable densities in terms of Fox's H functions, see \cite{fox1961g}. Other reparameterizations to obtain representations include \cite{hoffmann1993stable} who obtain representations with respect to incomplete hypergeometric functions. So in general for symmetric stable settings one may choose between a SMiN representation or a series expansion. A summary of the series and integral expansions for $\alpha$-stable models is given by \ref{lemma:LevySeverityStablePdfCdf}, see \cite{zolotarev1983univariate} for details. 

\begin{lemma}[$\alpha$-Stable Density and Distribution Representations]\textit{\label{lemma:LevySeverityStablePdfCdf}
w.l.o.g. the density function of an $\alpha$-Stable distribution (standardized such that $\gamma = 1$ and $\delta = 0$) can be evaluated pointwise according to
\begin{equation}
f_X(x;\alpha,\beta,1,0;S(0)) = \begin{cases}
\frac{1}{\pi}\mathrm{Re}\left\{\int_{0}^{\infty}\exp\left(-itx - t^{\alpha}\exp\left(-i\frac{\pi}{2}\beta K(\alpha)\right) \right) dt\right\}, & \; \mathrm{if} \; \alpha \neq 1,\\
\frac{1}{\pi}\mathrm{Re}\left\{\int_{0}^{\infty}\exp\left(-itx - \frac{\pi}{2}t - i\beta t \log t \right) dt\right\}, & \; \mathrm{if} \; \alpha = 1.\\
\end{cases}
\end{equation}
Alternatively via the series expansions \cite{zolotarev1983univariate} [Equation 2.4.6, p. 89]
\begin{equation}
f_X(x;\alpha,\beta,1,0;S(0)) = \begin{cases}
\frac{1}{\pi}\sum_{n=1}^{\infty} (-1)^{n-1}\frac{\Gamma(\frac{n}{\alpha} + 1)}{\Gamma(n+1)}\sin(n\pi\rho)x^{n-1}, & \; \mathrm{if} \; \alpha > 1, \beta \in [-1,1], x\in \mathbb{R},\\
\frac{1}{\pi}\sum_{n=1}^{\infty}(-1)^{n-1}nb_n x^{n-1}, & \; \mathrm{if} \; \alpha = 1, \beta \in (0,1], x\in \mathbb{R},\\
\frac{1}{\pi}\frac{\Gamma(n\alpha + 1)}{\Gamma(n+1)}\sin(n\pi\rho\alpha)x^{-n\alpha-1}, & \; \mathrm{if} \; \alpha < 1, \beta \in [-1,1], x\in \mathbb{R}^+,\\
\end{cases}
\end{equation}
where the coefficients $b_n$ are given by
\begin{equation}
b_n = \frac{1}{\Gamma(n+1)}\int_{0}^{\infty} \exp\left(-\beta u \ln u\right) u^{n-1} \sin\left[(1+\beta) u \frac{\pi}{2}\right]du.
\end{equation}
In addition, the distribution function of an $\alpha$-Stable model can be evaluated pointwise according to
\begin{equation}
F_X(x;\alpha,\beta,1,0;S(0))= \begin{cases}
C(\alpha,\theta) + \frac{\epsilon(\alpha)}{2}\int_{-\theta}^1 \exp\left(-x^{\frac{\alpha}{\alpha-1}}U_{\alpha}(\varphi,\theta)\right)d\varphi, & \; \mathrm{if} \; \alpha \neq 1 \; \mathrm{and} \; x > 0,\\
\frac{1}{2}\int_{-1}^{1}\exp\left(-\exp\left(-\frac{x}{\beta}\right)U_1(\varphi,\beta) \right)d\varphi, & \; \mathrm{if} \; \alpha = 1, \; \mathrm{and} \; \beta > 0,
\end{cases}
\end{equation}
otherwise in all other cases it suffices to utilise the duality principle of infinitely divisible stable distributions which has the consequence that 
\begin{equation}
F_X(-x;\alpha,\beta,1,0;S(0)) + F_X(x;\alpha,-\beta,1,0;S(0)) = 1. 
\end{equation}
Note, the notation of \cite{zolotarev1983univariate} [page 74] is adopted above in which
\begin{equation}
\begin{split}
\epsilon(\alpha) &= \mathrm{sgn}(1-\alpha), \; K(\alpha) = \alpha - 1 + \mathrm{sgn}(1-\alpha)\\
\theta &= \beta\frac{K(\alpha)}{\alpha}, \; C(\alpha,\theta) = 1 - \frac{1}{4}(1+\theta)\left(1+\epsilon(\alpha)\right),\\
U_{\alpha}\left(\varphi,\theta\right) &= \left(\frac{\sin\frac{\pi}{2}\alpha(\varphi + \theta)}{\cos \frac{\pi}{2}\varphi}\right)^{\frac{\alpha}{1-\alpha}}\frac{\cos \frac{\pi}{2}\left((\alpha-1)\varphi + \alpha \theta\right)}{\cos \frac{\pi}{2}\varphi}
\end{split}
\end{equation}
}
\end{lemma}
To proceed we demonstrate when each approach will be of utility in the setting of interference modelling in wireless communications.
\bigskip

\begin{definition}{\label{Defn:SameType} \textit{Distributions of the same type will be defined to be those that differ only in location and scale.}}
\end{definition}

\bigskip
Without loss of generality, consider $\tilde{X}_{\alpha} \sim \mathcal{S}_{\alpha}\left(0,1,0\right)$, symmetric scaled $\alpha$-stable random variables.
Variables $\tilde{X}$ will have distributions which are of the same type as random variable $X$ given in Definition \ref{defn_0}. In the following specific cases we may replace the infinite scaled mixture SMiN representation with analytic functions representations as alternative models to evaluate the density of a symmetric $\alpha$-stable random variate:
\begin{lemma}
\label{lemma_3b}
\textit{In the symmetric stable case, under two particular choices of $\alpha \in \left\{\frac{2}{3},\frac{3}{2}\right\}$, the distribution of the stable model can be analytically represented via known functions according to:
\begin{enumerate}
\item{\textbf{Finite Mean Interference Model ($\alpha =3/2$):} the Holtsmark density \cite{garoni2002levy} is given for random variable denoted $\tilde{X}_{3/2}$ which has an analytic density which is represented according to hypergeometric distributions as follows,
\begin{equation}
\begin{split}
f_{X}\left(x\right) &= \frac{1}{\pi}\Gamma(5/3)_2F_3\left(\frac{5}{12},\frac{11}{12};\frac{1}{3},\frac{1}{2},\frac{5}{6};-\frac{2^2x^6}{3^6}\right) - \frac{x^2}{3\pi}\, _3F_4\left(\frac{3}{4},1,\frac{5}{4};\frac{2}{3},\frac{5}{6},\frac{7}{6},\frac{4}{3};-\frac{2^2x^6}{3^6}\right)\\
& \; + \frac{7x^4}{3^4\pi}\Gamma(4/3)_2F_3\left(\frac{13}{12},\frac{19}{12};\frac{7}{6},\frac{3}{2},\frac{5}{3};-\frac{2^2x^6}{3^6} \right), -\infty < x < \infty,
\end{split}
\end{equation}
with $\alpha = \frac{3}{2},\beta = 0, \gamma = 1, \delta = 0$ and:
\begin{equation}	
_pF_q\left(a_1,\ldots,a_p;b_1,\ldots,b_q;z\right)=\sum_{n=0}^{+\infty}\frac{(a_1)_n\ldots(a_p)_n}{(b_1)_n\ldots(b_q)_n}\frac{z^n}{n!},
\label{eq:pFq}
\end{equation}
with $(a)_n=a(a+1)\ldots(a+n-1)$ and $(a)_0=1$.
}
\item{\textbf{Infinite Mean Interference Model ($\alpha =2/3$):}  the Whittaker function density representation \cite{uchaikin1999chance} is given for a random variable denoted $\tilde{X}_{2/3}$ which has an anlaytic denisty which is represented according to Whittaker functions, see \cite{abramowitz1964handbook} as follows,
\begin{equation}
\begin{split}
f_{X}\left(x\right) &= \frac{1}{2\sqrt{3\pi}}|x|^{-1}\exp\left(\frac{2}{27}x^{-2}\right)W_{-1/2;1/6}\left(\frac{4}{27}x^{-2}\right), -\infty < x < \infty,
\end{split}
\end{equation}
with $\alpha = \frac{2}{3},\beta = 0, \gamma = 1, \delta = 0$ and
\begin{equation}
W_{\lambda;\mu}\left(z\right)=\frac{z^{\lambda}\exp(-z/2)}{\Gamma\left(\mu - \lambda + 1/2\right)} \int_{0}^{\infty}\exp(-t)t^{\mu - \lambda - 1/2}\left( 1+\frac{t}{z} \right)^{\mu - \lambda - 1/2}dt. \\
\label{eq:W}
\end{equation}
We can also express $W_{\lambda;\mu}\left(z\right)$ in terms of confluent hypergeometric functions:
\begin{equation}
W_{\lambda;\mu}\left(z\right)=\exp(-z/2)z^{(\mu+1/2)}U(\mu-\lambda+1/2,1+2\mu;z)
\end{equation}
with
\begin{equation}
U(a,b;z)=\frac{\Gamma(1-b)}{\Gamma(a-b+1)}M(a,b;z)+\frac{\Gamma(b-1)}{\Gamma(a)}z^{1-b}M(a-b+1,2-b;z)
\end{equation}
and
\begin{equation}
M(a,b;z)=\;_1F_1(a,b;z)
\end{equation}
}
\end{enumerate}
}
\end{lemma}

\bigskip

The combination of Lemma \ref{lemma_1}, Lemma \ref{lemma_2} and Lemma \ref{lemma_3} will be directly utilized in results obtained in Theorem \ref{thm_pois_smin}. 
In addition the results in Lemma \ref{lemma_3b} are of relevance in the practical examples developed to illustrate the theoretical results obtained. 

\bigskip
\subsubsection{Multivariate $\alpha$-Stable Models} 
It is also relevant to consider some background on multivariate $\alpha$-stable random variables, in particular the symmetric isotropic bi-variate $\alpha$-Stable random variable.
Multivariate $\alpha$-Stable models are covered in detail in \cite{nolan:2012,nolan2001estimation,adler1997practical,press1972multivariate}. Here we first present the joint CF for an elliptically contoured multivariate $\alpha$-Stable distribution, see \cite[Proposition 2.5.8]{taqqu1994stable}, 
\begin{equation}
\varphi_{\bm{X}} = \mathbb{E}\left[j\bm{\omega}^T\bm{X}\right] = \exp\left(-\left(\bm{\omega}^T\Sigma\bm{\omega}\right)^{\alpha/2} + j\bm{\omega}^T\bm{\delta}\right)
\end{equation}
for some positive definite matrix $\Sigma$ and translation vector $\bm{\delta}$. Furthermore, there is a generalization of the result of Lemma \ref{lemma_3} for the multivariate settings given in Lemma \ref{lemma_4} below.
%%%%%%%%%%%%%%%%%%%%%%%%%%%%%%%%%%%%%%%%%%%%%%%%%%%%%%%%%%%%%%%%%%%%%%%%%%%%%%%%%%%%%%%%%%%%%%%%
\begin{lemma}
\label{lemma_4}
\textit{
According to \cite[Proposition 2.5.8]{samorodnitsky1994stable}, consider $\bm{G} \sim N(\bm{0},\Sigma)$ such that $\bm{Y} \in \mathbb{R}^d$ and $\lambda \sim \mathcal{S}_{\alpha/2}(1,\gamma,0;0)$ as an independent univariate $\alpha$-Stable random variable with $0<\alpha<2$. Then the transformed random vector $\bm{X} = \sqrt{\lambda}\bm{Y}$ is $\alpha$-Stable and elliptically contoured with CF,
\begin{equation}
\varphi_{\bm{X}} = \mathbb{E}\left[j\bm{\omega}^T\bm{X}\right] =
\exp\left(\left(\frac{\gamma}{2}\right)^{\alpha/2}\sec\left(\frac{\pi \alpha}{4} \right) \left(\bm{\omega}^T\Sigma\bm{\omega}\right)^{\alpha/2}\right).
\end{equation}
This substable family reduces in the isotropic (radially symmetric) case, in which $\Sigma$ is diagonal, to:
\begin{equation}
\varphi_{\bm{X}} = \mathbb{E}\left[j\bm{\omega}^T\bm{X}\right] =
\exp\left(\left(\gamma_0\right)^{\alpha}\left|\bm{\omega}\right|^{\alpha} + j\bm{\omega}^T\bm{\delta}\right),
\end{equation}
for scale parameter $\gamma_0 > 0$. Furthermore, in this case the spectral measure is a uniform distribution on the unit-sphere and we obtain,
\begin{equation}
\bm{X} \sim N\left(\bm{\delta},\lambda\Sigma\right).
\end{equation}
}
\end{lemma}

%%%%%%%%%%%%%%%%%%%%%%%%%%%%%%%%%%%%%%%%%%%%%%%%%%%%%%%%%%%%%%%%%%%%%%%%%%%%%%%%%%%%%%%%%%%%%%%%
In addition, to the result in Lemma \ref{lemma_4} for the generalization of the SMiN reprsentation to isotropic symmetric multivariate $\alpha$-Stable random vectors, it is also relevant to present the general properties for the density and distribution functions. %According to \cite[Chapter 6]{nolan:2012} and \cite{nolan2006multivariate} we can also present, in Lemma \ref{lemma_5}, the distribution function for the multivariate isotropic symmetric $\alpha$-Stable random vector $\bm{X}$, under a change of co-ordinates from Cartesian to polar, according to the amplitude distribution.
%%%%%%%%%%%%%%%%%%%%%%%%%%%%%%%%%%%%%%%%%%%%%%%%%%%%%%%%%%%%%%%%%%%%%%%%%%%%%%%%%%%%%%%%%%%%%%%%

\section{System Model and Assumptions} 
\label{SystemDescription}
In this section we complete the system model introduced in the introduction and detail the required statistical assumptions that will be used throughout the paper:
\begin{enumerate}

%%%%%%%%%%%%%%%%%%%%%%%%%%%%%%%%%%%1
\item We assume a wireless network with $N(t)$ transmitters at time $t$, distributed on a region $\mathcal{R}$ with area $A_{\mathcal{R}}$, at locations indexed by $\L(t)=\left\{ L^{(i)}(t) \right\}_{i=1\ldots N}$. %,\ldots, L^{(N)}(t)
Furthermore, we assume that the number of transmitters varies over time and space stochastically according to an inhomogeneous spatial-temporal Poisson process with intensity parameter $\lambda(x,y,t)$.

%%%%%%%%%%%%%%%%%%%%%%%%%%%%%%%%%%%2
\item  The $i$-th potential interferer transmits an i.i.d wide band signal, represented by:
\begin{equation}
S^{(i)}(t) = \sum_{k=1}^K X^{(i)}_k \exp\left\{2j\pi f_kt\right\},
\label{transmitted_signal}
\end{equation}
where $f_k$ is the subcarrier frequency and $X^{(i)}_k$ the source symbol from interferer $i$ on subcarrier $k$. 

%$A^{(i)}_k$ is the i.i.d random amplitude and $\Phi_k^{(i)}$ the i.i.d random phase of the information source of the potential $i$-th interferer on the $k$-th frequency band. We do not assume any specific distribution for $A_k^{(i)}$ (which makes our model valid even with power control and a random $g^(i)(t)$) and since we assume no \textit{a priori} knowledge of the phase, we consider $\Phi_k^{(i)} \sim U \left[0 , 2\pi \right]$.

%%%%%%%%%%%%%%%%%%%%%%%%%%%%%3
\item The bandwidth, as quantified by the number of frequency carriers $K$, will be considered as stochastic with a truncated Poisson distribution given by $K \sim \text{Pois} \left(\lambda_K\right)\; \forall K \in \left\{1,\ldots, K_{\text{max}}\right\}$. For example, this is practical if one considers that for all potential wideband interferers we assume one of the following two scenarios: all interferers transmit in the same bandwidth, but this bandwidth is unknown to the receiver and modelled according to a truncated Poisson distribution given by $K \sim \text{Pois} \left(\lambda_K\right)\; \forall K \in \left\{1,\ldots, K_{\text{max}}\right\}$; alternatively, 
all interferers utilize the same total bandwidth, but the frequencies occupied by any given user may not overlap, however, the total  bandwidth per user is unknown to the receiver and modelled according to a truncated Poisson distribution given by $K \sim \text{Pois} \left(\lambda_K\right)\; \forall K \in \left\{1,\ldots, K_{\text{max}}\right\}$. In deriving the theoretical results, the expressions we obtain are interpretable in either of these practical scenarios.

%%%%%%%%%%%%%%%%%%%%%%%%%%%%%%%%%%%4
\item The random distance of the $i$-th potential interferer from the receiver is denoted by $R_i$ and is given by: 
\begin{align}
R_i = \left\| L^{(i)} - l^{R}\right\|,
\end{align}
where $L^{(i)}$ is a random location of the $i$-th potential interferer and $l^{R}$ is a known location of the receiver in region $\mathcal{R}$.
In a first part of the paper we assume that the location of the $i$-th potential interferer for the case of Models 1 and 2 will be uniformly distributed in space, such that the \textit{circular interference domain} is given by:
\begin{align}
\label{uniform_space}
f_{R|N}\left(r|N\right) = 
\begin {cases}
\begin{split}
    &\frac{2 r}{ r_T^2}, \; \text{if} \;\;0\leq r \leq r_T\\
    &0, \; \text{otherwise}.
\end{split}
\end {cases}
\end{align}
where $r_T$ is the maximal distance in which an interfere can have a non-negligible contribution to the interference.

%%%%%%%%%%%%%%%%%%%%%%%%%%%%%5
\item For the $i$-th potential interferer, the low pass representation of the channel experienced by the symbol $X_k^{(i)}$ is given by $\frac{A_k^{(i)}e^{j\Phi_k^{(i)}}}{R_i^{-\sigma/2}}$. 
The path loss experienced by the $i$-th potential interferer is given by $R_i^{-\frac{\sigma}{2}}$, where $\sigma$ is the attenuation coefficient, a deterministic and known parameter reflecting the physical environment in which transmission is occurring. $A_k^{(i)}e^{j\Phi_k^{(i)}}$ is a complex coefficient that contains the shadowing and multipath fading (the amplitude distribution is not important and the phase $\Phi_k^{(i)}$ is uniformly distributed over $[0,2\pi]$).

%%%%%%%%%%%%%%%%%%%%%%%%%%%%%6
\item After the adapted filter at the receiver side, the resulting total interference is given by:
\begin{equation}
\label{Total_interference2}
Y= \sum_{i=1}^N \frac{1}{R_i^{-\sigma/2}} \sum_{k=1}^{K} A_k^{(i)}X^{(i)}_kc^{(i)}_ke^{j\Phi_k^{(i)}}= \sum_{k=1}^{K} \sum_{i=1}^N  Y_{I}^{(k,i)}+
 j \sum_{k=1}^{K}  \sum_{i=1}^N  Y_{Q}^{(k,i)}
\end{equation}
where $c^{(i)}_k$ is a random variable resulting from the filtering adapted for subcarrier $k$ and depends on the system parameters.
%%%%%%%%%%%%%%%%%%%%%%%%

\end{enumerate}

\section{Analytic Distributional Results for the Total Interference in homogeneous PNSC$(\alpha)$ Interference Models}
\label{sec:analytic}
In this section we present the distributional results for the total interference given in (\ref{Total_interference2}). 
We first detail explicitly the results for Model I in which we consider a homogeneous spatial and temporal intensity pattern for the distribution of interferers in the plane. Furthermore, we also consider the second order homogeneity of the process in which the intensity of the utilization of frequencies and therefore the number of occupied frequencies by each of the potential interferers is temporally and spatially homogeneous. This is significantly extended in section \ref{ModelIIModelIII} where we present abridged generalizations of these results for the inhomogeneous in time and space settings. % for Model II and Model III for both the distributions of interferers in the plane and the number of frequencies they utilize in transmission (their bandwidths of transmission).

The results we develop in this section include Theorem \ref{thm_1} which proves that at a given frequency of transmission, the total interference at the receiver experienced by interference from an unknown number of randomly distributed interferers can be shown to be an isotropic $\alpha$-stable model in $\mathbb{C}^2$. 
This result is known, however it is informative to the models and extensions to present clearly a complete derivation in Appendix \ref{theorem_1_proof}.
We then significantly extend this result to provide three possible equivalent analytic representations of the distribution and density functions of the total interference at the receiver. In Theorem \ref{thm_pois_smin} we detail the PNSC$(\alpha)$ model for a wideband system in which an unknown number of interferers, distributed randomly in the plane and transmitting on an unknown number of carrier frequencies is considered. This is achieved through development of representations of the resulting process as a Cox process, which in such models produces a Poisson-SMiN mixture representation. We provide two special examples of members from the family %of such models that we develop 
which characterize two important cases of infinite %mean 
and finite mean interference models. %In Theorem \ref{thm_3} we further study these Cox process models comprised of Poisson-SMiN mixtures and demonstrate analytic results for the tail probabilities of the distribution for the total interference. 

\subsection{Model I: distributional results for total interference under homogeneous Poisson field of 
interferers in a circular domain}
In this section we consider the first model defined according to the following assumptions on the intensity of interferers in the field.

\textbf{Model 1:} In this case we assume the intensity parameter $\lambda(x,y,t) = \lambda$, in other words, the mean number of transmitters does not change over time or space with distribution
\begin{align}
\mathbb{P} \left(N(t) = n\right)  =\frac{\left(\lambda A_{\mathcal{R} }\right)^n }{n !} \exp^{-\lambda A_{\mathcal{R} }}.
\end{align}

Given the system model for the total interference we now present the main result, which is to derive novel representations of the density and distribution in closed form for the total interference in Equation (\ref{Total_interference2}). 

Given a homogeneous spatial Poisson process with intensity parameter $\lambda$, to define the distribution for the stochastic number of interferers in a given region of space,
\begin{align}
\Omega\left(A_{\mathcal{R}} \right) = \left\{\x \in \mathbb{R}^2: 0\leq r\leq r_T \right\}.
\end{align}
In this setting we show the total interference has closed form analytic density represented according to the results derived in Theorem \ref{thm_1}.

We begin by presenting a distributional result for the $k$-th transmission frequency of the $i$-th potential user, based on previously derived results for narrow-band systems in \cite{Pin06}. %\textcolor{blue}{Ido please add eng refs}. 
In this case we can write the characteristic function based on in-phase and quadrature-phase components presented in (\ref{transmitted_signal}) according to definition \ref{Def:CharFn}.

%%%%%%%%%%%%%%%%%%%%%%%%%%%%%%%%%%%%%%%%%%%%%%%%%%%%%%%%%%%%%%%%%%%%%%%%%%%%%%%%%%%%%%%%%%%%%%%%
\begin{definition}
\textsl{The characteristic function of the interference at the $k$-th transmission frequency, from $i$-th potential interferer, is given by:}
\begin{align}
\label{eq:CharFn_Y}
\varphi_{Y_{I}^{(k,i)},Y_{Q}^{(k,i)}}\left(\omega^{(k,i)}_I,\omega^{(k,i)}_Q\right) = 
\exE_{_{Y_{I}^{(k,i)},Y_{Q}^{(k,i)}}}\left[\exp\left(j \omega^{(k,i)}_I  Y_{I}^{(k,i)}+j \omega^{(k,i)}_Q  Y_{Q}^{(k,i)} \right) \right].
\end{align}
\label{Def:CharFn}
\end{definition}
Using \eqref{eq:CharFn_Y}
%this characteristic function 
we can define for the $k$-th transmission frequency the characteristic function for the total interference, given an unknown number of independent potential interferes $N$ in Lemma \ref{lemma_5}.

%%%%%%%%%%%%%%%%%%%%%%%%%%%%%%%%%%%%%%%%%%%%%%%%%%%%%%%%%%%%%%%%%%%%%%%%%%%%%%%%%%%%%%%%%%%%%%%%
\bigskip
\begin{lemma}
\label{lemma_5}
\textit{
The characteristic function of the total interference for the $k$-th transmission frequency, for a random number of $N\left(A_{\mathcal{R}}\right)$ potential interferers, is given by:}
\begin{align}
\begin{split}
&\varphi_{Y_{I}^{(k)},Y_{Q}^{(k)}}\left(\omega^{(k)}_I,\omega^{(k)}_Q\right) 
 = \exE_{\R, \bm{c}_k, \A_k ,\bm{\Phi}_k,N}
\left[
\exp
\left( 
j \sqrt{\left(\omega^{(k)}_I\right)^2+\left(\omega^{(k)}_Q\right)^2 }
\right.
\right. \nonumber \\ 
& \hspace{3cm}  
\left.
\left.
\times\sum_{i=1}^N R_i^{-\sigma/2} A_k^{(i)} c_k^{(i)}  
\cos \left(\Phi_k^{(i)}-\arctan\left(\frac{\omega^{(k)}_Q}{\omega^{(k)}_I}\right) \right) 
\right)
\right]
%\exE_{Y_{I}^{(k)},Y_{Q}^{(k)}}
%\left[\exp\left(j \omega^{(k)}_I   Y_{I}^{(k)}+j \omega^{(k)}_Q  Y_{Q}^{(k)} \right) \right]\\
%%%%%%%%%%%%%%%%%%%%%%%%5
%&=\exE_{\R, \H_k, \A_k ,\bm{\Phi}_k,\bm{\Theta}_k,N}
%\left[\exp\left(j\omega^{(k)}_I \sum_{n=1}^N   R_n^{-\sigma/2} H_k^{(n)} A_k^{(n)}  \cos\left(\Phi_k^{(n)}+\Theta_k^{(n)}  \right)
%+j \omega^{(k)}_Q\sum_{n=1}^N   R_n^{-\sigma/2} H_k^{(n)} A_k^{(n)}   \sin \left(\Phi_k^{(n)}+\Theta_k^{(n)}\right) \right) \right]\\
%%%%%%%%%%%%%%%%%%%%%%%%5
%%%%%%%%%%%%%%%%%%%%%%%%%
\end{split}
\end{align}

\end{lemma}

\bigskip

Given the expression for the CF in Lemma \ref{lemma_5} we marginalize over the unknown number of interferers in region $A_{\mathcal{R}}$ to obtain the result in Lemma \ref{lemma_6}.

%%%%%%%%%%%%%%%%%%%%%%%%%%%%%%%%%%%%%%%%%%%%%%%%%%%%%%%%%%%%%%%%%%%%%%%%%%%%%%%%%%%%%%%%%%%%%%%%
\bigskip
\begin{lemma}
\label{lemma_6}
\textit{
The characteristic function, marginalized with respect to the random number of unknown potential interferers, $N$, at the $k$-th transmission frequency, is given by:}
\begin{align}
\begin{split}
&\varphi_{Y_{I}^{(k)},Y_{Q}^{(k)}}\left(\omega^{(k)}_I,\omega^{(k)}_Q\right)
=\sum_{N=0}^\infty \mathbb{P} \left(N(A_{\mathcal{R} })\right) 
\exE_{\R, \A_k, \bm{c}_k ,\bm{\Phi}_k}
\left[\exp\left(j\sqrt{\left(\omega^{(k)}_I\right)^2+\left(\omega^{(k)}_Q\right)^2 }
\right.\right.  \\
%%%%%%%%%%%%%%%%%%%%%%%%%5
& \hspace{1cm}
\left.
\left. 
\times \sum_{i=1}^N R_i^{-\sigma/2} A_k^{(i)} c_k^{(i)} \cos\left(\Phi_k^{(i)} -
 \arctan\left(\frac{\omega^{(k)}_Q}{\omega^{(k)}_I}\right) \right)
 \right)
   \left.\right| N \in A_{\mathcal{R}}
\right]
%%%%%%%%%%%%%%%%%%%%%%%%%5
%%%%%%%%%%%%%%%%%%%%%%%%%%
\end{split}
\end{align}

\end{lemma}
\normalsize
%%%%%%%%%%%%%%%%%%%%%%%%%%%%%%%%%%%%%%%%%%%%%%%%%%%%%%%%%%%%%%%%%%%%%%%%%%%%%%%%%%%%%%%%%%%%%%%%

%According to assumption 
Equation \eqref{uniform_space} %we have 
gives the spatial distribution of the interferers, %spatially, 
conditional on the number of interferers present at the $k$-th frequency. Furthermore, in many situations (asynchronism, no power control), the signal strengths (included in the term $c_k^{(i)}$) of the interferers can be assumed independent and identically distributed. Hence, the CF for the total interference at frequency $k$ in Lemma \ref{lemma_6} is expressed according to Lemma \ref{lemma_7}.

%%%%%%%%%%%%%%%%%%%%%%%%%%%%%%%%%%%%%%%%%%%%%%%%%%%%%%%%%%%%%%%%%%%%%%%%%%%%%%%%%%%%%%%%%%%%%%%%

\begin{lemma}
\label{lemma_7}
\textit{
Under the assumption that, given $N$ potential interferers in region $A_{\mathcal{R}}$, the spatial distribution of the locations for each of the interferers is given by density in (\ref{uniform_space}) %as uniformly distributed. Therefore
we can express the CF for the total interference at the $k$-th frequency according to:}
%\footnotesize
\begin{equation}
\label{log_CF}
\begin{split}
& \psi_{Y_{I}^{(k)},Y_{Q}^{(k)}}\left(\omega^{(k)}_I,\omega^{(k)}_Q\right) \triangleq 
\log \left(\varphi_{Y_{I}^{(k)},Y_{Q}^{(k)}}\left(\omega^{(k)}_I,\omega^{(k)}_Q\right)\right) \\
%%%%%%%%%%%%%%%%%%%%%%%%%5
& \qquad =
\lambda \pi r_T^2
\left(\exE_{\R, \A_k, \bm{c}_k ,\bm{\Phi}_k}
	\left[ \exp
		\left(j  R^{-\sigma/2} A_k c_k 
		\sqrt{\left(\omega^{(k)}_I\right)^2+\left(\omega^{(k)}_Q\right)^2}
		\right.\right.\right. \nonumber \\
& \qquad \qquad \qquad \qquad \qquad \qquad \left.\left.\left.  
\times 
     \cos\left(\Phi_k - \arctan\left(\frac{\omega^{(k)}_Q}{\omega^{(k)}_I}\right)  \right) 
     \right)   
	\right]-1
\right).
%%%%%%%%%%%%%%%%%%%%%%%%%5
\end{split}
\end{equation}
%\normalsize

\end{lemma}

\bigskip
%%%%%%%%%%%%%%%%%%%%%%%%%%%%%%%%%%%%%%%%%%%%%%%%%%%%%%%%%%%%%%%%%%%%%%%%%%%%%%%%%%%%%%%%%%%%%%%%
\begin{proof}
See Appendix \ref{log_CF_proof}.
\end{proof}
%%%%%%%%%%%%%%%%%%%%%%%%%%%%%%%%%%%%%%%%%%%%%%%%%%%%%%%%%%%%%%%%%%%%%%%%%%%%%%%%%%%%%%%%%%%%%%%%

\bigskip
In Lemma \ref{lemma_8} we re-express the argument of the expectation as a complex series expansion in terms of Bessel functions and then marginalize over the random variable  $\Phi_k$ in the expectation operator. % for random variables $\Phi_k$ and $\Theta_k$.

%%%%%%%%%%%%%%%%%%%%%%%%%%%%%%%%%%%%%%%%%%%%%%%%%%%%%%%%%%%%%%%%%%%%%%%%%%%%%%%%%%%%%%%%%%%%%%%%
\begin{lemma} 
\label{lemma_8}
\textit{
The log CF representation of the total interference at the $k$-th frequency, after marginalizing the random variable for $\Phi_k$, %and $\Theta_k$ 
is given by:
%\footnotesize
%\small
\begin{align}
\begin{split}
&\psi_{Y_{I}^{(k)},Y_{Q}^{(k)}}\left(\omega^{(k)}_I,\omega^{(k)}_Q\right)  =
\lambda \pi r_T^2\left(
\exE_{\R, \A_k, \bm{c}_k}
\left[
J_0\left(R^{-\sigma/2} A_k c_k \sqrt{\left(\omega^{(k)}_I\right)^2+\left(\omega^{(k)}_Q\right)^2 }\right)
\right]
-1\right).
%%%%%%%%%%%%%%%%%%%%%%%%%5
%%%%%%%%%%%%%%%%%%%%%%%%%%
\end{split}
\end{align}
}
\end{lemma}
%\normalsize
%%%%%%%%%%%%%%%%%%%%%%%%%%%%%%%%%%%%%%%%%%%%%%%%%%%%%%%%%%%%%%%%%%%%%%%%%%%%%%%%%%%%%%%%%%%%%%%%
\bigskip
\begin{proof}
See Appendix \ref{lemma_8_proof}
\end{proof}

%%%%%%%%%%%%%%%%%%%%%%%%%%%%%%%%%%%%%%%%%%%%%%%%%%%%%%%%%%%%%%%%%%%%%%%%%%%%%%%%%%%%%%%%%%%%%%%%
\bigskip

Without loss of generality, we assume that the receiver, from which we measure the distances of each potential interferer, lies at the center of region $A_{\mathcal{R}}$. Utilizing Lemma \ref{lemma_5} through to Lemma \ref{lemma_8}, we can now state the following result in Theorem \ref{thm_1} for the CF of the total interference at the $k$-th frequency. 
%%%%%%%%%%%%%%%%%%%%%%%%%%%%%%%%%%%%%%%%%%%%%%%%%%%%%%%%%%%%%%%%%%%%%%%%%%%%%%%%%%%%%%%%%%%%%%%%
\begin{theorem} 
\label{thm_1}
\textit{
The log CF, for the total interference at the $k$-th frequency, for a random number of potential interferers $N$ in a region $A_{\mathcal{R}}$, expressed in Lemma \ref{lemma_8}, can be re-expressed in the form of a CF representing the family of isotropic bivarite $\alpha$-stable distributions $\mathcal{S}(\alpha,0,\gamma,\delta;0)$,
\begin{align}
\psi_{Y_{I}^{(k)},Y_{Q}^{(k)}}\left(\omega^{(k)}_I,\omega^{(k)}_Q\right)
= -\gamma \left|\sqrt{\left(\omega^{(k)}_I\right)^2+\left(\omega^{(k)}_Q\right)^2 }\right|^{\alpha}.
\end{align}
}
\end{theorem}
%%%%%%%%%%%%%%%%%%%%%%%%%%%%%%%%%%%%%%%%%%%%%%%%%%%%%%%%%%%%%%%%%%%%%%%%%%%%%%%%%%%%%%%%%%%%%%%%

\bigskip
\begin{proof}
See Appendix \ref{theorem_1_proof}
\end{proof}
\bigskip
%%%%%%%%%%%%%%%%%%%%%%%%%%%%%%%%%%%%%%%%%%%%%%%%%%%%%%%%%%%%%%%%%%%%%%%%%%%%%%%%%%%%%%%%%%%%%%%%

\begin{remark}
\textit{Theorem \ref{thm_1} presents the bivariate $\log$ characteristic function of the complex random variable  $Y^{(k)}=\sum_{i=1}^N(Y_I^{(k,i)}+Y_Q^{(k,i)})$ in (\ref{Total_interference2}). The resulting characteristic function is a member of the elliptic family of stable distributions %characteristic functions 
for all $0 <\alpha<2$ and $\gamma >0$. In this case we get $\alpha=\frac{4}{\sigma}$ and $\gamma = \lambda \pi 
\exE_{\A_k, \bm{c}_k}
\left[
\left(A_k c_k\right)^{\frac{4}{\sigma}}
\right]
\int_{0}^{\infty}\frac{J_1\left(x\right)}{x^{\frac{4}{\sigma}}} \text{d}x$.
}
\end{remark}
%%%%%%%%%%%%%%%%%%%%%%%%%%%%%%%%%%%%%%%%%%%%%%%%%%%%%%%%%%%%%%%%%%%%%%%%%%%%%%%%%%%%%%%%%%%%%%%%

\bigskip
A consequence of this result is that we may represent the density and distribution functions for $Y^{(k)}$ as a real random vector with the first component corresponding to its real part %of $Y^{(k)}$ 
and the second %component corresponding 
component its imaginary part. % of $Y^{(k)}$.
We can then exploit one of the three common representations of the density for an isotropic bivariate stable vector, given in Lemma \ref{alpha_representation}.
%%%%%%%%%%%%%%%%%%%%%%%%%%%%%%%%%%%%%%%%%%%%%%%%%%%%%%%%%%%%%%%%%%%%%%%%%%%%%%%%%%%%%%%%%%%%%%%%
\begin{lemma}
\label{alpha_representation}
\textit{
Consider a multivariate isotropic $\alpha$-stable random vector $\Y$ of dimension $d$ with scale $\gamma$ and location $\bm{\delta}$. % and $d=\text{dim}\left(\Y\right)$.
%Then, the 
Its density can be represented as:\\ %equivalently as:\\
%\textbf{Representation I - Radial [equation 13, p. 6 \cite{nolan2006multivariate}]:}
%\begin{align}
%f_{\Y}\left(\y\right) = \frac{1}{\gamma^{d}} h\left(\frac{\left|\y-\bm{\delta}\right|}{\gamma}\left.\right| \alpha \right),
%\end{align}
%where $h\left(\cdot\right)$ is a radial function given by 
%\begin{align}
%h\left(r | \alpha\right) = 
%\begin{cases}
%\frac{\Gamma\left(\frac{d}{2}\right)}
%{2 \pi ^{d/2}}r^{1-d} f_{R}\left(r|\alpha, \gamma=1,d\right)&, r >0 \\
%%%%%%%%%%%%%%%%%%%%%%%%%%%%%%%%%%%%%%%%%%%%%%%%%%%%%%%%%%%%%%
%\frac{\Gamma\left(\frac{d}{\alpha}\right)}
%{2^{d-1} \alpha  \pi^{d/2} \Gamma\left(d/2\right)^2}&, r =0 ,
%\end{cases}
%\end{align}
%where we define $R = \left\|\Y\right\|$, and the density $f_{R}\left(r|\alpha, \gamma=1,d\right)$ is given by the expression from Zolotarev \cite{zolotarev1983univariate}
%\begin{align}
%f_R\left(r\right) = \frac{2}{2^{d/2} \Gamma\left(\frac{d}{2}\right)}
%\int_0^{\infty}
%\left(rt\right)^{d/2} J_{d/2-1}\left(r t\right)\exp^{-\gamma^\alpha t^\alpha } \text{d} t,
%\end{align}
%where $J_{v}\left(x\right)$ is the Bessel function of order $v$.\\
%%%%%%%%%%%%%%%%%%%%%%%%%%%%%%%%%%%%%%%%%%%%%%%%%%%%%%%%%%%%%%%%%%
\textbf{Representation I - SMiN}
%As detailed in Lemma \ref{lemma_8} 
The density $f_{\Y}\left(\y\right)$ is given by:
\begin{align}
f_{\Y|\lambda}\left(\y\right) =  N\left(\y; \bm{\delta},\lambda \Sigma \right) ,
\label{eq:repI}
\end{align}
%%%%%%%%%%%%%%%%%%%%%%%%%%%%%%%%%%%%%%%%%%%%%%%%%%%%%%%%%%%%%%%%%
with $\lambda \sim \mathcal{S}_{\alpha/2}\left(\lambda;1,\gamma,0;0 \right)$ which is the $\mathcal{S}_{0}$ parameterisation of Nolan \cite{nolan2001estimation}. \\
%%%%%%%%%%%%%%%%%%%%%%%%%%%%%%%%%%%%%%%%%%%%%%%%%%%%%%%%%%%%%%%%%
\textbf{Representation II - Projection:} \cite{nolan2001estimation}
For every vector $\u \in \mathbb{R}^d$, the one-dimensional projection $\left\langle \u,\Y\right\rangle$ is a univariate $\alpha$-stable symmetric RV with stability index $\alpha$. As detailed in \cite{nolan:2012,nolan2001estimation}, %and \cite{}, 
the projection onto vector $\u$ in the isotropic case is given by the stable univariate random vector:
\begin{align}
\left\langle \u,\Y\right\rangle \sim 
\mathcal{S}_{\alpha}\left(0,\gamma\left(\u\right),\delta \left(\u\right);0 \right).
\label{eq:repII}
\end{align}
By Cramer-Wold these univariate projections characterize the joint distribution, where
$\gamma\left(\cdot\right)$ and $\delta \left( \cdot \right)$ are called projection parameter functions, see definitions in \cite{samorodnitsky1994stable,zolotarev1983univariate} and \cite[Section 2.1]{nolan2001estimation}. In the special case of the isotropic multi-variate $\alpha$-stable model one gets $\forall \u \in \mathbb{R}^d$ the simplification $\gamma\left(\u\right)=\gamma$. 
}
\end{lemma}
%%%%%%%%%%%%%%%%%%%%%%%%%%%%%%%%%%%%%%%%%%%%%%%%%%%%%%%%%%%%%%%%%%%%%%%%%%%%%%%%%%%%%%%%%%%%%%%%
\bigskip
Given \eqref{eq:repI} and \eqref{eq:repII}, we derive the %closed form 
expression for the density and distribution function of the total interference for an unknown number of interferers with unknown bandwidth for a few possible scenarios relating to how each potential interferer utilises the carrier frequencies available. To achieve this we utilise one of the three possible representations discussed in Lemma \ref{alpha_representation}. This is complicated by the fact that we will be working with the Poisson model under a temporal Cox process setting for the bandwidth utilised by each of the potential interferers. %To do this, w
We first need to derive expressions for the distribution and density functions for the $k$-fold convolution integrals obtained from linear combinations of random vector $\widetilde{\Y}^K = \sum_{k=1}^K \Y^{(k)}$:
\begin{align}
\begin{split}
f_{\widetilde{\Y}^K} \left(\y\right) &=
f_{\widetilde{\Y}}^{(K)\ast} \left(\y\right)=
\left(	f_{\Y^1} \ast f_{\Y^2}\ast \cdots \ast f_{\Y^K} 	\right) \left(\y\right)
= \left(f^{(K-1)\ast}_{\Y} \ast f_{\Y}\right)\left(\y\right)\\
&= \int 
f^{(K-1)\ast}_{\Y}\left(\y-\tau\right)  f_{\Y}\left(\tau\right) \text{d} \tau,
\end{split}
\end{align}
and the CDF is expressed according to:
\begin{align}
\begin{split}
F_{\widetilde{\Y}^K} \left(\y\right) &=
F_{\widetilde{\Y}}^{(K)\ast} \left(\y\right)=
\left(	F_{\Y^1} \ast F_{\Y^2}\ast \cdots \ast F_{\Y^K} 	\right) \left(\y\right)
= \left(F^{(K-1)\ast}_{\Y} \ast F_{\Y}\right)\left(\y\right)\\
&= \int 
F^{(K-1)\ast}_{\Y}\left(\y-\tau\right)  f_{\Y}\left(\tau\right) \text{d} \tau.
\end{split}
\end{align}
%%%%%%%%%%%%%%%%%%%%%%%%%%%%%%%%%%%%%%%%%%%%%%%%%%%%%%%%%%%%%%%%%%%%%%%%%%%%%%%%%%%%%%%%%%%%%%%%
\bigskip

\begin{remark}
\textit{
Analytic solutions for these convolution integrals can be obtained since we are working with a special class of RVs characterized by the stable laws. This is a key advantage of stable RVs detailed in the univariate case in Lemma \ref{lemma_2}. Generalizing this result to the bivariate setting developed in Theorem \ref{thm_1} and representations in Lemma \ref{alpha_representation} is non-trivial. %for the first and second representations. 
In order to do it, we %choose to 
work under the projection representation.
}
\end{remark}
%%%%%%%%%%%%%%%%%%%%%%%%%%%%%%%%%%%%%%%%%%%%%%%%%%%%%%%%%%%%%%%%%%%%%%%%%%%%%%%%%%%%%%%%%%%%%%%%

In Theorem \ref{thm_pois_smin} we evaluate the distribution of the total interference in (\ref{Total_interference2}).

%%%%%%%%%%%%%%%%%%%%%%%%%%%%%%%%%%%%%%%%%%%%%%%%%%%%%%%%%%%%%%%%%%%%%%%%%%%%%%%%%%%%%%%%%%%%%%%%
\begin{theorem}{\textbf{\textnormal Compound Poisson PNSC$(\alpha)$ Interference Model}}\\
\textit{ 
\label{thm_pois_smin}
The PNSC$(\alpha)$ Interference Model can be expressed by the distribution of the total interference (indexed by $\alpha$), in the case in which the unknown bandwidth of each potential interferer 
$K \sim \text{Pois} \left(\lambda_K\right),\; \forall K \in \left\{1,\ldots, K_{\text{max}}\right\}$, 
\begin{align}
Y =  \sum_{k=1}^{K} \sum_{n=1}^N  Y_{I}^{(k,n)}+  j \sum_{k=1}^{K}  \sum_{n=1}^N  Y_{Q}^{(k,n)},
\end{align}
can be obtained uniquely by projection onto the real axes (radial symmetry argument).
Therefore, $\forall \u \in \mathcal{R}^2$ the density and distribution of the projection RV $Y\left(\u\right) \triangleq \left\langle Y,\u \right\rangle$, are represented by
\begin{align}
\begin{split}
f_{Y\left(\u\right) }\left(y\right)  
&=
c^{-1}_{K_{\text{max}}}
\sum_{k=1}^{ K_{\text{max}}} \mathbb{P}\left(K=k\right)
f_{\widetilde{\Y}^K\left(\u\right) }\left(y\right)
\mathbb{I}\left(k \leq K_{\text{max}}\right)\\
&=
c^{-1}_{K_{\text{max}}}
\sum_{k=1}^{ K_{\text{max}}}
\exp\left(-\lambda_K\right) \frac{\lambda_K^k}{k!}
\mathcal{S}_{\alpha}\left( \widetilde{\beta}_k\left(\u\right),\widetilde{\gamma}_k\left(\u\right), \widetilde{\delta}_k\left(\u\right);0 \right)
\mathbb{I}\left(k \leq K_{\text{max}}\right),
\end{split}
\label{genericCompoundProc}
\end{align}
%%%%%%%%%%%%%%%%%%%%%%%%%%%%%%%%%%%%
where in the model defined in Theorem \ref{thm_1}, and using Lemma \ref{lemma_2}, the following holds:\\
\begin{equation}
\begin{split}
& \widetilde{\beta}_k\left(\u\right)=\widetilde{\delta}_k\left(\u\right)=0, \; \forall \u \in \mathbb{R}^d \; \text{and} \; k \in \left\{1, \ldots, K_{\text{max}}\right\}, \\
 %$\widetilde{\gamma}_k^{0.5}\left(\u\right) =
 %\sum_{i=1}^{k}\gamma_i^{0.5} =k \gamma^{0.5} $, 
& \widetilde{\gamma}_k^{\alpha}\left(\u\right)=\sum_{i=1}^{k}\gamma_i^{\alpha} =k \gamma^{\alpha} \; \text{and} \; \gamma = \lambda \pi \exE_{\A_k, \bm{c}_k} %{ H_k, A_k}
\left[ \left(A_k c_k\right)^{\frac{4}{\sigma}} \right]
\int_{0}^{\infty}\frac{J_1\left(x\right)}{x^{\frac{4}{\sigma}}} \text{d}x, \\
& c_{K_{\text{max}}} \triangleq \sum_{k=1}^{ K_{\text{max}}} \mathbb{P}\left(K=k\right). \\
\end{split}
%\nonumber
\label{eq:parameters}
\end{equation}
This representation admits the following closed form expressions, for the distribution and density function, according to Lemma \ref{lemma_3}, given by:
\begin{enumerate}
	\item The conditional density of $Y\left(\u\right) | \lambda \in \mathbb{R}$ is given by the following finite mixture of Gaussians,
\begin{align}
\begin{split}
f_{Y\left(\u\right)|\kappa }\left(y\right)  =
c^{-1}_{K_{\text{max}}}
\sum_{k=1}^{ K_{\text{max}}}
\exp\left(-\lambda_K\right) \frac{\lambda_K^k}{k!}
\frac{1}{2 \pi  \sqrt{ \widetilde{\gamma}_k\left(\u\right) \kappa}} \exp\left(-\frac{1}{2  \widetilde{\gamma}_k\left(\u\right) \kappa }
y^2
\right)
\mathbb{I}\left(k \leq K_{\text{max}}\right),
\end{split}
\label{PoissDensityHomogeneous}
\end{align}	
where $\kappa \sim \mathcal{S}_{\alpha/2}\left(0,1,1;0\right).$
%%%%%%%%%%%%%%%%%5
	\item The conditional distribution function of $Y\left(\u\right) | \kappa \in \mathbb{R}$ is given by the following finite mixture of Gaussians,
\begin{align}
\begin{split}
F_{Y\left(\u\right) |\kappa}\left(y\right)  
&= \mathbb{P} \left(Y\left(\u\right) < y |\kappa\right)=
c^{-1}_{K_{\text{max}}}
\sum_{k=1}^{ K_{\text{max}}} 
\exp\left(-\lambda_K\right) \frac{\lambda_K^k}{k!}
\Phi\left(\frac{y}{\sqrt{ \widetilde{\gamma}_k\left(\u\right) \kappa}} |\kappa \right)
\mathbb{I}\left(k \leq K_{\text{max}}\right).
\end{split}
\label{PoissDistributionHomogeneous}
\end{align}
\end{enumerate}
}
\end{theorem}
%%%%%%%%%%%%%%%%%%%%%%%%%%%%%%%%%%%%%%%%%%%%%%%%%%%%%%%%%%%%%%%%%%%%%%%%%%%%%%%%%%%%%%%%%%%%%%%%
\bigskip
\begin{proof}
the proof of \eqref{PoissDensityHomogeneous} and \eqref{PoissDistributionHomogeneous} is obtained in two steps. %of the result for the representation of the distribution and density of the total interference 
First, we take the spherically symmetric bivariate stable distribution derived in Theorem \ref{thm_pois_smin} and utilize the closure under convolution result presented in Lemma \ref{lemma_2} for each Poisson mixture component to obtain the result in (\ref{genericCompoundProc}). 
Then we %of the density in (\ref{PoissDensityHomogeneous}) and distribution function in (\ref{PoissDensityHomogeneous}) 
%are obtained 
use the projection and SMiN representations in Lemma \ref{alpha_representation} to obtain a parametric closed form representation which can be evaluated.
\end{proof}

\begin{example}[\textbf{Distribution of the PNSC($\alpha$) interference model}]
\textsl{We illustrate in Figure \ref{fig:PNSC_05_15} the model with the probability density function when the mean interfering bandwidth increases. We select a useful bandwidth given by $K_{max}=64$, and we choose different mean occupied bandwidths for interferers and plot the corresponding probability density functions (Subplots \textit{a}, \textit{b} and \textit{c}). The dispersion parameter is set to $\gamma=1$ for this illustration.}
\begin{figure}[htbp]
	\centering
		\includegraphics[width=0.99\textwidth]{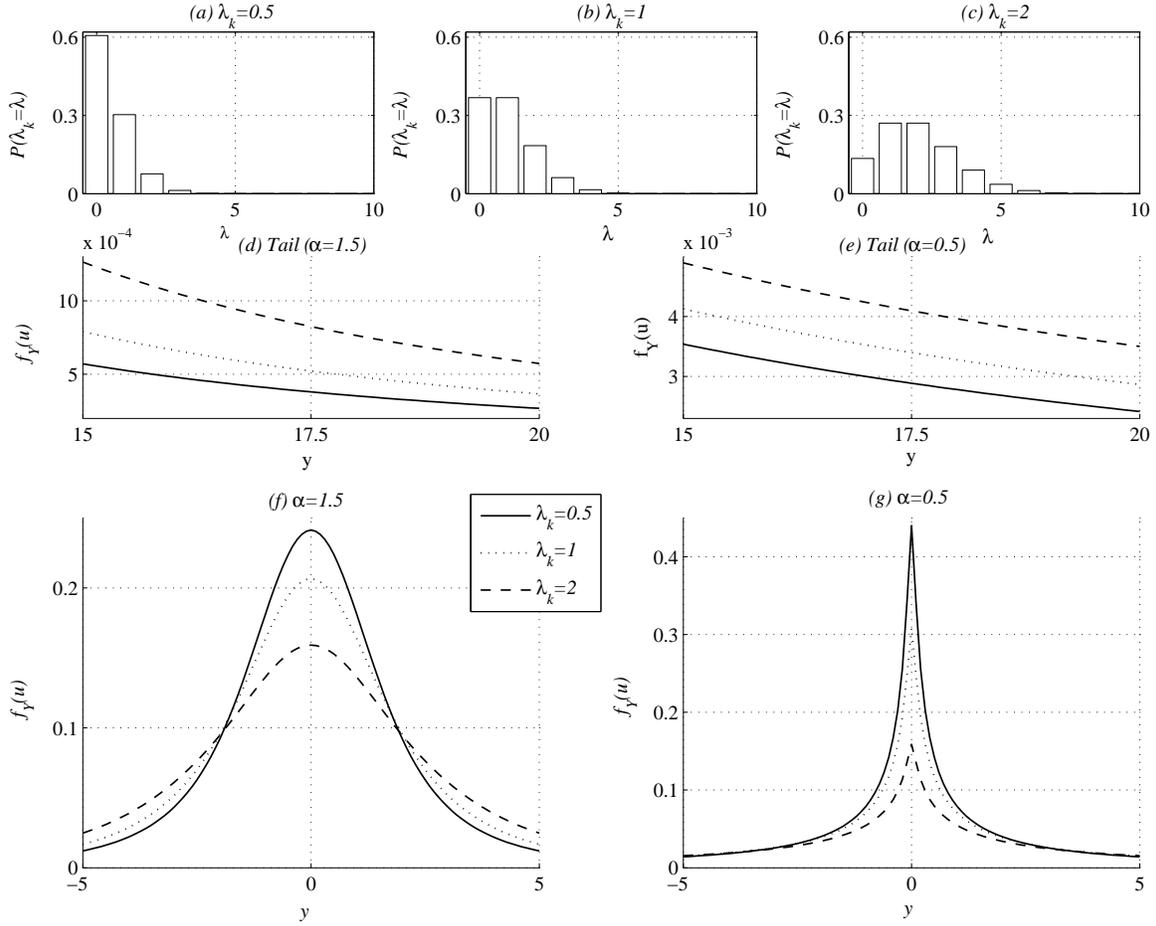}
	\caption{PNSC($\alpha$) model for $\alpha=1.5$ and $\alpha=0.5$ with a useful bandwith determined by the number of carrier frequencies $K_{max}=64$. The subplots \textit{a} to \textit{c} give the distribution of the number of subcarriers used by the interferers. The larger $\lambda_k$, the more bandwidth is used. Then we represent the tails and the center of the interference distribution for $\alpha=1.5$ (subplots \textit{d} and \textit{f}) and for $\alpha=0.5$ (subplots \textit{e} and \textit{g}).}
	\label{fig:PNSC_05_15}
\end{figure}
\end{example}
\textsl{When the bandwidth occupied by interferers increases, the importance of strong users increases as can be seen on figure \ref{fig:PNSC_05_15}, subplots \textit{d} and \textit{e}. As a consequence the peak at the center ($y=0$) is reduced. The impulsiveness of the interference is increased if interfering users (for instance secondary users in cognitive radio) are allowed to use more bandwidth.}

%%%%%%%%%%%%%%%%%%%%%%%%%%%%%%%%%%%%%%%%%%%%%%%%%%%%%%%%%%%%%%%%%%%%%%%%%%%%%%%%%%%%%%%%%%%%%%%%
\begin{remark}
\textit{ 
The result developed in Theorem \ref{thm_pois_smin} is significant as it provides a closed form analytic expression for both the density and distribution functions of the total interference for an unknown number of users and unknown bandwidth which were derived as a compound process. It is well known that in general closed form expressions for the distribution or density functions are generally non-analytic for general compound processes. Typically, evaluation of such distributions can be calculated pointwise via computationally expensive approximations, in the univariate case, via either Monte Carlo approximation or Panjer recursions, see discussions in \cite{peters2011analytic}.  
}
\end{remark}

\bigskip

Next we present two examples for the total interference based on %the analytic representations provided in L
lemma \ref{lemma_3b} which provide results %of illustrative cases of a Poisson number of interferers in the plan, with a 
when the %combined total 
interference %which has 
for each frequency in the bandwidth has either an infinite mean %interference model %in which we consider 
(case $\alpha =2/3$) or a finite mean %interference model 
($\alpha =3/2$).
We assume that the number of interferers per frequency carrier is Poisson distributed with an intensity $\lambda$ and that the total number of frequencies occupied by all users, denoted $k \in \left\{1,2,\ldots,K_{max}\right\}$ is truncated Poisson with intensity $\lambda_K$. 

\begin{example}[\textbf{Compound Poisson PNSC$(3/2)$ Holtsmark Interference Model}]
%\textsl{\\%Here we present a novel interference model which is a member of the family of the PNSC$(\alpha)$ models developed in Theorem \ref{thm_pois_smin}. %In the case in which w
%We assume that the number of interferers per frequency carrier is Poisson distributed with an intensity $\lambda$ and that the total number of frequencies occupied by all users, denoted $k \in \left\{1,2,\ldots,K_{max}\right\}$ is truncated Poisson with intensity $\lambda_K$. Furthermore, f
\textsl{For any projection $\u \in \mathbb{R}^d$ the resulting doubly stochastic Poisson-Stable process is comprised of each mixture component as defined in (\ref{genericCompoundProc}) in which we consider the case:
\begin{align}
%Y(\u)|K=k \sim f_{Y\left(\u\right)|K=k}\left(y\right) &=\mathcal{S}_{2/3}\left(y; 0, \left|c_k\right|\sqrt{k}\gamma,0;0 \right)
Y(\u)|K=k \sim f_{Y\left(\u\right)|K=k}\left(y\right) &=\mathcal{S}_{3/2}\left(y; 0,\sqrt{k}\gamma,0;0 \right)
 \end{align}
with $\gamma$ given in \eqref{eq:parameters}.
%$\gamma = \lambda \pi \exE_{\A_k, \bm{c}_k} %{ H_k, A_k}
%\left[\left(A_k c_k\right)^{\frac{4}{\sigma}}\right] \int_{0}^{\infty}\frac{J_1\left(x\right)}{x^{\frac{4}{\sigma}}} \text{d}x$.
Now defining the transformed random variable for $Y(\u)$ given $K=k$ according to $Z_k = 
%\left|\frac{1}{\left|c_k\right|\sqrt{k}\gamma}\right|Y(\u)|K=k$ 
\left|\frac{1}{\sqrt{k}\gamma}\right|Y(\u)|K=k$ 
and utilizing Lemma \ref{lemma_1} and Lemma \ref{lemma_3b} one obtains the analytic result for the density of $Z$ expressed as a Poisson weighted mixture of Holtsmark densities: 
%\begin{equation}
%\begin{split}
%f_{Z}\left(z\right) &= c^{-1}_{K_{\text{max}}}
%\sum_{k=1}^{ K_{\text{max}}}
%\exp\left(-\lambda_K\right) \frac{\lambda_K^k}{k!} 
%\left\{
%\frac{1}{\pi}\Gamma(5/3)_2F_3\left(\frac{5}{12},\frac{11}{12};\frac{1}{3},\frac{1}{2},\frac{5}{6};-\frac{2^2z^6}{3^6}\right) \right.\\
%& \; \left. -\frac{z^2}{3 \pi} _3F_4\left(\frac{3}{4},1,\frac{5}{4};\frac{2}{3},\frac{5}{6},\frac{7}{6},\frac{4}{3};-\frac{2^2z^6}{3^6}\right) + \frac{7z^4}{3^4\pi}\Gamma(4/3)_2F_3\left(\frac{13}{12},\frac{19}{12};\frac{7}{6},\frac{3}{2},\frac{5}{3};-\frac{2^2z^6}{3^6} \right) \right\}. %\mathbb{I}\left(k \leq K_{\text{max}}\right).
%\end{split}
%\end{equation}
\begin{equation}
\begin{split}
f_{Z}\left(z\right) &= c^{-1}_{K_{\text{max}}}
\sum_{k=1}^{ K_{\text{max}}}
\exp\left(-\lambda_K\right) \frac{\lambda_K^k}{k!} 
\left\{
\frac{1}{\pi}\Gamma(5/3)_2F_3\left(\frac{5}{12},\frac{11}{12};\frac{1}{3},\frac{1}{2},\frac{5}{6};-\frac{2^2z^6}{3^6}\right) 
\right.\\
&
\left. -\frac{z^2}{3 \pi} _3F_4\left(\frac{3}{4},1,\frac{5}{4};\frac{2}{3},\frac{5}{6},\frac{7}{6},\frac{4}{3};-\frac{2^2z^6}{3^6}\right) + \frac{7z^4}{3^4\pi}\Gamma(4/3)_2F_3\left(\frac{13}{12},\frac{19}{12};\frac{7}{6},\frac{3}{2},\frac{5}{3};-\frac{2^2z^6}{3^6} \right) \right\}. %\mathbb{I}\left(k \leq K_{\text{max}}\right).
\end{split}
\end{equation}
}
\end{example} 
\vspace{0.5cm}
\begin{example}[\textbf{Compound Poisson PNSC$(2/3)$ Whittaker Interference Model}]
%\textsl{\\%Here we present a novel interference model which is a member of the PNSC$(\alpha)$ family of models developed in Theorem \ref{thm_pois_smin}. In the case in which w
%We assume that the number of interferers per frequency carrier is Poisson distributed with an intensity $\lambda$ and that the total number of frequencies occupied by all users, denoted $k \in \left\{1,2,\ldots,K_{max}\right\}$ is truncated Poisson with intensity $\lambda_K$. 
%Furthermore, f
\textsl{\\For any projection $\u \in \mathbb{R}^d$ the resulting doubly stochastic Poisson-Stable process is comprised of each mixture component as defined in \eqref{genericCompoundProc} in which we consider the case:
\begin{align}
%Y(\u)|K=k \sim f_{Y\left(\u\right)|K=k}\left(y\right) &=\mathcal{S}_{2/3}\left(y; 0, \left|c_k\right|\sqrt{k}\gamma,0;0 \right)
Y(\u)|K=k \sim f_{Y\left(\u\right)|K=k}\left(y\right) &=\mathcal{S}_{2/3}\left(y; 0,\sqrt{k}\gamma,0;0 \right)
\end{align}
with $\gamma$ given in \eqref{eq:parameters}.
%$\gamma = \lambda \pi \exE_{\A_k, \bm{c}_k} %_{ H_k, A_k}
%\left[\left(A_k c_k\right)^{\frac{4}{\sigma}}\right] \int_{0}^{\infty}\frac{J_1\left(x\right)}{x^{\frac{4}{\sigma}}} \text{d}x$,
Now defining the transformed random variable for $Y(\u)$ given $K=k$ according to $Z_k = \left|\frac{1}{%\left|c_k\right|
\sqrt{k}\gamma}\right|Y(\u)|K=k$ and utilizing Lemma \ref{lemma_1} and Lemma \ref{lemma_3b} one obtains the analytic result for the density of $Z$ expressed as a Poisson weighted mixture of densities constructed from Whittaker functions 
\begin{equation}
\begin{split}
f_{Z}\left(z\right) &= c^{-1}_{K_{\text{max}}}
\sum_{k=1}^{ K_{\text{max}}}
\exp\left(-\lambda_K\right) \frac{\lambda_K^k}{k!} \frac{1}{2\sqrt{3\pi}}|z|^{-1}\exp\left(\frac{2}{27}z^{-2}\right)W_{-1/2;1/6}\left(\frac{4}{27}z^{-2}\right),
\end{split}
\end{equation}
with $W_{\lambda;\mu}\left(z\right)$ given in \eqref{eq:W}.
%\begin{equation}
%W_{\lambda;\mu}\left(z\right)=\frac{z^{\lambda}\exp(-z/2)}{\Gamma\left(\mu - \lambda + 1/2\right)} \int_{0}^{\infty}\exp(-t)t^{\mu - \lambda - 1/2}\left( 1+\frac{t}{z} \right)^{\mu - \lambda - 1/2}dt, \\
%\end{equation}
}
\end{example}

%%%%%%%%%%%%%%%%%%%%%%%%%%%%%%%%%%%%%%%%%%%%%%%%%%%%%%%%%%%%%%%%%%%%%%%%%%%%%%%%%%%%%%%%%%%%%
Next we present a result for the univariate $\alpha$-stable distribution, for an analytic representation of the tail probability of the stable distribution in Lemma \ref{tailProb}. 
\begin{lemma}
\label{tailProb}
\textit{Given a random variable  $X \sim S(\alpha,\beta,\gamma,\delta;0)$ then as $x \rightarrow \infty$ one can write the limiting tail distribution
\begin{equation}
\begin{split}
\mathbb{P}(X>x) &\sim \gamma^{\alpha}c_{\alpha}(1+\beta)x^{-\alpha}, \; \text{as x $\rightarrow \infty$} \\
f_{X}(x|\alpha,\beta,\gamma,\delta;0) &\sim \alpha\gamma^{\alpha}c_{\alpha}(1+\beta)x^{-(\alpha+1)}, \; \text{as x $\rightarrow \infty$.}
\end{split}
\end{equation}
where $c_{\alpha} = \sin\left(\frac{\pi \alpha}{2}\right)\frac{\Gamma(\alpha)}{\pi}$. This result follows from (\cite{nolan:2012},Theorem 1.12)}
\end{lemma}
We can utilise this result to obtain analytic expressions for the tail probability of the Poisson Process for the total interference defined in (\ref{Total_interference2}). This analytic representation, presented in Theorem \ref{thm_pois_tail}, is particularly useful for evaluation of tail probabilities and tail expectations, as may be relevant for calculations of BER, capacity and outage probabilities.
%%%%%%%%%%%%%%%%%%%%%%%%%%%%%%%%%%%%%%%%%%%%%%%%%%%%%%%%%%%%%%%%%%%%%%%%%%%%%%%%%%%%%%%%%%%%%%
 ido to fix
\begin{theorem}
\label{thm_pois_tail}
\textit{ 
The representation of the tail probability of the total interference, in the case in which the unknown bandwidth of each potential interferer 
$K \sim \text{Pois} \left(\lambda_K\right),\; \forall K \in \left\{1,\ldots, K_{\text{max}}\right\}$, defined by the probability 
\begin{align}
\mathbb{P}(Y >  y) =  \mathbb{P}\left(\sum_{k=1}^{K} \sum_{n=1}^N  Y_{I}^{(k,n)}+  j \sum_{k=1}^{K}  \sum_{n=1}^N  Y_{Q}^{(k,n)} > y\right),
\end{align}
can be obtained uniquely by projection onto the real axes (radial symmetry argument).
Therefore, $\forall \u \in \mathbb{R}^2$, the tail of the density and tail probability of the projected RV $Y\left(\u\right) \triangleq \left\langle Y,\u \right\rangle$, are represented as $Y\left(\u\right) \rightarrow \infty$ by,
{\small{
\begin{align}
\begin{split}
f_{Y\left(\u\right) }\left(y\right)  
&\sim c^{-1}_{K_{\text{max}}}
\sum_{k=1}^{ K_{\text{max}}} \exp\left(-\lambda_K\right) \frac{\lambda_K^k}{k!}
\alpha\left(\widetilde{\gamma}_k\left(\u\right)\right)^{\alpha}
c_{\alpha}\left[1+\left(
\widetilde{\beta}_k\left(\u\right)\right)\right]
\left(y\right)^{-(\alpha+1)},
\end{split}
\label{genericCompoundProcTail}
\end{align}
}}
%%%%%%%%%%%%%%%%%%%%%%%%%%
and we may analytically present the tail probability by the limiting survival function given as $Y\left(\u\right) \uparrow \infty$ by,
\begin{align}
\begin{split}
\mathbb{P}(Y\left(\u\right) >  y) &=  \mathbb{P}\left(\sum_{k=1}^{K} \sum_{n=1}^N  Y_{I}^{(k,n)}+  j \sum_{k=1}^{K}  \sum_{n=1}^N  Y_{Q}^{(k,n)} > y\right)\\
&= c^{-1}_{K_{\text{max}}} \sum_{k=1}^{ K_{\text{max}}} \exp\left(-\lambda_K\right) \frac{\lambda_K^k}{k!}
\left( \widetilde{\gamma}_k\left(\u\right)\right)^{\alpha} 
c_{\alpha}(1+\left(\widetilde{\beta}_k\left(\u\right)\right))
y^{-\alpha},
\end{split}
\label{PoissDensityHomogeneousTail}
\end{align}	
%%%%%%%%%%%%%%%%%%%%%%%%%%%%%%%%%%%
where in the model defined in Theorem \ref{thm_1}, and using Lemma \ref{lemma_2}, the following holds:\\
$ \widetilde{\beta}_k\left(\u\right) =  \widetilde{\delta}_k\left(\u\right) = 0, \; \forall \u \in \mathbb{R}^d$, and $k \in \left\{1, \ldots, K_{\text{max}}\right\}$, $\widetilde{\gamma}_k^{\alpha}\left(\u\right) = \sum_{i=1}^{k}\gamma_i^{\alpha} =k \gamma^{\alpha} $, \\
$\gamma = 
\lambda \pi 
\exE_{ H_k, A_k}
\left[
\left(H_k A_k\right)^{\frac{4}{\sigma}}
\right]
\int_{0}^{\infty}\frac{J_1\left(x\right)}{x^{\frac{4}{\sigma}}} \text{d}x$,
$c_{K_{\text{max}}} \triangleq \sum_{k=1}^{ K_{\text{max}}} \mathbb{P}\left(K=k\right)$ and $c_{\alpha} = \sin\left(\frac{\pi \alpha}{2}\right)\frac{\Gamma(\alpha)}{\pi}$.
}
\end{theorem}

%%%%%%%%%%%%%%%%%%%%%%%%%%%%%%%%%%%%%%%%%%%%%%%%%%%%%%%%%%%%%%%%%%%%%%%%%%%%%%%%%%%%%%%%%%%%%%
\begin{proof}
The proof of the result for the representation of the distribution and density of the tail of the total interference follows by taking the spherically symmetric bivariate stable distribution derived in Theorem \ref{thm_1} and utilising the closure under convolution result presented in Lemma \ref{lemma_2} for each Poisson mixture component. Next, the representation of the tail density in (\ref{genericCompoundProcTail}) and tail distribution (limiting survival function) in (\ref{PoissDensityHomogeneousTail}) are obtained using the projection and SMiN representations in Lemma \ref{alpha_representation} to obtain a parametric closed form representation which, in the limit, allows us to apply Lemma \ref{tailProb}. 

\end{proof}

\section{Analytic Distributional Results for the Total Interference in inhomogeneous PNSC$(\alpha)$ Interference Models}
\label{sec:analytic2}
In this section %we present the distributional results for the total Interference given in (\ref{Total_interference2}). We first detail explicitly the results for Model I in which 
we consider a significant extension of the %class of models 
model presented in the previous section to the Doubly Stochastic Poisson-SMiN models. %, which 
It can be considered as stochastic intensity models in the Cox process setting. 
The results in Theorem \ref{thm_pois_gamma_smin} present analytic distribution and density functions for the total interference in the case in which one models an unknown number of potential interferers, distributed uniformly in the plane and each occupying an unknown bandwidth in which the mean bandwidth of the potential interferers is stochastic and given by a Poisson-Gamma conjugacy. We then extend to the case of Model II and Model III %are presented 
in Theorem \ref{thm_5}, in which we consider %generalize the assumptions relating to 
that the intensity of the interferers in the plane and in time %to the cases of 
can be inhomogeneous. % intensity models for space and time in the Poisson process.

%Next we present a result for the case in which we allow the occupied bandwidth of each potential interferer to vary stochastically over time.
%%%%%%%%%%%%%%%%%%%%%%%%%%%%%%%%%%%%%%%%%%%%%%%%%%%%%%%%%%%%%%
%%%%%%%%%%%%%%%%%%%%%%%%%%%%%%%%%%%%%%%%%%%%%%%%%%%%%%%%%%%%%%
\subsection{Model I: stochastic intensity for the bandwidth occupied by interferers.}

\begin{theorem}
\label{1-4}
{\textbf{\textnormal [Doubly Stochastic Poisson-Gamma PNSC$(\alpha)$ %Interference 
Model]}}
\label{thm_pois_gamma_smin} %Here the analytic expressions for the total interference in an Doubly Stochastic Poisson-Gamma-PNSC$(\alpha)$ interference model is developed. 
\textit{\\
Consider the case in which the bandwidth occupied by each potential interferer, quantified by the number of carrier frequencies $K$, is a random variable from inhomogeneous Poisson process: 
$K \sim \text{Pois} \left(\lambda_K\right)\; \forall K \in \left\{1,\ldots, K_{\text{max}}\right\}$, where $\lambda_K \sim \Ga \left(a, b\right)$. 
The distribution of the total interference follows a temporal Cox process (doubly stochastic Poisson process) represented by a compound process given in \eqref{Total_interference2}.
%\begin{align}
%Y =  \sum_{k=1}^{K} \sum_{n=1}^N  Y_{I}^{(k,n)}+  j \sum_{k=1}^{K}  \sum_{n=1}^N  Y_{Q}^{(k,n)}.
%\end{align}
Under any projection, ($\forall \u \in \mathbb{R}^2$), the density and distribution of the projection $Y\left(\u\right) = \left\langle Y,\u \right\rangle$, are represented by:
%\small
\begin{align}
f_{Y\left(\u\right) }\left(y\right)  
=
d^{-1}_{K_{\text{max}}}
\sum_{k=1}^{ K_{\text{max}}} 
&\frac{\left(a+k-1\right)!}{\left(a-1\right)! k!}
\left(\frac{b}{1+b}\right)^a
\left(\frac{1}{1+b}\right)^k  \mathcal{S}_{\alpha}\left(y; 
\widetilde{\beta}_k\left(\u\right), 
\widetilde{\gamma}_k\left(\u\right), 
\widetilde{\delta}_k\left(\u\right);0 \right).
%&\times \mathcal{S}_{\alpha}\left(y; \text{sgn}\left(c_k\right)
%\widetilde{\beta}_k\left(\u\right), \left|c_k\right|\widetilde{\gamma}_k\left(\u\right), c_k \widetilde{\delta}_k\left(\u\right);0 \right).
%\mathbb{I}\left(k \leq K_{\text{max}}\right).
\label{genericCoxProc1}
\end{align}
%\normalsize
%%%%%%%%%%%%%%%%%%%%%%%%%%%%%%%%%%%%
where the parameters were defined in \eqref{eq:parameters} except:
%
%in the model defined in Theorem \ref{thm_1}, and using Lemma \ref{lemma_2}, the following holds:
\begin{equation}
%& \widetilde{\beta}_k\left(\u\right) =  \widetilde{\delta}_k\left(\u\right) = 0, \forall \u \in \mathbb{R}^d \quad \text{and} \quad k \in \left\{1, \ldots, K_{\text{max}}\right\}, \nonumber \\
%& \widetilde{\gamma}_k^{\alpha}\left(\u\right) = \sum_{i=1}^{k}\gamma_i^{\alpha} =k \gamma^{\alpha}, \nonumber \\
%& \gamma = 
%\lambda \pi 
%\exE_{\A_k, \bm{c}_k} %{ H_k, A_k}
%\left[
%\left(A_k c_k\right)^{\frac{4}{\sigma}}
%\right]
%\int_{0}^{\infty}\frac{J_1\left(x\right)}{x^{\frac{4}{\sigma}}} \text{d}x, 
%\nonumber \\
d_{K_{\text{max}}} = 1- \sum_{k=1}^{ K_{\text{max}}} \mathbb{P}\left(K=k\right)
=1- \sum_{k=1}^{ K_{\text{max}}} \frac{\left(a+k-1\right)!}{\left(a-1\right)! k!}
\left(\frac{b}{1+b}\right)^a
\left(\frac{1}{1+b}\right)^k 
\end{equation}
%%%%%%%%%%%%%%%%%%%%%%%%%%%%%%%%%%%%
%%%%%%%%%%%%%%%%%%%%%%%%%%%%%%%%%%%%
%%%%%%%%%%%%%%%%%%%%%%%%%%%%%%%%%%%%
%%%%%%%%%%%%%%%%%%%%%%%%%%%%%%%%%%%%
%%%%%%%%%%%%%%%%%%%%%%%%%%%%%%%%%%%%
%%%%%%%%%%%%%%%%%%%%%%%%%%%%%%%%%%%%
%%%%%%%%%%%%%%%%%%%%%%%%%%%%%%%%%%%%
This representation admits the following closed form expressions, for the distribution and density function, according to Lemma \ref{lemma_7}, given by:
\begin{enumerate}
	\item The conditional density of $Y\left(\u\right) | \lambda \in \mathbb{R}$ is given by	the following finite mixture of normals:
%	\small
\begin{align}
\label{PoissDensityCox}
f_{Y\left(\u\right)|\lambda }\left(y\right)  =
d^{-1}_{K_{\text{max}}}
\sum_{k=1}^{ K_{\text{max}}} 
\frac{\left(a+k-1\right)!}{\left(a-1\right)! k!}
& \left(\frac{b}{1+b}\right)^a
\left(\frac{1}{1+b}\right)^k 
 %\frac{1}{2 \pi  \sqrt{\left|c_k\right| \widetilde{\gamma}_k\left(\u\right) \lambda}} \exp\left(-\frac{1}{2 \left|c_k\right| \widetilde{\gamma}_k\left(\u\right) \lambda} y^2 \right),
\frac{1}{2 \pi \sqrt{\widetilde{\gamma}_k\left(\u\right) \lambda}} \exp\left(-\frac{1}{2\widetilde{\gamma}_k\left(\u\right) \lambda }
y^2 \right),
\end{align}	
%\normalsize
where $\lambda \sim \mathcal{S}_{\alpha/2}\left(0,1,1;0\right).$
%%%%%%%%%%%%%%%%%5
	\item The conditional distribution function of $Y\left(\u\right) | \lambda \in \mathbb{R} $ is given by	the following finite mixture of normals,
%	\small
\begin{align}
F_{Y\left(\u\right)| \lambda} & \left(y\right)  
= \mathbb{P} \left(Y\left(\u\right) < y |\lambda\right) \nonumber \\
&=
d^{-1}_{K_{\text{max}}}
\sum_{k=1}^{ K_{\text{max}}} 
\frac{\left(a+k-1\right)!}{\left(a-1\right)! k!}
\left(\frac{b}{1+b}\right)^a
\left(\frac{1}{1+b}\right)^k
\Phi\left(\frac{y}{\sqrt{\widetilde{\gamma}_k\left(\u\right) \lambda}} |\lambda \right),
%\Phi\left(\frac{y}{\sqrt{\left|c_k\right| \widetilde{\gamma}_k\left(\u\right) \lambda}} |\lambda \right),
\label{PoissDistributionCox}
\end{align}
%\normalsize
with 
$F_{Y\left(\u\right) |\lambda}\left(0\right) = \mathbb{P} \left(Y\left(\u\right) =0|\lambda\right)=\exp\left(-\lambda\right)$.
\end{enumerate}
}
\end{theorem}
%%%%%%%%%%%%%%%%%%%%%%%%%%%%%%%%%%%%%%%%%%%%%%%%%%%%%%%%%%%%%%%%%%%%%%%%%%%%%%%%%%%%%%%%%%%%%%%%
\begin{proof}
See Appendix \ref{theorem_4}.
\end{proof}

%%%%%%%%%%%%%%%%%%%%%%%%%%%%%%%%%%%%%%%%%%%%%%%%%%%%%%%%%%%%%%%%%%%%%%%%%%%%%%%%%%%%%%%%%%%%%%%%
\begin{remark}
\textit{ 
The result developed in Theorem \ref{thm_pois_smin} is significant as it provides a closed form analytic expression for both the density and distribution functions of the total interference for an unknown number of users and unknown bandwidth.
It is well known that %in general 
closed form expressions for the distribution or density functions of such doubly stochastic compound process are generally non-analytic. Typically, %evaluation of 
such distributions can be %calculated 
evaluated point wise via computationally expensive approximations, in the univariate case, via Monte Carlo approximations.
}
\end{remark}

\bigskip

As special cases of the general analytic results for the distribution of the total interference presented in Theorem \ref{thm_pois_gamma_smin}, %two examples 
we present in the following analytic channel model results for %distributional members corresponding to 
the infinite mean %interference case 
and the finite mean interference models in which the number of interferers in domain $\mathcal{A}_R$ is stochastic and Poisson distributed and furthermore the number of carrier frequencies they occupy is also stochastic and modeled according to a doubly stochastic Poisson-Gamma-Stable process (i.e. a Cox-Gamma-Stable Process - %) which is 
a special form of a renewal process): % in which we obtain analytic representations for the channel model for the total interference in each case.
the number of interferers per frequency carrier is Poisson distributed with an intensity parameter $\lambda$ and that the total number of frequencies occupied by all users, denoted $k \in \left\{1,2,\ldots,K_{max}\right\}$ is truncated Poisson with a stochastic intensity parameter $\lambda_K \sim \mathcal{G}a(a,b)$.

\begin{example}[\textbf{Doubly Stochastic %Poisson-Gamma-
PNSC$(2/3)$-Holtsmark Interference}]
%\textsl{Here we present a novel channel model which is a member of the family of models developed in Theorem \ref{thm_pois_smin}. In the case in which we assume that the number of interferers per frequency carrier is Poisson distributed with an intensity parameter $\lambda$ and that the total number of frequencies occupied by all users, denoted $k \in \left\{1,2,\ldots,K_{max}\right\}$ is truncated Poisson with a stochastic intensity parameter $\lambda_K \sim \mathcal{G}a(a,b)$. Furthermore, f
\textsl{\\For any projection $\u \in \mathbb{R}^d$ the resulting doubly stochastic Poisson-Stable process is comprised of each mixture component as defined in \eqref{genericCompoundProc} in which we consider the case:
\begin{align}
%Y(\u)|K=k \sim f_{Y\left(\u\right)|K=k}\left(y\right) &=\mathcal{S}_{2/3}\left(y; 0, \left|c_k\right|\sqrt{k}\gamma,0;0 \right)
Y(\u)|K=k \sim f_{Y\left(\u\right)|K=k}\left(y\right) &=\mathcal{S}_{2/3}\left(y; 0,\sqrt{k}\gamma,0;0 \right) 
\end{align}
with $\gamma$ given in \eqref{eq:parameters}.
%$\gamma = \lambda \pi \exE_{\A_k, \bm{c}_k}\left[\left(A_k c_k\right)^{\frac{4}{\sigma}}\right] \int_{0}^{\infty}\frac{J_1\left(x\right)}{x^{\frac{4}{\sigma}}} \text{d}x$. 
Defining the transformed random variable for $Y(\u)$ given $K=k$ according to %$Z_k = \left|\frac{1}{\left|c_k\right|\sqrt{k}\gamma}\right|Y(\u)|K=k$ 
$Z_k = \left|\frac{1}{\sqrt{k}\gamma}\right|Y(\u)|K=k$ and utilizing Lemma \ref{lemma_1} and Lemma \ref{lemma_3b} one obtains the analytic result for the density of $Z$ expressed as a Poisson weighted mixture of Holtsmark densities 
%\begin{align}
%f_{Z}\left(z\right) = c^{-1}_{K_{\text{max}}}
%\sum_{k=1}^{ K_{\text{max}}}
%& \frac{\left(a+k-1\right)!}{\left(a-1\right)! k!}
%\left(\frac{b}{1+b}\right)^a
%\left(\frac{1}{1+b}\right)^k \nonumber \\
%& \qquad \times 
%\left\{
%\frac{1}{\pi} \Gamma(5/3)_2F_3\left(\frac{5}{12},\frac{11}{12};\frac{1}{3},\frac{1}{2},\frac{5}{6};-\frac{2^2z^6}{3^6}\right) \right.  \nonumber \\
%& \qquad \qquad -
%\frac{z^2}{3 \pi} 
%\; _3F_4\left(\frac{3}{4},1,\frac{5}{4};\frac{2}{3},\frac{5}{6},\frac{7}{6},\frac{4}{3};-\frac{2^2z^6}{3^6}\right)
%\nonumber \\
%& \qquad \qquad \left. + \frac{7z^4}{3^4\pi}\Gamma(4/3)_2F_3\left(\frac{13}{12},\frac{19}{12};\frac{7}{6},\frac{3}{2},\frac{5}{3};-\frac{2^2z^6}{3^6} \right) \right\} %\mathbb{I}\left(k \leq K_{\text{max}}\right).
%\end{align}
\begin{align}
\begin{split}
f_{Z}\left(z\right) &= c^{-1}_{K_{\text{max}}}
\sum_{k=1}^{ K_{\text{max}}}
\frac{\left(a+k-1\right)!}{\left(a-1\right)! k!}
\left(\frac{b}{1+b}\right)^a
\left(\frac{1}{1+b}\right)^k 
\left\{
\frac{1}{\pi} \Gamma(5/3)_2F_3\left(\frac{5}{12},\frac{11}{12};\frac{1}{3},\frac{1}{2},\frac{5}{6};-\frac{2^2z^6}{3^6}\right) \right.\\
&\left.
-
\frac{z^2}{3 \pi} 
\; _3F_4\left(\frac{3}{4},1,\frac{5}{4};\frac{2}{3},\frac{5}{6},\frac{7}{6},\frac{4}{3};-\frac{2^2z^6}{3^6}\right)
+ \frac{7z^4}{3^4\pi}\Gamma(4/3)_2F_3\left(\frac{13}{12},\frac{19}{12};\frac{7}{6},\frac{3}{2},\frac{5}{3};-\frac{2^2z^6}{3^6} \right) \right\} %\mathbb{I}\left(k \leq K_{\text{max}}\right).
\end{split}
\end{align}
}
\end{example}

\begin{example}[\textbf{Doubly Stochastic %Poisson-Gamma-
PNSC$(3/2)$-Whittaker Interference}]
%\textsl{\\Here we present a novel channel model which is a member of the family of models developed in Theorem \ref{thm_pois_smin}. In the case in which we assume that the number of interferers per frequency carrier is Poisson distributed with an intensity $\lambda$ and that the total number of frequencies occupied by all users, denoted $k \in \left\{1,2,\ldots,K_{max}\right\}$ is truncated Poisson with intensity $\lambda_K$. Furthermore, f
\textsl{\\For any projection $\u \in \mathbb{R}^d$ the resulting doubly stochastic Poisson-Stable process is comprised of each mixture component as defined in \eqref{genericCompoundProc} in which we consider the case:
\begin{align}
%Y(\u)|K=k \sim f_{Y\left(\u\right)|K=k}\left(y\right) &=\mathcal{S}_{3/2}\left(y; 0, \left|c_k\right|\sqrt{k}\gamma,0;0 \right)
Y(\u)|K=k \sim f_{Y\left(\u\right)|K=k}\left(y\right) &=\mathcal{S}_{3/2}\left(y; 0,\sqrt{k}\gamma,0;0 \right)
\end{align}
with $\gamma$ given in \eqref{eq:parameters}.
%with 
%$\gamma = \lambda \pi \exE_{\A_k, \bm{c}_k}\left[\left(A_k c_k\right)^{\frac{4}{\sigma}}\right] \int_{0}^{\infty}\frac{J_1\left(x\right)}{x^{\frac{4}{\sigma}}} \text{d}x$. 
Defining the transformed random variable for $Y(\u)$ given $K=k$ according to 
%$Z_k = \left|\frac{1}{\left|c_k\right|\sqrt{k}\gamma}\right|Y(\u)|K=k$ 
$Z_k = \left|\frac{1}{\sqrt{k}\gamma}\right|Y(\u)|K=k$ 
and utilizing Lemma \ref{lemma_1} and Lemma \ref{lemma_3b} one obtains the analytic result for the density of $Z$ expressed as a Poisson weighted mixture of densities constructed from Whittaker functions 
\begin{equation}
\begin{split}
f_{Z}\left(z\right)= c^{-1}_{K_{\text{max}}}
\sum_{k=1}^{ K_{\text{max}}}
& \frac{\left(a+k-1\right)!}{\left(a-1\right)! k!}
\left(\frac{b}{1+b}\right)^a
\left(\frac{1}{1+b}\right)^k \\
& \times
\frac{1}{2\sqrt{3\pi}}|z|^{-1}\exp\left(\frac{2}{27}z^{-2}\right)W_{-1/2;1/6}\left(\frac{4}{27}z^{-2}\right),
\end{split}
\end{equation}
with $W_{\lambda;\mu}\left(z\right)$ given in \eqref{eq:W}.
%
%with
%\begin{equation}
%W_{\lambda;\mu}\left(z\right)=\frac{z^{\lambda}\exp(-z/2)}{\Gamma\left(\mu - \lambda + 1/2\right)} \int_{0}^{\infty}\exp(-t)t^{\mu - \lambda - 1/2}\left( 1+\frac{t}{z} \right)^{\mu - \lambda - 1/2}dt, \\
%\end{equation}
}
\end{example}

%%%%%%%%%%%%%%%%%%%%%%%%%%%%%%%%%%%%%%%%%%%%%%%%%%%%%%%%%%%%%%%%%%%%%%%%%%%%%%%%%%%%%%%%%%
\subsection{Models II \& III: distributional results for total interference under inhomogeneous Poisson field of interferers in a circular domain} 
\label{ModelIIModelIII}

In this section we generalize the model assumptions related to the functional form of the distribution of the interferers in time and space according to one of the following two model choices.

\textbf{Model 2:} we assume the intensity parameter is $\lambda(x,y,t) = \lambda(t)$. In other words, the mean number of transmitters is inhomogeneous in time and homogeneous in space with distribution:
\begin{align}
\mathbb{P} \left(\left[N(t)- N(t-\tau)\right]= n\right) = \frac{\left(\lambda_{\left[t-\tau,t\right]} A_{\mathcal{R} }\right)^n }{n !} e^{-\lambda_{\left[t-\tau,t\right]} A_{\mathcal{R} }}
\end{align}
where $\lambda_{\left[t-\tau,t\right]} = \int_{t-\tau}^t \lambda(t) dt$.\\
	
\textbf{Model 3:} we assume the intensity parameter is $\lambda(x,y,t) = \lambda(x,y)$, in other words, the mean number of transmitters is homogeneous in time and inhomogeneous in space.	
%%%%%%%%%%%%%%%%%%%%%%%
We consider the Borel $\sigma$-field of $\mathbb{R}^2$ in order to define a spatial Poisson inhomogeneous point process (SPIP). 
This allows one to define on a measurable space $S \subseteq \mathbb{R}^2$, a random countable subset $\Pi$. %of $S$. 
It will be governed by a stochastic mechanism that induces two properties for random variables $N(A) =\left|  \Pi \cap A\right|$,
i.e., the number of points of $\Pi$ lying in measurable subsets $A$ of $S$. Specifically, for any
finite collection $A_1, \ldots,A_k$ of pairwise disjoint measurable subsets of $S$, the random variables
$N(A_1),\ldots, N(A_k)$ are independent and for any measurable subset $A$ of $S$, $N(A)$ follows
a Poisson distribution with mean $\int_{A} \lambda(\x) d\x$. Here, $\lambda(\x)$ is the intensity function of the
spatial inhomogeneous Poisson process, a non-negative measurable function defined on $S$ such that $\int_{A} \lambda(\x) d\x <1$
for all bounded subsets $A$ of $S$. Thus the distributional properties of a SPIP
are determined by its intensity function $\lambda(\cdot)$, or, equivalently, by the mean measure of
the process $\Lambda(A)$,  defined for all measurable subsets $A$ of $S$. For theoretical
background on spatial Poisson processes, see, for instance, \cite{rathbun1994space,kingman1993poisson,daley2007introduction}. 
This results in a %\textit{spatial Poisson inhomogeneous point process} 
SPIP distribution given by:
%%%%%%%%%%%%%%%%%%%%%%%%
\begin{align}
\mathbb{P} \left(N(A_{\mathcal{R} }) = n\right) = \frac{\left(\Lambda(A_{\mathcal{R} }) \right)^n }{n !} e^{-\Lambda (A_{\mathcal{R} }) }.
\end{align}

%%%%%%%%%%%%%%%%%%%%%%%%%%%%%%%%%%%%
%%%%%%%%%%%%%%%%%%%%%%%%%%%%%%%%%%%%

Given the system model for the total interference we now present the main result, which is to derive novel representations of the density and distribution in closed form for the total interference in \eqref{Total_interference2}. 

Consider a setting in which we have a \textbf{inhomogeneous} temporal or spatial Poisson process with intensity parameter $\lambda(x,y,t)$ which defines the stochastic number of interferers in a given region of space at a given time interval. In the first instance we consider the parametric interference space defined by:
\begin{align}
\Omega\left(A_{\mathcal{R}} \right) = \left\{\x \in \mathbb{R}^2: 0\leq r\leq r_T \right\}.
\end{align}

Under the assumptions of Model II and Model III, %in Section \ref{ModelIIModelIII}, 
we generalize the analytic distributional results for the total interference developed for the homogeneous Poisson model derived in Theorem \ref{1-4}.
We utilize a generic property of Poisson processes provided in Lemma \ref{lemma_poisson_mapping} to re-obtain analogous distribution results under inhomogeneous Poisson models.

\begin{lemma}
\label{lemma_poisson_mapping}
\textit{
According to the mapping Theorem of \cite[pp. 18]{kingman1993poisson}, any inhomogeneous Poisson processes can be made homogeneous in space or time via a suitable monotonic transformation.
Consider such inhomogeneous Poisson point process ($\Pi$) in $\mathbb{R}^d \times \mathbb{R}^+$ with mean intensity measure $\lambda (\cdot)$ defined over a spatial region $A_R \in \mathbb{R}^d$ and time interval $\tau \in \mathbb{R}^+$ with $\lambda\left(A_R,\tau\right) = \int_{\tau}\int_{A_R} \lambda(x,y,t) \; \text{d} t \; \text{d} x \; \text{d} y$  and define a mapping $f: \mathbb{R}^d \times \mathbb{R}^+ \rightarrow \mathcal{T}$.
Given a smooth bijective function $f$ then the transformation of the Poisson process $f\left(\Pi\right) $ is itself a Poisson process on $\mathcal{T}$ with induced intensity measure $\lambda^*$ given for any $B \subseteq \mathcal{T}$
\begin{align}
\lambda^*\left(B\right) = \lambda\left(f^{-1}\left(B\right)\right).
\end{align}
}
\end{lemma}

Under Model II in which the intensity function is time inhomogeneous as specified by $\lambda\left(x,y,t\right)=\lambda\left(t\right)$,
one may utilise the result in Lemma \ref{lemma_poisson_mapping} to transform the unknown number of interferers which is time varying into a time homogeneous Poisson process. The same is true for Model III in which the intensity function is space inhomogeneous as specified by $\lambda\left(x,y,t\right)=\lambda\left(x,y\right)$.
We illustrate this concept for Model III with the results presented in the following Theorem:
\begin{theorem}
%\label{model_2}
\label{thm_5}
\textit{
Consider Model III in which the unknown number of potential interferers is a spatially inhomogeneous Poisson process with spatial intensity function given by one of the following two scenarios:}
\begin{enumerate}
	\item
	\textit{\textbf{Scenario I:} consider a mean spatial intensity for the number of potential interferers that decays according to a power law as a function of distance from the base-station. Therefore working in polar coordinates with $\lambda\left(r,\phi\right)=\lambda_0 r^{\beta-2}$, for some coefficient of decay $\beta$, base intensity measure $\lambda_0$ and $r > 0$ . Defining $\omega =r^{\beta}$ we have the mapping from polar to Cartesian coordinates given by $x=\omega^{1/\beta} \cos\left(\phi\right)$ and $y=\omega^{1/\beta} \sin\left(\phi\right)$.
Therefore when the inhomogeneous Poisson process, with intensity function $\lambda\left(r\right)$, is mapped to the plane $\left(\omega, \phi\right)$, we obtain the intensity measure in the plane given by:}
\begin{align}
\begin{split}
\lambda^* &= \int \int \lambda\left(r\right) \text{d} x \;\text{d} y =
\int \int \frac{1}{\beta} r^{\beta-2} \omega^{-\frac{\beta-2}{\beta}} \text{d} \omega \;\text{d} \phi.
\end{split}
\end{align}
\textit{Therefore ignoring $\phi$ the homogeneous Poisson process $r^{\beta}$ has rate $\lambda^*=\lambda_0\left(\frac{2 \pi}{\beta}\right)$.}
%%%%%%%%%%%%%%%%
\item{\textbf{Scenario II:} \textit{consider situations where we have directional antennas. In this case we work with a restricted portion (sector) of the plane in which potential interferers may be present. Furthermore, assume that the interferers are Poisson distributed with mean intensity $\lambda$ restricted to this sector. Working in polar coordinates we consider this sector of the plane traced out by angle $\phi$, and we map the Poisson process in this sector to homogeneous Poisson process on the whole plane with rate:} 
\begin{align}
\lambda^*= \lambda\left(\frac{1}{2 \pi}\int_0^{\phi} \text{d}\phi\right)=
\lambda\left(\frac{\phi}{2 \pi}\right).
\end{align}}
\end{enumerate}
 %%%%%%%%%%%%%
\textit{Therefore, working under this transformed Poisson process for the potential number of interfrers in the plane, we obtain the $\log$ CF for the total interference at the $k$-th frequency, given by:}
\begin{align}
\begin{split}
&\psi_{Y_{I}^{(k)},Y_{Q}^{(k)}}\left(\omega^{(k)}_I,\omega^{(k)}_Q\right) \\
& \quad =
\lambda^* \pi r_T^2\left(
\exE_{\R,\A_k,\bm{c}_k}
\left[
J_0\left(R^{-\frac{\sigma}{2}} A_kc_k\sqrt{\left(\omega^{(k)}_I\right)^2+\left(\omega^{(k)}_Q\right)^2 }\right)
\right]
-1\right).
%%%%%%%%%%%%%%%%%%%%%%%%%5
%%%%%%%%%%%%%%%%%%%%%%%%%%
\end{split}
\end{align}

\end{theorem}
%%%%%%%%%%%%%%%%%%%%%%%%%%%%%
\begin{proof} The proof of this result proceeds directly according to the result in Theorem \ref{theorem_1_proof}. \end{proof}

\begin{corollary}
According to Theorem \ref{thm_5}, %\ref{model_2}, 
the results stated in Theorems \ref{thm_pois_smin} %, \ref{thm_pois_tail} 
and \ref{thm_pois_gamma_smin} can be easily derived under the transformed homogeneous Poisson process for the number of potential interferers in the plane.
\end{corollary}

%%%%%%%%%%%%%%%%%%%%%%%%%%%%%%%%%%%%%%%%%%%%%%%%%%%%%%%%%%%%%%%%%%%%%%%%%%%%%%%%%%%%%%%%%%%%%%%%%%%%%%%%%%%%%%%%%%%%%%%%%%%%%%%%%%%%%%%%%%%%%%%%%%%%%%%%%%%%%%%%%%%%%%%%%%%%%%%%%%%%%%%%%%%%%%%%%%%%%%%%%%%%%%%%%%%
\section{Generalised SNR in PNSC$(\alpha)$ Interference Models}
\label{sec:analyticCapacity}
%%%%%%%%%%%%%%%%%%%%%%%%%%%%%%%%%%%%%%%%%%%%%%%%%%%%%%%%%%%%%%%%%%%%%%%%%%%%%%%%%%%%%%%%%%%%%%%%%%%%%%%%%%%%%%%%%%%%%%%%%%%%%%%%%%%%%%%%%%%%%%%%%%%%%%%%%%%%%%%%%%%%%%%%%%%%%%%%%%%%%%%%%%%%%%%%%%%%%%%%%%%%%%%%%%%
In this section we consider an extension to the model proposed in \cite{ben2010asymptotic} which develops the concept of an Additive White Symmetric $\alpha$-stable Noise (AWS$\alpha$SN) model. In this section we consider extending the analyse they performed to the generalized model developed in this paper. We now allow for an unknown number of potential interferers who are each transmitting on an unknown random bandwidth and interterfering from uniformly and unknown locations in the plane as specified in the results developed in Theorem \ref{thm_1} and Theorem \ref{thm_pois_smin}.  Again throughout this section we consider the results in the case of projection $Y(\u)$, which hold for all projections $\u \in \mathbb{C}$.

\subsection{Generalised Geometric SNR in PNSC$(\alpha)$ Interference Models}

To begin the definition of the Geometric Signal-to-Noise Ratio (GSNR) for the PNSC$(\alpha)$ interference model in a wideband system. We first define the following properties of a stable distributed random variable $Y$ known as the fractional moments as defined for symmetric $\alpha$-stable models as follows.

\begin{definition}
\label{lemma:FLOMs}
\textbf{$\alpha$-Stable Fractional Lower Order Moments (S(0)-FLOM's):}\\
\textit{
Given a sub-exponential $\alpha$-Stable distribution $Y \sim S_{\alpha}\left(y;\beta,\gamma,\delta;S(0)\right)$ 
then the FLOM can be evaluated analytically according to the following moments or Fractional Lower Order Moment conditions for $p \in (-1,2)$, 
\begin{equation}
\mathbb{E}\left[|Y|^p\right] = \begin{cases} 
+\infty, & \; p > \alpha\\
2\Gamma(1+p)\sin\frac{\pi}{2}p\int_{0}^{\infty}(1-\mathrm{Re}\left\{\Phi(t)\right\})t^{p-1}dt, & \; 0 < p < \alpha < 1, \beta \neq 0\\
C(p,\alpha)\gamma^{\frac{p}{\alpha}}, & \; -1 < p < \alpha < 2, \beta = 0,
\end{cases}
\end{equation}
where $\Phi(t)$ represents the characteristic function of r.v. $Y$ and
$$C(p,\alpha) = \frac{2^{p+1}\Gamma\left(\frac{p+1}{2}\right)\Gamma\left(-\frac{p}{\alpha}\right)}{\alpha \sqrt{\pi} \Gamma\left(-\frac{p}{2}\right)},$$ 
and the following special cases of the first integer moment given by
\begin{equation}
\mathbb{E}\left[Y\right] = \begin{cases}
\delta - \beta \gamma \tan \frac{\pi \alpha}{2}, & \; p = 1, 1 < \alpha < 2\\
\delta, & \; p = 1, \alpha = 2.
\end{cases}
\end{equation}
}
\end{definition}

An alternative representation of the FLOM's results are derived for the parameterization $S(1)$ in \cite{samorodnitsky1994stable} and gives the results for the FLOM's according to Lemma \ref{lemma:FLOMs}.

\begin{definition}
\label{lemma:FLOMs}
\textbf{$\alpha$-Stable Fractional Lower Order Moments (S(1)-FLOM's):}\\
\textit{
Given a sub-exponential $\alpha$-Stable distribution \\ $X \sim S_{\alpha}\left(x;\beta,\gamma,\delta;S(1)\right)$ in which $\alpha \in (0,2)$ and $\beta=0$ when $\alpha = 1$. Then the FLOM can be evaluated analytically according to the following moments or Fractional Lower Order Moment conditions for $p \in (0,\alpha)$ according to 
\begin{equation}
\mathbb{E}\left[|X|^p\right] = C\left(p,\alpha,\beta;S(1)\right)^p\gamma
\end{equation}
where \cite[Property 1.2.17]{samorodnitsky1994stable} showed that 
\begin{equation}
C\left(p,\alpha,\beta;S(1)\right) = \frac{2^{p-1}\Gamma\left(1-\frac{p}{\alpha}\right)}{p\int_{0}^{\infty}u^{-p-1}\sin^2u du}\left(1 + \beta^2\tan^2\frac{\alpha \pi}{2}\right)^{\frac{p}{2\alpha}}\cos\left(\frac{p}{\alpha}\arctan\left(\beta \tan\frac{\alpha \pi}{2} \right)\right).
\end{equation}
}
\end{definition}

Therefore we can derive the following Theorem \ref{ThmMoments} for the FLOM's of the PNSC$(\alpha)$ Total interference model in three cases, the homogeneous case, the doubly stochastic Poisson-Gamma-PNSC$(\alpha)$ interference model and the general inhomogeneous in time or space Inhomogeneous-Poisson-PNSC$(\alpha)$ model.

\begin{theorem}\textit{\label{ThmMoments}The FLOM's of the total interference in the PNSC$(\alpha)$ homogeneous interference model in Theorem \ref{thm_pois_smin} are given by
\begin{equation}
\begin{split}
\mathbb{E}\left[\left|Y\right|^p\right] = \begin{cases} \infty & p \geq \alpha, \\ 
C(p,\alpha) c^{-1}_{K_{max}}\sum_{k=1}^K \exp\left(-\lambda_K\right) \frac{\lambda_K^k}{k!} \left(\sqrt{K}\gamma\right)^{\frac{p}{\alpha}} & 0 < p < \alpha. \end{cases}
\end{split}
\end{equation}
In the case of the doubly stochastic Poisson-Gamma-PNSC$(\alpha)$ interference model the FLOM's are given by
\begin{equation}
\begin{split}
\mathbb{E}\left[\left|Y\right|^p\right] = \begin{cases} \infty & p \geq \alpha, \\ 
C(p,\alpha) d^{-1}_{K_{max}}\sum_{k=1}^K \frac{\left(a+k-1\right)!}{\left(a-1\right)! k!}
\left(\frac{b}{1+b}\right)^a
\left(\frac{1}{1+b}\right)^k \left(\sqrt{K}\gamma\right)^{\frac{p}{\alpha}} & 0 < p < \alpha. \end{cases}
\end{split}
\end{equation}
In the case of the inhomogenous in space or time PNSC$(\alpha)$ interference models, replace $\lambda_K$ with $\lambda^*$ in Theorem \ref{lemma_poisson_mapping}.
}
\end{theorem}
\begin{proof} These results follow trivially from application of the mixture representations of the interference models.
\end{proof} 

Having derived the FLOM's for the PNSC$(\alpha)$ Total interference models in wideband transmission, we now consider the definition of the Geometric SNR in the PNSC$(\alpha)$ model. As discussed in \cite{ben2010asymptotic} in an $\alpha$-stable based interference model, the notion of noise power is mathematically undefined and the standard "signal-to-noise ratio" (SNR) is not a well-defined strength measure. 

The solution proposed involves a new indicator of the process strength, which is the geometric noise power. Since this estimator is a scale parameter it can be used as an indicator of process strength or "power" in situations where second-order methods are inadequate. The power is defined in \cite[Equation 2]{ben2010asymptotic}  to be given in the total interference model for the PNSC$(\alpha)$ interference models according to Proposition \ref{PropS0}.

\begin{definition} \textit{ \label{PropS0} For the interference $Y$ the general definition of the geometric noise power which provides a measure of process strength is given by
\begin{equation}
S0 = S0(Y) = \exp\left(\mathbb{E}\left[\log|Y|\right]\right).
\end{equation}
}
\end{definition}

We note that using the FLOM's an upper bound on the noise power can be obtained via Jensen's inequality as given in Lemma \ref{lemmBnd}.

\begin{lemma} \textit{ \label{lemmBnd} Consider $Y$ as the total interference in the homogeneous PNSC$(\alpha)$ or doubly stochastic Poisson-Gamma-PNSC$(\alpha)$ models. Then the following bound on the noise power applies for $p=1$ and $\alpha \in (1,2)$,
\begin{equation}
S0 \leq \mathbb{E}\left[|Y|\right].
\end{equation}
with $\mathbb{E}\left[|Y|\right]$ as given in Theorem \ref{ThmMoments}.
}
\end{lemma}

In addition, we can also evaluate the noise power when $Y$ is an $\alpha$-stable random variable from the PNSC$(\alpha)$ homogeneous total interference model or the doubly stochastic Poisson-Gamma-PNSC$(\alpha)$ model as shown in Lemma \ref{LemmaGenS0}.

\begin{lemma}\textit{\label{LemmaGenS0}
Consider $Y$ as the total interference in the homogeneous PNSC$(\alpha)$ or doubly stochastic Poisson-Gamma-PNSC$(\alpha)$ models. Then the following evaluation on the geometric noise power applies 
\begin{equation}
S0 = \frac{c^{-1}_{K_{max}}\sum_{k=1}^K \exp\left(-\lambda_K\right) \frac{\lambda_K^k}{k!}  \sqrt{k}\gamma C_g^{\frac{1}{\alpha}}}{C_g}
\end{equation}
where $C_g$ is the exponential of the Euler constant. In the case that we consider $Y$ as an $\alpha$-stable RV from the doubly stochastic Poisson-Gamma-PNSC$(\alpha)$ homogeneous total interference model the total geometric power is given by 
\begin{equation}
S0 = \frac{d^{-1}_{K_{max}}\sum_{k=1}^K \frac{\left(a+k-1\right)!}{\left(a-1\right)! k!}
\left(\frac{b}{1+b}\right)^a
\left(\frac{1}{1+b}\right)^k \sqrt{k}\gamma C_g^{\frac{1}{\alpha}}}{C_g}
\end{equation}
In the case of the inhomogenous in space or time PNSC$(\alpha)$ interference models, replace $\lambda_K$ with $\lambda^*$ in Theorem \ref{lemma_poisson_mapping}.
}
\end{lemma}

We can then define the GNSR for each of the PNSC$(\alpha)$ total interference models in the wideband system according to the definition in \cite{ben2010asymptotic} [Equation 4] is given in Proposition \ref{PropGNSR}.

\begin{lemma} \textit{\label{PropGNSR}
For the interference $Y$ the general definition is given by
\begin{equation}
GNSR = \frac{1}{2 C_g} \left(\frac{A}{S0}\right)^2
\end{equation}
Therefore when $Y$ is an $\alpha$-stable RV from the PNSC$(\alpha)$ homogeneous total interference model the total GNSR is given by 
\begin{equation}
GNSR = \frac{1}{2 C_g} \left(\frac{AC_g}{c^{-1}_{K_{max}}\sum_{k=1}^K \exp\left(-\lambda_K\right) \frac{\lambda_K^k}{k!}  \sqrt{k}\gamma C_g^{\frac{1}{\alpha}}}\right)^2.
\end{equation}
where A is the amplitude of a modulated signal. In the case that we consider $Y$ as an $\alpha$-stable RV from the doubly stochastic Poisson-Gamma-PNSC$(\alpha)$ homogeneous total interference model the total geometric power is given by 
\begin{equation}
GNSR = \frac{1}{2 C_g} \left(\frac{AC_g}{d^{-1}_{K_{max}}\sum_{k=1}^K \frac{\left(a+k-1\right)!}{\left(a-1\right)! k!}
\left(\frac{b}{1+b}\right)^a
\left(\frac{1}{1+b}\right)^k \sqrt{k}\gamma C_g^{\frac{1}{\alpha}}}\right)^2.
\end{equation}
In the case of the inhomogenous in space or time PNSC$(\alpha)$ interference models, replace $\lambda_K$ with $\lambda^*$ in Theorem \ref{lemma_poisson_mapping}.
}
\end{lemma}

\begin{remark}\textit{As noted in \cite{ben2010asymptotic} the normalizing constant $2C_g$ ensures
that for the Gaussian case $(\alpha = 2)$, the GSNR becomes the standard SNR for the PNSC$(\alpha)$ model}
\end{remark}

In Example \ref{ExmpGSNRExample} we present a study of the characteristic of the GNSR for the total interference under the PNSC$(\alpha)$ model developed. This is important to analyze, since the properties of the $\alpha$-stable model are known to be complex functions of $\alpha$, $\beta$, $\gamma$ and $K$. In this case the $\alpha$-stable model is symmetric removing $\beta=0$.

\begin{example} \label{ExmpGSNRExample} In the following simulations we consider the PNSC$(\alpha)$ homogeneous model with a range of $\alpha \in \left[0.1,1.9\right]$ values including three of which correspond to the special cases of the PNSC$(\alpha)$ Interference model developed which are the Compound Poisson PNSC$(3/2)$-Holtsmark Interference Model, the Compound Poisson PNSC$(1)$-Cauchy Interfence Model and the Compound Poisson PNSC$(2/3)$-Whittaker Interference Model. We set the GNSR according to the scale parameter given by $\gamma = \left\{0.1,10,250,500,750,1000\right\}\times 10^{-5}$, $A=1$, $K=64$ (64 carriers) and $\lambda_K = 10$. The results are provided in Figures \ref{fig:GSNR} which demonstrate that there is a clearly non-linear relationship between $\alpha$ and the GSNR, with a linear relationship between $\gamma$ and GNSR as expected from the derive GNSR in Lemma \ref{PropGNSR}.
\begin{figure}[htbp]
	\centering
		\includegraphics[width=\textwidth, height = 10cm]{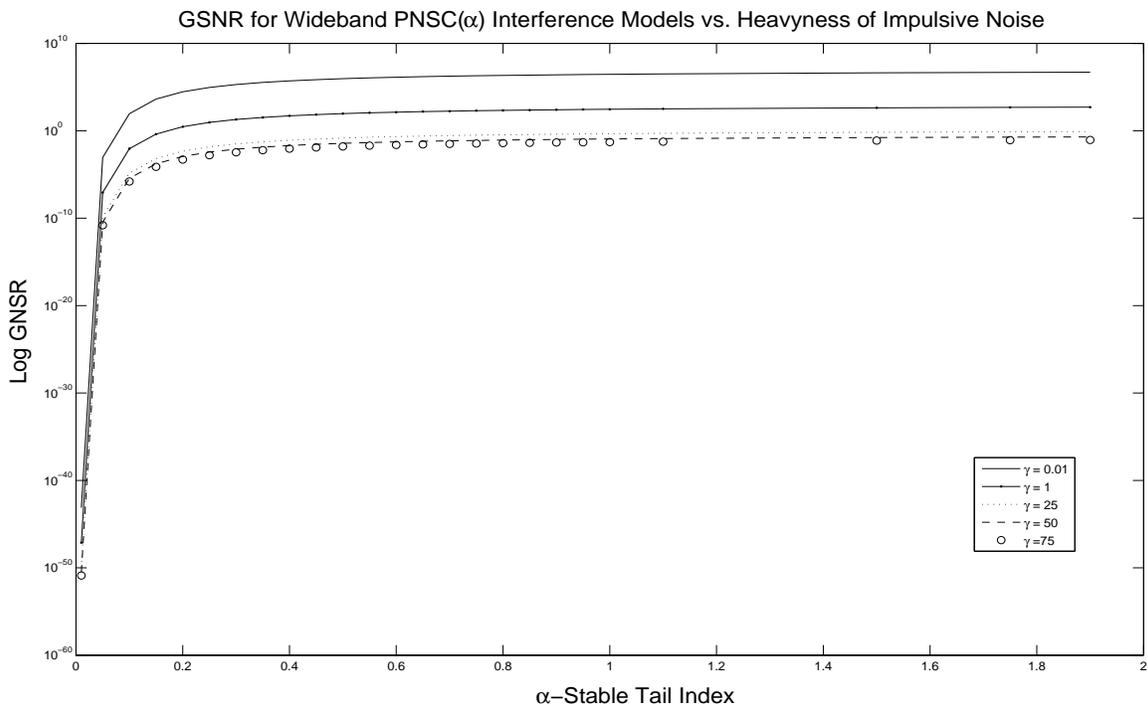}
	\caption{Relationship between GSNR,Tail Index $\alpha$ and scale $\gamma$ in the Compound Poisson PNSC Interference Model.}
	\label{fig:GSNR}
\end{figure}
\end{example}

\section{Likelihood Ratio in %Capacity of binary input memoryless channel with 
additive PNSC($\alpha$) interference}
\label{sec:analyticCapacity}
We consider that the input $X$ comes from a BPSK modulation and is affected by an impulsive noise $Y$ modeled by the results developed in Theorems \ref{thm_1} and \ref{thm_pois_smin}. 
Furthermore, we develop results below conditional on a fixed bandwidth, ie. fixed $k$, in the transmission and we assume that the user transmits in the presence of an unknown number of potential interferers who are uniformly distributed in the field of transmission. 

In this scenario we consider two possible hypothesis for the received sample $k$, denoted by %$R_k^{(j)} = X^{j} + Y_k$
$R_k = X_k + Y_k$ with $Y_k=\sum_{i=1}^N  Y_{I}^{(k,i)}+j\sum_{i=1}^NY_{Q}^{(k,i)}$, given, according to Lemma \ref{lemma_1} and Theorem \ref{thm_pois_smin}, by the following:
\begin{equation}
\label{BPSKHyp}
\left\{
\begin{split}
&\mathcal{H}_0: %R^{(j)}_k = X^{j}(\mathcal{H}_0) + Y_k %= 1 + \sum_{i=1}^N  Y_{I}^{(k,i)}+j \sum_{i=1}^N  Y_{Q}^{(k,i)} 
R_k = X(\mathcal{H}_0) + Y_k
 \sim S_{\alpha}\left(0,\widetilde{\gamma}_k,1; S(0)\right) \\   
&\mathcal{H}_1: %R^{(j)}_k = X^{j}(\mathcal{H}_1) + Y_k %= -1 + \sum_{i=1}^N  Y_{I}^{(k,i)}+ j \sum_{i=1}^N  Y_{Q}^{(k,i)} 
R_k = X(\mathcal{H}_1) + Y_k
\sim S_{\alpha}\left(0,\widetilde{\gamma}_k,-1; S(0)\right).  
\end{split}
\right.
\end{equation}

%Now utilizing the definition of the capacity in \cite[Equation 6]{Ben10} one may develop the following definition of capacity in the PNSC$(\alpha)$ homogeneous interference model as given in theorem \ref{ThmCapacityBasic}.
Based on the previously presented representation of the distribution, we give analytical expressions for the likelihood ratio in Theorem \ref{ThmCapacityBasic}.

\begin{theorem}

\textit{\label{ThmCapacityBasic} 
Consider the BPSK transmitted signal in the presence of the homogeneous PNSC$(\alpha)$ interference model. %, then the capacity is given by: 
%\begin{equation}
%\mathbb{C}\left(\frac{E_b}{N_0}\right) = 1-\mathbb{E}\left[\log_2\left(1+e^{-\widetilde{\Lambda}(r)}\right)\right] 
%\end{equation}
%with t
The log likelihood ratio %where $\widetilde{\Lambda}(r) = \log 
$\Lambda(r)$ is given by:
\begin{equation}
    \Lambda(r) = \frac{ S_{\alpha}\left(0,\widetilde{\gamma}_k,x%^{j}
    (\mathcal{H}_0); S(0)\right) }{ S_{\alpha}\left(0,\widetilde{\gamma}_k,x%^{j}
    (\mathcal{H}_1); S(0)\right) }
\end{equation}
which can be expressed according to the following forms depending on the tail index $\alpha$:
\begin{enumerate}
\item{As the tail exponent of the stable distribution approaches 1, the distribution of the test statistic for any value of received signal $r \in \mathbb{R}$ converges in distribution to the Cauchy receiver with test statistic of \cite{Gha10}, given by
\begin{equation}
\lim_{\alpha \rightarrow 1} \Lambda(r) \rightarrow \frac{\widetilde{\gamma}_k + \left(r - x%^{j}
(\mathcal{H}_0) \right)^2}{\widetilde{\gamma}_k + (r- x%^{j}
(\mathcal{H}_1))^2}.
\label{eq:LRTCauchy}
\end{equation}
}
\item{In the infinite mean interference models characterized by the Compound Poisson Holtsmark Interference Model, under an affine transformation of the received signal $Z_k = \left|\frac{1}{%\left|c_k\right|
\sqrt{k}\gamma}\right|R_k$ for the null and alternative hypothesis, %. Then 
the test statistic for the LRT is analytically evaluated according to
%\small{
\begin{equation}
\begin{split}
\lim_{\alpha \rightarrow 2/3} \Lambda(z) \rightarrow 
\frac
{\frac{1}{\pi}\Gamma(5/3)\widetilde{F}_{2,3}\left(z-\widetilde{x}(\mathcal{H}_0)\right)
-\frac{(z-\widetilde{x}(\mathcal{H}_0))^2}{3\pi}\widetilde{F}_{3,4}\left(z-\widetilde{x}(\mathcal{H}_0)\right)}
{\frac{1}{\pi}\Gamma(5/3)\widetilde{F}_{2,3}\left(z-\widetilde{x}(\mathcal{H}_1)\right)
-\frac{(z-\widetilde{x}(\mathcal{H}_1))^2}{3\pi}\widetilde{F}_{3,4}\left(z-\widetilde{x}(\mathcal{H}_1) \right)} \\
\frac
{+\Gamma(4/3)\frac{7(z-\widetilde{x}(\mathcal{H}_0))^4}{3^4\pi}\widetilde{F}_{2,3}\left(z-\widetilde{x}(\mathcal{H}_0)\right)}
{+\Gamma(4/3)\frac{7(z-\widetilde{x}(\mathcal{H}_1))^4}{3^4\pi}\widetilde{F}_{2,3}\left(z-\widetilde{x}^j(\mathcal{H}_1) \right)} 
\end{split}
\label{eq:LRTHoltsmark}
\end{equation}
with $\widetilde{x}(\mathcal{H}_{\cdot}) = \left|\frac{1}{%\left|c_k\right|
\sqrt{k}\gamma}\right|x%^j
(\mathcal{H}_{\cdot})$, $x%^j
(\mathcal{H}_{\cdot}) \in \left\{-1,1\right\}$,
$\widetilde{F}_{2,3}(z)=\;_2F_3\left(\frac{5}{12},\frac{11}{12};\frac{1}{3},\frac{1}{2},\frac{5}{6};-\frac{2^2z^6}{3^6}\right)$ and $\widetilde{F}_{3,4}(z)=\;_3F_4\left(\frac{3}{4},1,\frac{5}{4};\frac{2}{3},\frac{5}{6},\frac{7}{6},\frac{4}{3};-\frac{2^2z^6}{3^6}\right)$.
}
\item{In the finite mean interference models characterized by the Compound Poisson PNSC$(\alpha)$ Interference Channel Model, the test statistic for the LRT is analytically evaluated according to:
\begin{equation}
\begin{split}
\lim_{\alpha \rightarrow 3/2} \Lambda(r) \rightarrow 
\frac{|r - x(\mathcal{H}_0)|^{-1}%\exp
e^{\frac{2}{27}(r - x(\mathcal{H}_0))^{-2}}W_{-1/2;1/6}\left(\frac{4}{27}(r-x(\mathcal{H}_0))^{-2}\right)}
{\left|r - x(\mathcal{H}_1)\right|^{-1}
e^{\frac{2}{27}\left(r - x(\mathcal{H}_1)\right)^{-2}}W_{-1/2;1/6}\left(\frac{4}{27}\left(r - x(\mathcal{H}_1)\right)^{-2}\right)}.
\end{split}
\label{eq:LRTWhit}
\end{equation}
}
\item{In the case that one considers a general $\alpha \in [0,2]$ then the test statistic for the LRT after standardization involving scaling by $\gamma^{-1/\alpha}$ is evaluated (via truncation of sums) according to series expansions provided in \cite[Sections 2.4.6, 2.4.8]{zolotarev1983univariate},
\begin{equation}
\Lambda(r) = \left\{
\begin{split} 
& \frac{\sum_{s=1}^{\infty}D_s^I \left(r-x%^j
(\mathcal{H}_0)\right)^{-\alpha s-1}}{\sum_{s=1}^{\infty}D_s^I \left(r-x%^j
(\mathcal{H}_1)\right)^{-\alpha s-1}}, & 0 < \alpha < 1,\\
& \frac{\left(r-x%^j
(\mathcal{H}_0)\right)^2+1}{\left(r-x%^j
(\mathcal{H}_1)\right)^2+1}, & \alpha = 1, \\
& \frac{\sum_{s=0}^{\infty}D_s^{II} (r-x%^j
(\mathcal{H}_0))^{2s}}{ \sum_{s=0}^{\infty}D_s^{II} \left(r-x%^j
(\mathcal{H}_1)\right)^{2s}}, & 1 < \alpha < 2,\\ 
& \frac{\exp\left(-\frac{(x-x%^j
(\mathcal{H}_0))^2}{4}\right)}{\exp\left(-\frac{(x-x%^j
(\mathcal{H}_1))^2}{4}\right)} & \alpha = 2. 
\end{split}
\right.
\label{eq:LRTgeneral}
\end{equation}
with $D_s^I = \frac{(-1)^{s-1}}{s!}\Gamma\left(\alpha s + 1\right)\sin\left(\frac{s\alpha\pi}{2}\right)$, $D_s^{II} = \frac{(-1)^{s-1}}{(2s)!}\Gamma\left(\frac{2s + 1}{\alpha}\right)$.
}
\end{enumerate}
}

\end{theorem}

\begin{proof}%The proof of this result follows from the definition of capacity in \cite[Equation 6]{Ben10}; t
The analytic LRT representations are derived using results from theorem \ref{thm_pois_smin} and theorem \ref{thm_1} for BPSK modulation. In addition, for item 4, the result follows from utilization of a series expansion representation of the $\alpha$-stable model. 
\end{proof}

\bigskip

In the following simulation we consider the PNSC$(\alpha)$ homogeneous model. 
We evaluate the log likelihood ratio (LRT) at a range of values $r \in \mathbb{R}$ versus tail index $\alpha$ under two settings. 
The first is an exhaustive simulation via Monte Carlo methods, which indicates the true estimated LRT. 
Then the analytic expressions derived for different $\alpha$ is done.
%In example \ref{ExmpTruncationLRT} we analyze the %behavior of 
%evaluation of the LRT for the four items.
The limitations come for the Compound Poisson Holtsmark Interference Model \eqref{eq:LRTHoltsmark} in the evaluation of functions $_pF_q(.)$ in \eqref{eq:pFq} and in the case of a general $\alpha$ \eqref{eq:LRTgeneral} due to the sum truncation.
%approximated as a function of truncation $S$, and tail index $\alpha$ for $\gamma = 1$. We found that the number of terms in such applications in the evaluation of the density in the numerator and denominator of the LRT can be accurately evaluated with $S > 100$ making the evaluation relatively efficient. 
%In addition, we note that under this parameterization of the $\alpha$-stable distribution one has a clear asymptote at $\alpha=1$, this is a well known feature of this parameterization. We note there are other parameterizations %which we presented in Section \ref{} 
%which admit densities which are absolutely continuous in all parameters, however, we work with this parameterization as it admits a bound on the capacity that we develop in Theorem \ref{ThmCapacityBound} through representation of the capacity via a bound on the expected LRT which we prove can be bounded as a function of the FLOM's.

\begin{example}[\textbf{Log likelihood ratio for $\alpha=1$ \eqref{eq:LRTCauchy}}]
\label{ExmpTruncationLRT}

\textsl{In this first example we plot the LRT in the well known case $\alpha=1$ (Figure \ref{fig:LRTcauchy}).
\begin{figure}[htbp]
	\centering
		\includegraphics[height = 6cm, width=12cm]{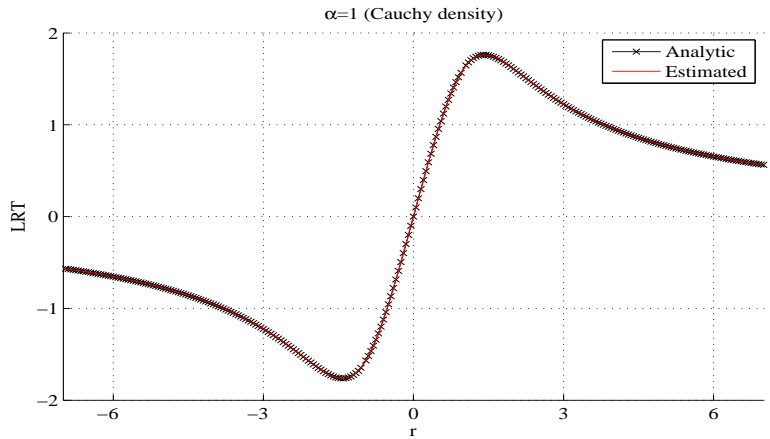}
	\caption{Comparison between exact expression \eqref{eq:LRTCauchy} of LRT versus Monte Carlo under the Cauchy model with $\alpha=1$.}
	\label{fig:LRTcauchy}
\end{figure}
As expected the fit is very good. This plot essentially allows to validate the accuracy of the Mote Carlo LRT evaluation.
}
\end{example}

\begin{example}[\textbf{Log likelihood ratio for $\alpha=3/2$ \eqref{eq:LRTHoltsmark}}]
\textsl{We plot the LRT in the case $\alpha=3/2$ (figure \ref{fig:LRTHolt}) corresponding to the finite mean case - Holtsmark density.
\begin{figure}[htbp]
	\centering
		\includegraphics[height = 6cm, width=12cm]{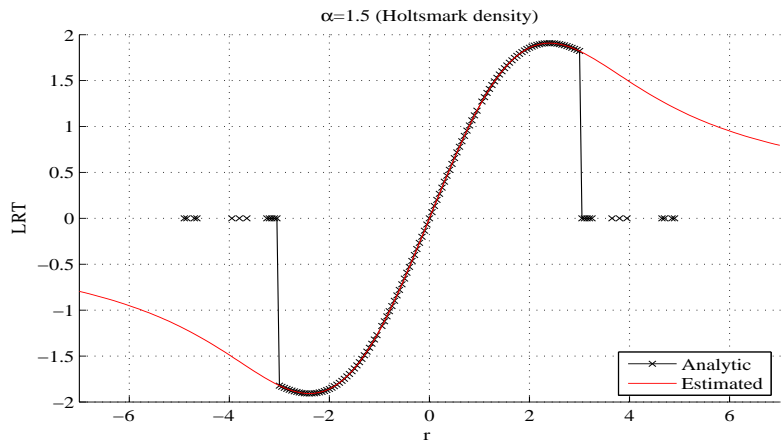}
	\caption{Comparison between exact expressions of LRT versus Monte Carlo under the PNSC$(3/2)$ model.}
	\label{fig:LRTHolt}
\end{figure}
The fit is very accurate in the range $[-3,3]$ but for larger values the calculation of the functions $_pF_q$ in \eqref{eq:LRTHoltsmark} can not be performed accurately. It can however be sufficient in practice because larger values will be clipped at least due to the non linearity of the circuits.
}
\end{example}

\begin{example}[\textbf{Log likelihood ratio for $\alpha=2/3$ \eqref{eq:LRTWhit}}]
\textsl{We plot the LRT in the case $\alpha=2/3$ (Figure \ref{fig:LRTWhit}) corresponding to the infinite mean case - Whittaker density.
\begin{figure}[htbp]
	\centering
		\includegraphics[height = 6cm, width=12cm]{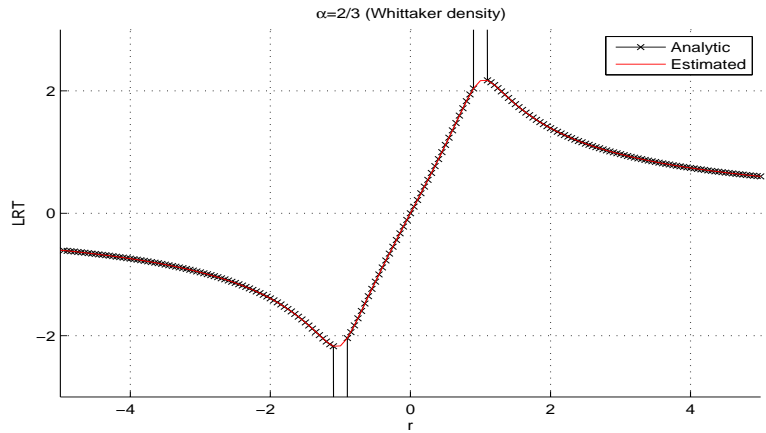}
	\caption{Comparison between exact expressions of LRT versus Monte Carlo under the PNSC$(2/3)$ model.}
	\label{fig:LRTWhit}
\end{figure}
The fit is very close except in very small intervals around $1$ and $-1$.
With a minimum care, this limitation is not a strong difficulty. 
Similarly to the Holtsmark density, this distribution can give very interesting insights in the impact of impulsive noise on communications and allow to derive appropriate solutions for detection and estimation problems.
} 
\end{example}

%moderately low truncation levels $S \in \left\{100,150,200\right\}$ with a scale parameter given by $\gamma = \left\{1\right\}$, $A=1$, $K=64$ (64 carriers) and $\lambda_K = 10$. The results in Figure \ref{fig:LRTTruncation} present the LRT. We aim to assess the efficiency and accuracy of the LRT derived expressions for decreasing $\alpha$, under low truncation ranges and compare to the true LRT in the PNSC$(\alpha)$ model (accurately estiamted via exhaustive Monte Carlo simulation). 
%
\begin{example}[\textbf{Log likelihood ratio for $1<\alpha<2$ \eqref{eq:LRTgeneral}}]
\textsl{We plot the LRT in the cases $\alpha=1.8$ and $\alpha=1.2$ (Figure \ref{fig:LRT14}) from the general series expansion case in \eqref{eq:LRTgeneral}.
\begin{figure}[htbp]
	\centering
		\includegraphics[height = 6cm, width=0.48\textwidth]{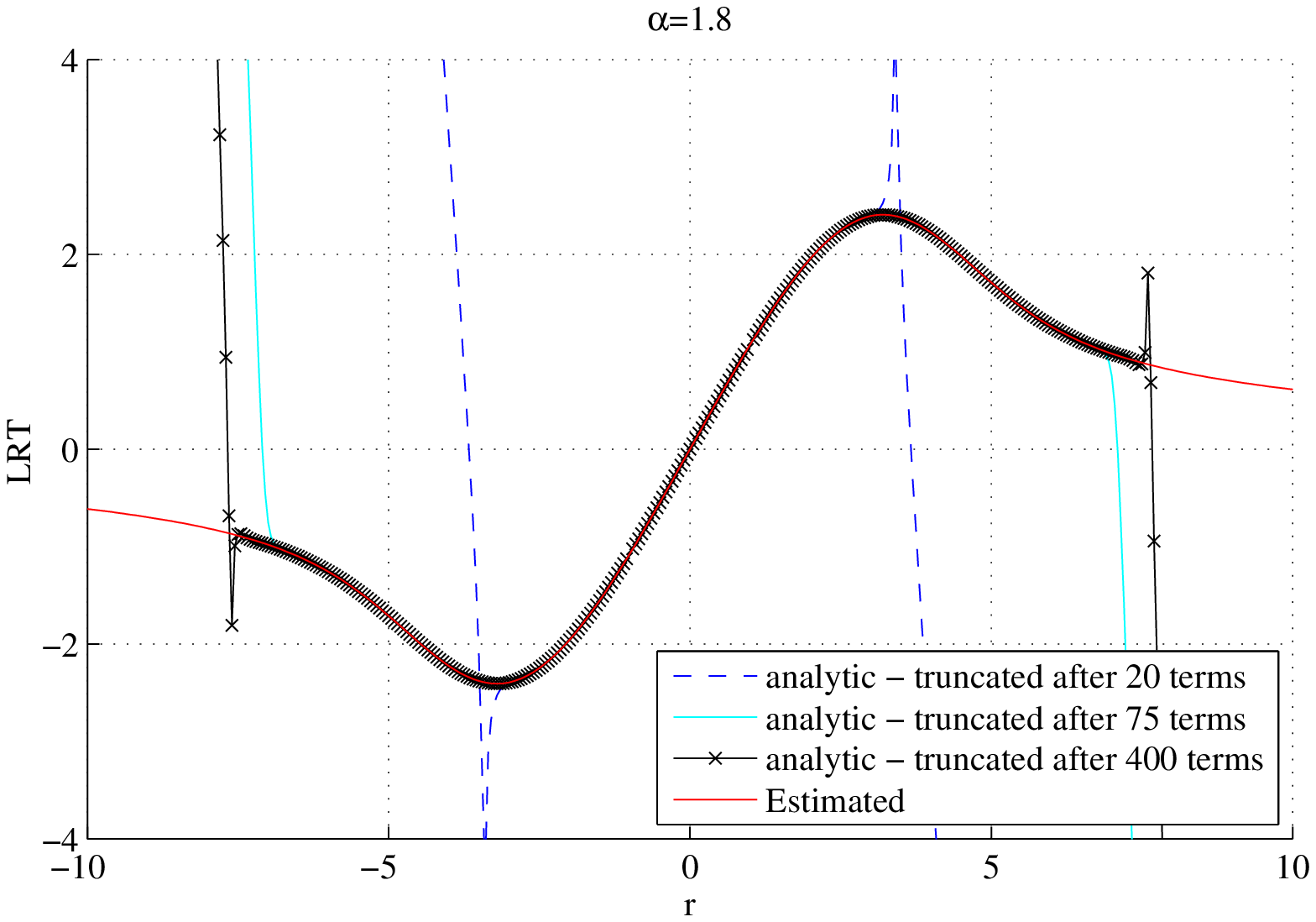}		
		\includegraphics[height = 6cm, width=0.48\textwidth]{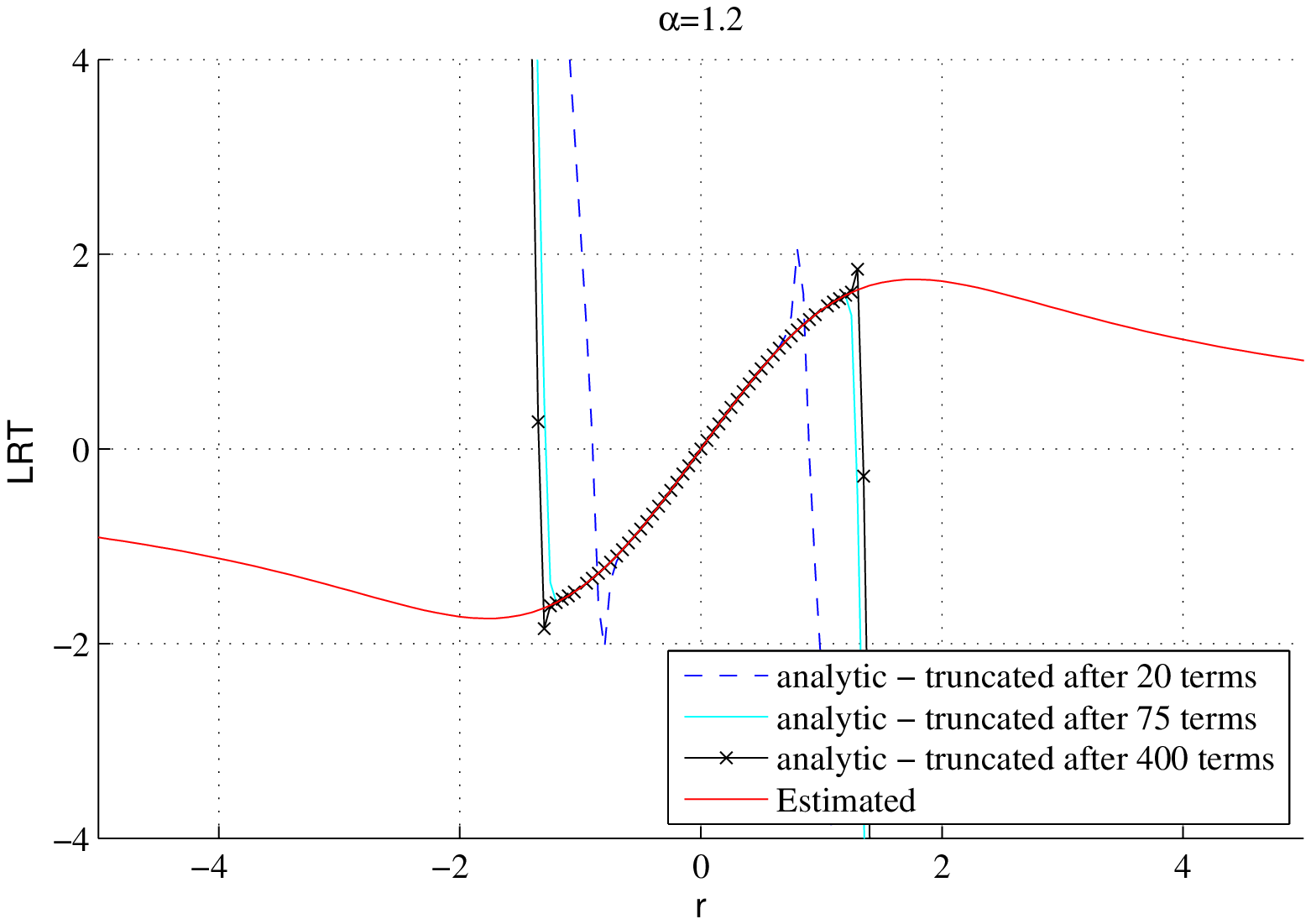}
	\caption{Comparison between an estimated LRT from truncated series expansions versus (Monte Carlo - exact) under the PNSC$(\alpha)$ model with $\alpha =1.9$ (left subplot) and $\alpha =1.4$ (right subplot). }
	\label{fig:LRT14}
\end{figure}
We see a very accurate fit between the real LRT and the truncated sum.
The range of validity however depends on the number of terms we consider in the sum.
For low values of $r$ it converges very quickly (20 terms are sufficient) but we reach the computer limitations for larger values of $r$ if we try to compute more than 400 terms in the sum. 
The range of convergence is also dependent on the value of $\alpha$. The larger $\alpha$ is, the larger the range.
When getting close to one, the series expansion (due to computer limitation) can only be used in a range smaller than $[-1,1]$. It can however have interest for instance for detection of small signals in a stable noise.
}
\end{example}

%(100 terms) in the range $r \in [-8,8]$ for $\alpha = 1.9$ after which the accuracy deteriorates unless more terms are incorporated in the series expansion. In comparison as the tails of the interference model get heavier, for $\alpha =1.4$, we see that the range over which we can consider 200 terms as reasonable between $r \in [-3,3]$. Note it is expected that the number of terms in the series expansion is inversely related to the tail index value. As it decreases away from two the number of terms required for accurate approximation increases, as we demonstrate here, or alternatively, the range of accuracy decreases. However, importantly these series expansion are very efficient to evaluate for large numbers of terms, and even with these low numbers of terms we see the most interesting range of the LRT test statistics values is covered. We next explore cases in which the interference has infinite mean characteristics.
%
\begin{example}[\textbf{Log likelihood ratio for $0<\alpha<1$ \eqref{eq:LRTgeneral}}]
\textsl{We finally plot the LRT in the cases $\alpha=0.8$ and $\alpha=0.2$ (figure \ref{fig:LRT04}) from the general series expansion case in \eqref{eq:LRTgeneral}.
\begin{figure}[htbp]
	\centering
		\includegraphics[height = 6cm, width=0.48\textwidth]{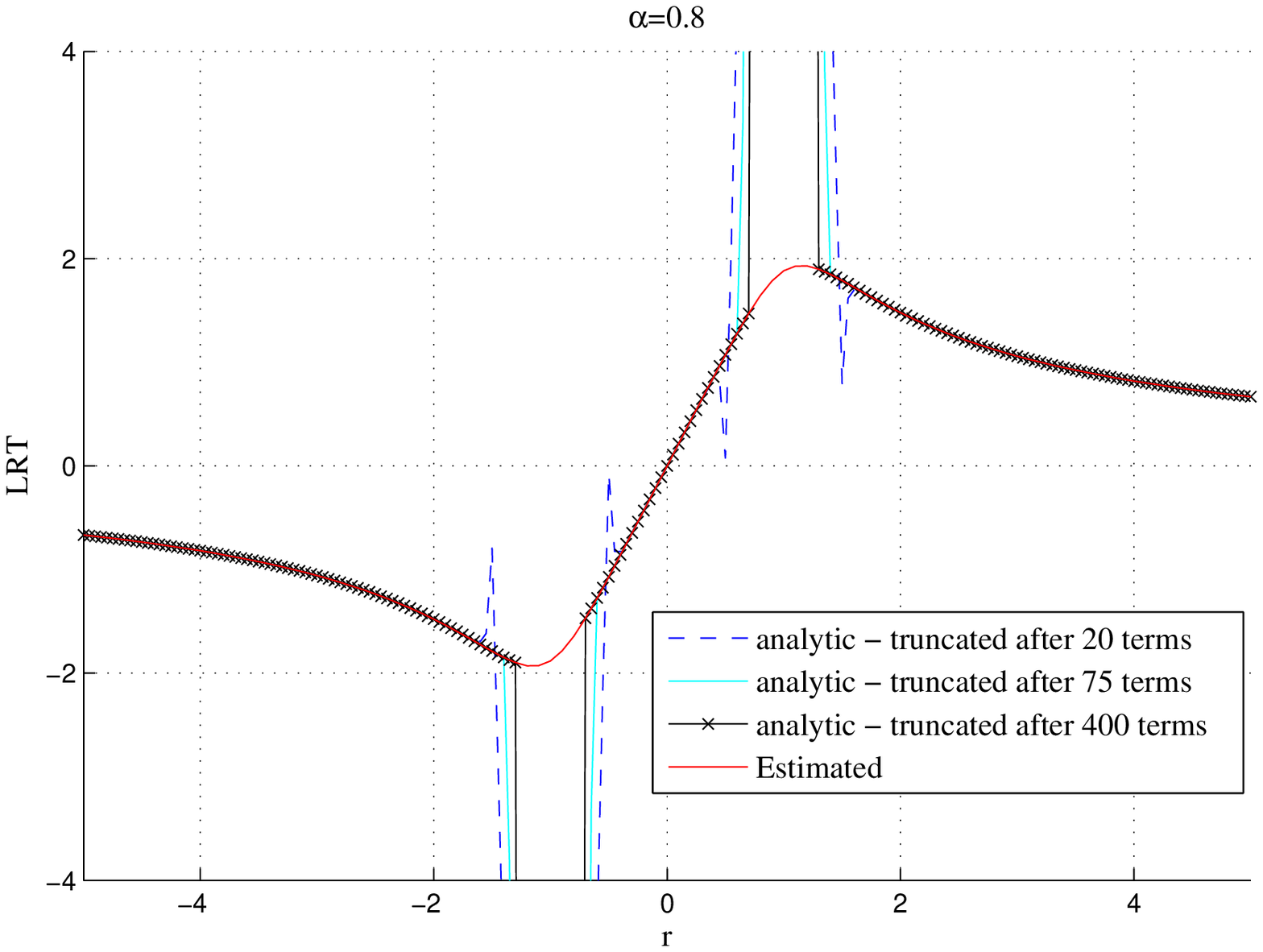}
		\includegraphics[height = 6cm, width=0.48\textwidth]{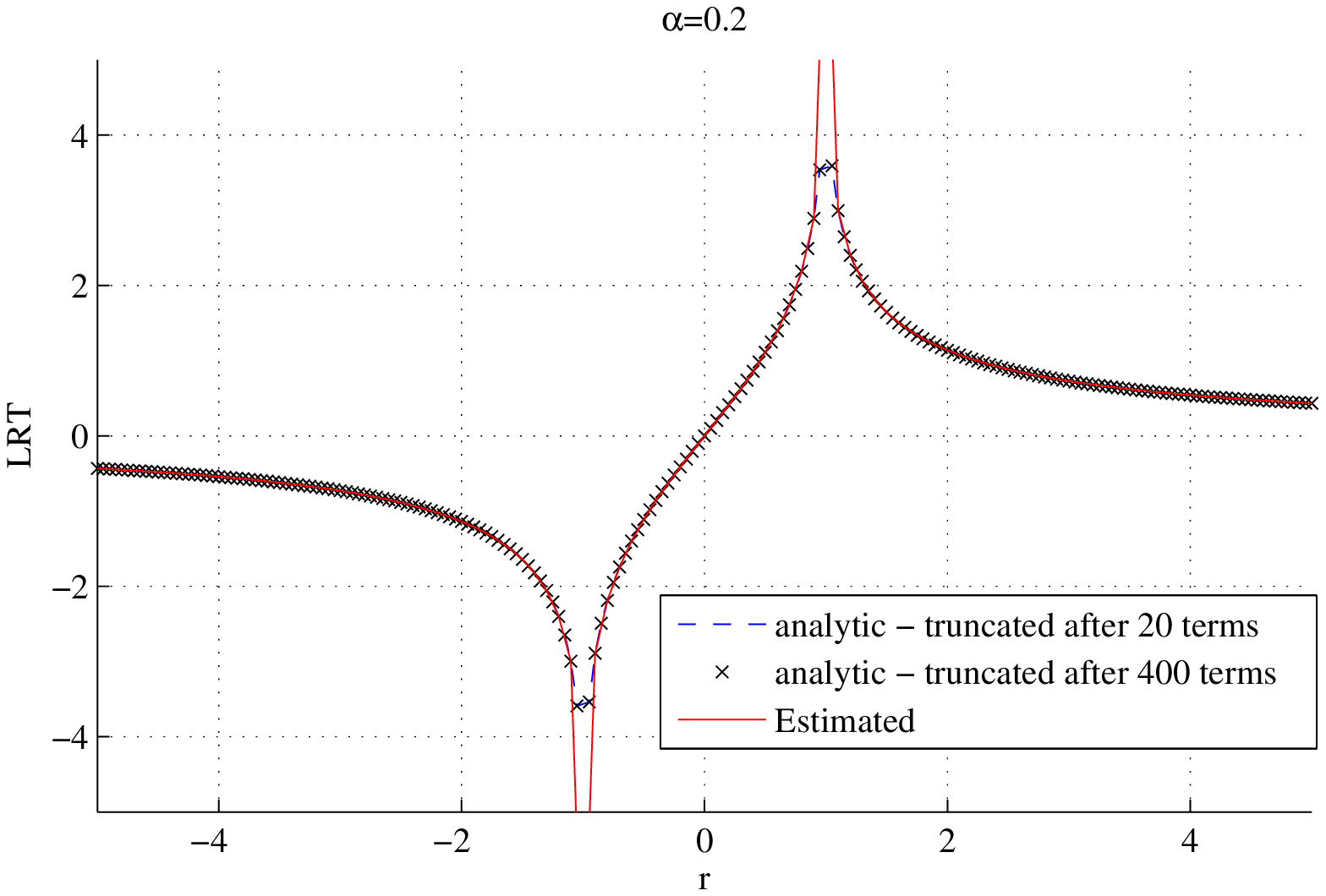}
	\caption{Comparison between an estimated LRT from truncated series expansions versus (Monte Carlo - exact) under the PNSC$(\alpha)$ model with $\alpha=1.2$ (left subplot) and $\alpha=0.2$ (right subplot).}
	\label{fig:LRT04}
\end{figure}
Again we see a very good fit between the real LRT and the truncated sum except for a small intervals around $1$ and $-1$.
The range where the model is not accurate gets larger as $\alpha$ gets closer to $1$. However it remains quite small and, out of this range, less than 20 terms in the sum give an accurate result.
}
\end{example}

\section{Conclusion}
This paper deals with interference model that can be useful for \textit{ad hoc} networks or cognitive radio. 
We consider a Poisson field of interferers with varying occupied bandwidth. 
This results in a doubly stochastic Poisson model. 
The resulting law is an $\alpha$-stable distribution.
Our main contributions are as follows:
\begin{enumerate}
	\item the extension of the model with a random number of interferers to a doubly stochastic model; this includes a system with a random number of interferers having a random transmission bandwidth. It can have a direct application to cognitive radio systems where interferers are secondary users that adapt their bit rate to the environment they sense and where the primary users use an OFDM transmission like in LTE or IEEE 802.11;
	\item proposals for analytical expression of the probability density function and the cumulative distribution based on series expansion. Such tools can be very useful when tackling other problems like the capacity evaluation of large networks or solutions for detection and estimation. 
	Our approach is based on two representations of the stable distributions: the SMiN representation, where the stable random variable is represented by a normal distribution with a random variance following a stable law, and the projection of a stable random vector which is a univariate stable random variable.
	\item we also include some special cases that can be obtained without the need to truncate the infinite sum of the series extension in two cases (other than Gauss and Cauchy which have exact expressions). We include a finite mean case ($\alpha=3/2$) based on Holtsmark distribution and an infinite mean case ($\alpha=2/3$) using Whittaker functions. Those expressions can be very useful in the sense they represent significant situations that could be used as reference for any evaluation of the considered scenario;
	\item the model is finally extended for inhomogeneous in time or space Poisson process. This allows to consider varying mean bandwidth occupation for interferers or spatially inhomogeneous positions of interferers.
\end{enumerate}

\bigskip

We give an illustration of the proposed representation through the evaluation of the likelihood ratio for a binary source in additive symmetric $\alpha$-stable interference, highlighting the accuracy of the approach, the rate of convergence of the involved series expansion but also its limits. 

Our results can have significant application linked to information theory, digital communications or signal processing when the noise has an impulsive nature.

\appendices

\section{Characteristic function}
\label{log_CF_proof}
\begin{proof}
%\textsl{
Under the assumption that, given $N$ potential interferers in region $A_{\mathcal{R}}$, the interferer locations are uniformly distributed, we can express the CF for the total interference at the $k$-th frequency according to, 
%\small
\begin{align*}
\begin{split}
&\varphi_{Y_{I}^{(k)},Y_{Q}^{(k)}}\left(\omega^{(k)}_I,\omega^{(k)}_Q\right)%\\
%%%%%%%%%%%%%%%%%%%%%%%%%5
%& \qquad  
=\sum_{N=0}^\infty 
\exE_{\R, \bm{c}_k, \A_k ,\bm{\Phi}_k}  %{R, H_k, A_k ,\Phi_k,\Theta_k}
\left[\exp\left(j R^{-\sigma/2} A_k c_k  \sqrt{\left(\omega^{(k)}_I\right)^2+\left(\omega^{(k)}_Q\right)^2 }
\right.\right.\\
& \qquad \qquad \qquad \qquad \qquad \qquad \left.\left.
\times \cos\left(\Phi_k -
 \arctan\left(\frac{\omega^{(k)}_Q}{\omega^{(k)}_I}\right)
   \right)\right)  
\right]^N \mathbb{P} \left(N(A_{\mathcal{R} })\right).
%%%%%%%%%%%%%%%%%%%%%%%%%5
%%%%%%%%%%%%%%%%%%%%%%%%%%
\end{split}
\end{align*}
%\normalsize

Using the Taylor series representation of an exponent, we rewrite the CF 
%\footnotesize
%\begin{align*}
%\begin{split}
%&\varphi_{Y_{I}^{(k)},Y_{Q}^{(k)}}\left(\omega^{(k)}_I,\omega^{(k)}_Q\right) = \\
%%%%%%%%%%%%%%%%%%%%%%%%%%5
%&=
%\exp\left(\lambda \pi r_T^2\left(
%\exE_{R, H_k, A_k ,\Phi_k,\Theta_k}
%\left[\exp\left(j  R^{-\sigma/2} H_k A_k \sqrt{\left(\omega^{(k)}_I\right)^2+\left(\omega^{(k)}_Q\right)^2 }
%      \cos\left(\Phi_k+\Theta_k -
% \arctan\left(\frac{\omega^{(k)}_Q}{\omega^{(k)}_I}\right)
%   \right)\right)   
%\right]-1\right)\right),
%%%%%%%%%%%%%%%%%%%%%%%%%%5
%%%%%%%%%%%%%%%%%%%%%%%%%%%
%\end{split}
%\end{align*}
%\normalsize
in the $\log$ domain, according to:
%
%\footnotesize
\begin{align*}
\label{log_CF}
\begin{split}
& \psi_{Y_{I}^{(k)},Y_{Q}^{(k)}}\left(\omega^{(k)}_I,\omega^{(k)}_Q\right) \triangleq 
\log \left(\varphi_{Y_{I}^{(k)},Y_{Q}^{(k)}}\left(\omega^{(k)}_I,\omega^{(k)}_Q\right)\right)\\
%%%%%%%%%%%%%%%%%%%%%%%%%5
& \qquad \qquad =
\lambda \pi r_T^2\left(
\exE_{\R, \bm{c}_k, \A_k ,\bm{\Phi}_k,} %{R, H_k, A_k ,\Phi_k,\Theta_k}
\left[\exp\left(j  R^{-\sigma/2} A_k c_k \sqrt{\left(\omega^{(k)}_I\right)^2+\left(\omega^{(k)}_Q\right)^2}\right.\right.\right.\\
%%%%%%%%%%%%%%%%%%%%%%%%%5
& \qquad \qquad \qquad \qquad \qquad \qquad \left.\left.\left.
\times \cos\left(\Phi_k - \arctan\left(\frac{\omega^{(k)}_Q}{\omega^{(k)}_I}\right)
   \right)\right)   
\right]-1\right).
%%%%%%%%%%%%%%%%%%%%%%%%%5
%%%%%%%%%%%%%%%%%%%%%%%%%%
\end{split}
\end{align*}%}
\end{proof}

\section{Proof of lemma 8}
\label{lemma_8_proof}
\begin{proof}
%\textsl{
First re-express the argument of the expectation in the log CF in Lemma \ref{lemma_6} using the complex series expansion based on Bessel functions, given by \cite{abramowitz1964handbook}
\begin{align*}
\exp\left(j a \cos\left(\theta\right)\right) = \sum_{s=1}^{\infty}
j^s \epsilon_s J_s\left(a\right)\cos\left(s \theta\right)
\end{align*}
where $\epsilon_0=1$ and $\epsilon_s = 2$ for all $s \geq 1$, and $J_s$  is the Bessel function of order $s$ defined by:
\begin{align*}
J_s\left(x\right)=\frac{1}{2\pi} \int_{-\pi}^{\pi} \exp\left(-j \left(s \tau -x \sin \left(\tau\right)\right)\right) d\tau.
\end{align*}
Applying this identity allows us to re-express the argument of the expectation in (\ref{log_CF}) according to,
\begin{align*}
\begin{split}
&\exp\left(j R^{-\frac{\sigma}{2}} A_k c_k \sqrt{\left(\omega^{(k)}_I\right)^2+\left(\omega^{(k)}_Q\right)^2}
    \cos\left(\Phi_k - \arctan\left(\frac{\omega^{(k)}_Q}{\omega^{(k)}_I}\right)
   \right)\right)   \\
&=\sum_{s=0}^{\infty}
j^s
\epsilon_s
J_s\left(R^{-\frac{\sigma}{2}} A_k c_k\sqrt{\left(\omega^{(k)}_I\right)^2+\left(\omega^{(k)}_Q\right)^2 }\right)
\cos\left(s \Phi_k-s \arctan\left(\frac{\omega^{(k)}_Q}{\omega^{(k)}_I}\right)\right)
\end{split}
\end{align*}
which allows us to write the $\log$ CF as:
%\footnotesize
\begin{align*}
\begin{split}
&\psi_{Y_{I}^{(k)},Y_{Q}^{(k)}}\left(\omega^{(k)}_I,\omega^{(k)}_Q\right) \\
%%%%%%%%%%%%%%%%%%%%%%%%%5
&=
\lambda \pi r_T^2\left(
\exE_{\R, \bm{c}_k, \A_k ,\bm{\Phi}_k}%{R, H_k, A_k ,\Phi_k,\Theta_k}
\left[
\sum_{s=0}^{\infty}
j^s
\epsilon_s
J_s\left(R^{-\frac{\sigma}{2}} A_k c_k\sqrt{\left(\omega^{(k)}_I\right)^2+\left(\omega^{(k)}_Q\right)^2 }\right)
\right.\right.\\
& \qquad \qquad \qquad \qquad \qquad \qquad \left.\left.
\times \cos\left(s \Phi_k -s \arctan\left(\frac{\omega^{(k)}_Q}{\omega^{(k)}_I}\right)\right)
\right]-1\right).
%%%%%%%%%%%%%%%%%%%%%%%%%5
%%%%%%%%%%%%%%%%%%%%%%%%%%
\end{split}
\end{align*}
\normalsize

Next, under model assumptions in Section \ref{SystemDescription} we have that, for the $k$-th frequency, the random variable $\Phi_k$ is uniformly distributed in $\left[0, 2\pi\right]$. Therefore: 
%
%\footnotesize
\begin{align*}
\begin{split}
&\psi_{Y_{I}^{(k)},Y_{Q}^{(k)}}\left(\omega^{(k)}_I,\omega^{(k)}_Q\right) \\
%%%%%%%%%%%%%%%%%%%%%%%%%5
& \qquad=
\lambda \pi r_T^2\left(
\sum_{s=0}^{\infty}
j^s
\epsilon_s
\exE_{\R, \bm{c}_k, \A_k}
\left[
J_s\left(R^{-\sigma/2} A_k c_k\sqrt{\left(\omega^{(k)}_I\right)^2+\left(\omega^{(k)}_Q\right)^2 }\right)
\right]
\right.\\
& \qquad \qquad \qquad \qquad \qquad \qquad \qquad \qquad \left.
\times\exE_{\bm{\Phi}_k}
\left[
\cos\left(s \Phi_k -s \arctan\left(\frac{\omega^{(k)}_Q}{\omega^{(k)}_I}\right)\right)
\right]
-1\right)\\
%%%%%%%%%%%%%%%%%%%%%%%%%5
& \qquad=
\lambda \pi r_T^2\left(
\exE_{\R, \bm{c}_k, \A_k}
\left[
J_0\left(R^{-\sigma/2} A_k c_k\sqrt{\left(\omega^{(k)}_I\right)^2+\left(\omega^{(k)}_Q\right)^2 }\right)
\right]
-1\right),
%%%%%%%%%%%%%%%%%%%%%%%%%5
%%%%%%%%%%%%%%%%%%%%%%%%%%
\end{split}
\end{align*}
%\normalsize
which holds due to the observation that $\exE_{\bm{\Phi}_k}
\left[
\cos\left(s \Phi_k-s \arctan\left(\frac{\omega^{(k)}_Q}{\omega^{(k)}_I}\right)\right)
\right]=0$ for $s \geq 1$.\\
%}
\end{proof}

\section{Proof of Theorem I}
\label{theorem_1_proof}

\begin{proof}
To derive this result for the CF as a representation of a isotropic bivariate $\alpha$-stable distribution we make use of \cite[Identity 3.12, p.152] {haenggi2009interference} given by:
\begin{equation}
\begin{split}
\lim_{a \rightarrow \infty} a^2 \left(\int_0^a \frac{2r}{a^2} \exp\left(j\omega l(r)\right)dr - 1\right) = \int_{0}^{\infty} \left(l^{-1}(x)\right)^2 j\omega \exp\left(j \omega x\right)dx
\end{split}
\end{equation}
Conditional on any given number of potential interferers, in the spatial region $A_{\mathcal{R}}$, we can marginalize the CF given in Lemma \ref{lemma_8}, with respect to the unknown spatial locations of these interferers. To achieve this we utilise the assumption on the spatial distribution of these interferers given in model assumptions in Section \ref{SystemDescription}. Hence, we integrate the log CF given in Lemma \ref{lemma_8} as follows,
%
%\footnotesize
\begin{align*}
\begin{split}
&\psi_{Y_{I}^{(k)},Y_{Q}^{(k)}}\left(\omega^{(k)}_I,\omega^{(k)}_Q\right) \\
%%%%%%%%%%%%%%%%%%%%%%%%%5
&=
\lambda \pi r_T^2\left(
\int_{0}^{r_T}
\exE_{\bm{c}_k,\A_k} %{A_k,c_k}
\left[
J_0\left(r^{-\sigma/2} A_k c_k\sqrt{\left(\omega^{(k)}_I\right)^2+\left(\omega^{(k)}_Q\right)^2 }\right)
\right]
\frac{2 r}{r_T^2} \text{d} r
-1\right)\\
%%%%%%%%%%%%%%%%%%%%%%%%%5
&=
\lambda \pi 
r_T^2\left(
\int_{0}^{r_T}
\left(
\exE_{\bm{c}_k,\A_k} %{ H_k, A_k}
\left[
J_0\left(r^{-\sigma/2} A_k c_k\sqrt{\left(\omega^{(k)}_I\right)^2+\left(\omega^{(k)}_Q\right)^2 }\right)
\right]
-1
\right)
\frac{2 r}{r_T^2} \text{d} r
\right)\\
%%%%%%%%%%%%%%%%%%%%%%%%%
&=
\lambda \pi 
r_T^2\left(
\int_{0}^{r_T}
\left(
\exE_{\bm{c}_k,\A_k} %{ H_k, A_k}
\left[
J_0\left(r^{-\sigma/2} A_k c_k\sqrt{\left(\omega^{(k)}_I\right)^2+\left(\omega^{(k)}_Q\right)^2 }\right)
-1
\right]
\right)
\frac{2 r}{r_T^2} \text{d} r
\right).
%%%%%%%%%%%%%%%%%%%%%%%%%5
\end{split}
\end{align*}
\normalsize

Next we integrate by parts as follows,
%
%%%%%%%%%%%%%%%%%%%%%%%%%5
%\footnotesize
\begin{align*}
\begin{split}
&\psi_{Y_{I}^{(k)},Y_{Q}^{(k)}}\left(\omega^{(k)}_I,\omega^{(k)}_Q\right) \\
%%%%%%%%%%%%%%%%%%%%%%%%%5
&=
\lambda \pi 
r_T^2\left(
%%%%%%%%%%%55
\exE_{\bm{c}_k,\A_k} %{ H_k, A_k}
\left[
J_0\left(r_T^{-\sigma/2} A_k c_k\sqrt{\left(\omega^{(k)}_I\right)^2+\left(\omega^{(k)}_Q\right)^2 }\right)-1
\right]
\int_{0}^{r_T}
\frac{2 r}{r_T^2}
\text{d} r \right.\\
&\left.-
%%%%%%%%%%%%%5
\int_{0}^{r_T}
\frac{d}{d r}
\exE_{\bm{c}_k,\A_k} %{ H_k, A_k}
\left[
J_0\left(r^{-\sigma/2} A_k c_k\sqrt{\left(\omega^{(k)}_I\right)^2+\left(\omega^{(k)}_Q\right)^2 }\right)
\right]
\text{d} r
\right)\\
%%%%%%%%%%%%%%%%%%%%%%%%%
%%%%%%%%%%%%%%%%%%%%%%%%%5
&=
\lambda \pi 
r_T^2\left(
%%%%%%%%%%%55
\exE_{\bm{c}_k,\A_k} %{ H_k, A_k}
\left[
J_0\left(r_T^{-\sigma/2} A_k c_k\sqrt{\left(\omega^{(k)}_I\right)^2+\left(\omega^{(k)}_Q\right)^2 }\right)-1
\right] \right.\\
&\left.-
%%%%%%%%%%%%%5
\int_{0}^{r_T}
\frac{d}{d r}
\exE_{\bm{c}_k,\A_k} %{ H_k, A_k}
\left[
J_0\left(r^{-\sigma/2} A_k c_k\sqrt{\left(\omega^{(k)}_I\right)^2+\left(\omega^{(k)}_Q\right)^2 }\right)
\right]
\text{d} r
\right).
%%%%%%%%%%%%%%%%%%%%%%%%%
\end{split}
\end{align*}
%\normalsize

To proceed, we next expand the region in which the interferers are distributed via the limit $r_T \rightarrow \infty$ and apply the aforementioned identity. 

\begin{align*}
\begin{split}
%%%%%%%%%%%%%%%%%%%%%%%%%
&\lim_{r_T \rightarrow \infty} 
\lambda \pi 
r_T^2\left(
%%%%%%%%%%%55
\exE_{\bm{c}_k,\A_k} %{ H_k, A_k}
\left[
J_0\left(r_T^{-\sigma/2} A_k c_k\sqrt{\left(\omega^{(k)}_I\right)^2+\left(\omega^{(k)}_Q\right)^2 }\right)-1
\right]
 \right.\\
&\left.-
%%%%%%%%%%%%%5
\int_{0}^{r_T}
\frac{d}{d r}
\exE_{\bm{c}_k,\A_k} %{ H_k, A_k}
\left[
J_0\left(r^{-\sigma/2} A_k c_k\sqrt{\left(\omega^{(k)}_I\right)^2+\left(\omega^{(k)}_Q\right)^2 }\right)
\right]
\text{d} r
\right)\\
%%%%%%%%%%%%%%%%%%%%%%%%%
%%%%%%%%%%%%%%%%%%%%%%%%%
%&=
%\lim_{r_T \rightarrow \infty} 
%\lambda \pi 
%r_T^2
%%%%%%%%%%%%55
%\exE_{ H_k, A_k}
%\left[
%J_0\left(r_T^{-\sigma/2} H_k A_k\sqrt{\left(\omega^{(k)}_I\right)^2+\left(\omega^{(k)}_Q\right)^2 }\right)-1
%\right]
%\\
%&-
%%%%%%%%%%%%%%%5
%\lim_{r_T \rightarrow \infty} 
%\lambda \pi 
%r_T^2
%\int_{0}^{r_T}
%\frac{d}{d r}
%\exE_{ H_k, A_k}
%\left[
%J_0\left(r^{-\sigma/2} H_k A_k\sqrt{\left(\omega^{(k)}_I\right)^2+\left(\omega^{(k)}_Q\right)^2 }\right)
%\right]
%\text{d} r\\
%%%%%%%%%%%%%%%%%%%%%%%%%%
%%%%%%%%%%%%%%%%%%%%%%%%%
&=
\lim_{r_T \rightarrow \infty} 
\lambda \pi 
r_T^2
%%%%%%%%%%%55
\exE_{\bm{c}_k,\A_k} %{ H_k, A_k}
\left[
J_0\left(r_T^{-\sigma/2} A_k c_k\sqrt{\left(\omega^{(k)}_I\right)^2+\left(\omega^{(k)}_Q\right)^2 }\right)-1
\right]
\\
&-
%%%%%%%%%%%%%%5
\lim_{r_T \rightarrow \infty} 
\lambda \pi 
r_T^2
\int_{0}^{r_T}
\frac{d}{d r}
\exE_{\bm{c}_k,\A_k} %{ H_k, A_k}
\left[
J_0\left(r^{-\sigma/2} A_k c_k\sqrt{\left(\omega^{(k)}_I\right)^2+\left(\omega^{(k)}_Q\right)^2 }\right)
\right]
\text{d} r.
%%%%%%%%%%%%%%%%%%%%%%%%%%
\end{split}
\end{align*}

We can now evaluate the limits in each term above. Starting with the first one, we utilize the result from \cite[Equation (12)]{Sou92} which allows us to state the following  equivalent limit expression for the characteristic function for the total interference at two extremes:
\begin{align*}
\begin{split}
%%%%%%%%%%%%%%%%%%%%%%%%%
&\lim_{r_T \rightarrow \infty} 
r_T^2
%%%%%%%%%%%55
\exE_{\bm{c}_k,\A_k} %{ H_k, A_k}
\left[
J_0\left(r_T^{-\sigma/2} A_k c_k\sqrt{\left(\omega^{(k)}_I\right)^2+\left(\omega^{(k)}_Q\right)^2 }\right)-1
\right]\\
%%%%%%%%%%%%%%%%%%%%%%%
&\qquad =\lim_{r_T \rightarrow 0} 
r_T^{-2}
%%%%%%%%%%%55
\exE_{\bm{c}_k,\A_k} %{ H_k, A_k}
\left[
J_0\left(r_T^{\sigma/2} A_k c_k\sqrt{\left(\omega^{(k)}_I\right)^2+\left(\omega^{(k)}_Q\right)^2 }\right)-1
\right].
%%%%%%%%%%%%%%%%%%%%%%%%%%
\end{split}
\end{align*}
Now we note that since $\underset{r_T \rightarrow 0}{\lim} r_T^2 =0$ and\\ $\underset{r_T \rightarrow 0}{\lim} 
\exE_{\bm{c}_k,\A_k} %{ H_k, A_k}
\left[
J_0\left(r_T^{\sigma/2} A_k c_k\sqrt{\left(\omega^{(k)}_I\right)^2+\left(\omega^{(k)}_Q\right)^2 }\right)-1
\right] =0  $, we can apply L'Hopitals rule and the identity $\frac{\text{d}}{\text{d}x } J_0\left(x\right)=-J_1\left(x\right)$ given in \cite{abramowitz1964handbook} in conjunction with the chain rule to obtain
\begin{align*}
\begin{split}
%%%%%%%%%%%%%%%%%%%%%%%%%
&\lim_{r_T \rightarrow 0} 
r_T^{-2}
%%%%%%%%%%%55
\exE_{\bm{c}_k,\A_k} %{ H_k, A_k}
\left[
J_0\left(r_T^{\sigma/2} A_k c_k\sqrt{\left(\omega^{(k)}_I\right)^2+\left(\omega^{(k)}_Q\right)^2 }\right)-1
\right]\\
%%%%%%%%%%%%%%%%%%%%%%%%%%
%&=
%%%%%%%%%%%%%%%%%%%%%%%%%%
%\lim_{r_T \rightarrow 0} 
%2 r_T^{-1}
%%%%%%%%%%%%55
%\exE_{ H_k, A_k}
%\left[
%-J_1\left(r_T^{\sigma/2} H_k A_k\sqrt{\left(\omega^{(k)}_I\right)^2+\left(\omega^{(k)}_Q\right)^2 }\right)
%\frac{\sigma}{2}r_T^{\sigma/2-1} H_k A_k \sqrt{\left(\omega^{(k)}_I\right)^2+\left(\omega^{(k)}_Q\right)^2 }
%\right]\\
&=
%%%%%%%%%%%%%%%%%%%%%%%%%
\lim_{r_T \rightarrow 0} 
\sigma r_T^{\frac{\sigma}{2}-2}
%%%%%%%%%%%55
\exE_{\bm{c}_k,\A_k} %{ H_k, A_k}
\left[
-J_1\left(r_T^{\frac{\sigma}{2}}A_k c_k\sqrt{\left(\omega^{(k)}_I\right)^2+\left(\omega^{(k)}_Q\right)^2 }\right)
A_kc_k \sqrt{\left(\omega^{(k)}_I\right)^2+\left(\omega^{(k)}_Q\right)^2 }
\right].
%%%%%%%%%%%%%%%%%%%%%%%%%%
\end{split}
\end{align*}

We note that for $\sigma >2$ this limit converges to $0$.
Working with the second term, we have the following
\begin{align*}
\begin{split}
&\lim_{r_T \rightarrow \infty} 
\lambda \pi 
r_T^2
\int_{0}^{r_T}
\frac{d}{d r}
\exE_{\bm{c}_k,\A_k} %{ H_k, A_k}
\left[
J_0\left(r^{-\sigma/2}A_k c_k\sqrt{\left(\omega^{(k)}_I\right)^2+\left(\omega^{(k)}_Q\right)^2 }\right)
\right]
\text{d} r\\
&=
\lim_{r_T \rightarrow \infty} 
\lambda \pi 
r_T^2
\int_{0}^{r_T}
\exE_{\bm{c}_k,\A_k} %{ H_k, A_k}
\left[
J_1\left(r^{-\sigma/2} A_k c_k\sqrt{\left(\omega^{(k)}_I\right)^2+\left(\omega^{(k)}_Q\right)^2 }\right)\right.\\
& \left. \qquad \qquad \qquad \qquad \qquad \qquad \qquad
\times \frac{\sigma}{2} r^{-\sigma/2-1} A_k c_k\sqrt{\left(\omega^{(k)}_I\right)^2+\left(\omega^{(k)}_Q\right)^2 }
\right]
\text{d} r.
\end{split}
\end{align*}

%\normalsize
Using the specified identity and noting the result of \cite[Equation 17, pp. 8]{gulati2010statistics}, we obtain
%\small
%
\begin{align}
\label{LogChalpha}
\begin{split}
& \psi_{Y_{I}^{(k)},Y_{Q}^{(k)}}\left(\omega^{(k)}_I,\omega^{(k)}_Q\right) \\
& \qquad \qquad=
-\lambda \pi \left(\left(\omega^{(k)}_I\right)^2+\left(\omega^{(k)}_Q\right)^2 \right)^{\frac{2}{\sigma}}
\exE_{\bm{c}_k,\A_k} %{ H_k, A_k}
\left[
\left(A_k c_k\right)^{\frac{4}{\sigma}}
\right]
\int_{0}^{\infty}\frac{J_1\left(x\right)}{x^{\frac{4}{\sigma}}} \text{d}x.
%%%%%%%%%%%%%%%%%%%%%%%%%5
\end{split}
\end{align}
%\normalsize
%

Equation (\ref{LogChalpha}) is the $\log$ characteristic function of an isotropic bivariate symmetric $\alpha$-stable distribution, where the characteristic exponent $\alpha = \frac{4}{\sigma}$, and the dispersion parameter 
%
%\small
\begin{align*}
\gamma = 
\lambda \pi 
\exE_{\bm{c}_k,\A_k} %{ H_k, A_k}
\left[
\left(A_k c_k\right)^{\frac{4}{\sigma}}
\right]
\int_{0}^{\infty}\frac{J_1\left(x\right)}{x^{\frac{4}{\sigma}}} \text{d}x.
\end{align*}
%\normalsize

\end{proof}

\section{Proof of Theorem IV}
\label{theorem_4}
\begin{proof}
The proof of the result for the representation of the distribution and density of the total interference follows by taking the spherically symmetric bivariate stable distribution derived in Theorem \ref{thm_1} and utilising the closure under convolution result presented in Lemma \ref{lemma_2} for each Poisson mixture component to obtain the result in (\ref{genericCoxProc1}). 
%
%\small
\begin{align}
\begin{split}
&f_{Y\left(\u\right) }\left(y\right)  
=
d^{-1}_{K_{\text{max}}}
\sum_{k=1}^{ K_{\text{max}}} 
\mathbb{P}\left(K=k\right)
f_{\widetilde{\Y}^K\left(\u\right) }\left(y\right)
%\mathbb{I}\left(k \leq K_{\text{max}}\right)
\\
%%%%%%%%%%%%%%%%%%%%%%
&=
d^{-1}_{K_{\text{max}}}
\sum_{k=1}^{ K_{\text{max}}} 
\int
\mathbb{P}\left(K=k|\lambda_K\right)
\Ga \left(\lambda_K;a, b\right)
f_{\widetilde{\Y}^K\left(\u\right) }\left(y\right)
\text{d} \lambda_K
%\mathbb{I}\left(k \leq K_{\text{max}}\right)
\\
%%%%%%%%%%%%%%%%%%%%%%
&=
d^{-1}_{K_{\text{max}}}
\sum_{k=1}^{ K_{\text{max}}} 
\int_{0}^{\infty}
\exp\left(-\lambda_K\right) \frac{\lambda_K^k}{k!}
\frac{b^a}{\Gamma\left(a\right)}\exp\left(-b \lambda_K\right) \lambda_K^{a-1} \text{d}\lambda_K
f_{\widetilde{\Y}^K\left(\u\right) }\left(y\right)
\text{d} \lambda_K
%\mathbb{I}\left(k \leq K_{\text{max}}\right)
\\
%%%%%%%%%%%%%%%%%%%%%5
%%%%%%%%%%%%%%%%%%%%%%
&=
d^{-1}_{K_{\text{max}}}
\sum_{k=1}^{ K_{\text{max}}} 
\frac{\left(a+k-1\right)!}{\left(a-1\right)! k!}
\left(\frac{b}{1+b}\right)^a
\left(\frac{1}{1+b}\right)^k
\\
& \qquad \qquad \qquad  \qquad \qquad \times
\mathcal{S}_{\alpha}\left(\text{sgn}\left(c_k\right)
\widetilde{\beta}_k\left(\u\right), \left|c_k\right|\widetilde{\gamma}_k\left(\u\right), c_k \widetilde{\delta}_k\left(\u\right);0 \right)
%\mathbb{I}\left(k \leq K_{\text{max}}\right).
\end{split}
\end{align}
\label{genericCoxProc2}
%\normalsize

Note, the mixture weight of the doubly stochastic compound processes (Cox process) for the Poisson-Gamma distribution is derived by considering Bayes Theorem and the conjugacy property of the Poisson-Gamma model \cite{peters2011analytic} from which we know $p(\lambda|k) = \Gamma(\lambda; \alpha + k, \beta + 1)$, hence:
\begin{equation}
\begin{split}
\mathbb{P}(k) 
&= \frac{\mathbb{P}(k|\lambda)p(\lambda)}{p(\lambda|k)} 
=\frac{Po(k;\lambda)\Gamma(\lambda;\alpha,\beta)}{\Gamma(\lambda; \alpha + k, \beta + 1)} 
\\
%&=\frac{\frac{\lambda^k e^{-\lambda}}{k!}\frac{\beta^{\alpha}}{\Gamma(\alpha)}\lambda^{\alpha-1}e^{-\beta\lambda}}{\frac{(\beta+1)^{\alpha + k}}{\Gamma(\alpha + k)}\lambda^{\alpha + k -1}e^{-(\beta + 1)\lambda}}\\
&=\left(\frac{\lambda^k e^{-\lambda}}{k!}\frac{\beta^{\alpha}}{\Gamma(\alpha)}\lambda^{\alpha-1}e^{-\beta\lambda}\right)\div
\left(\frac{(\beta+1)^{\alpha + k}}{\Gamma(\alpha + k)}\lambda^{\alpha + k -1}e^{-(\beta + 1)\lambda}\right)\\
&= \frac{\Gamma(\alpha+k)}{\Gamma(\alpha)k!}\frac{\beta^{\alpha}}{(1+\beta)^{\alpha + k}}
= \frac{(\alpha + k- 1)!}{(\alpha-1)!k!} \left(\frac{\beta}{1+\beta}\right)^{\alpha}\left(\frac{1}{1+\beta}\right)^{k}. 
\end{split}
\end{equation}
Note, this mixing weight is then a negative binomial probability with $r = \alpha$ and $p = \frac{1}{1+\beta}$ as presented in \cite{peters2011analytic}.

Next, the representation of the density in (\ref{PoissDensityCox}) and distribution function in (\ref{PoissDistributionCox}) are obtained using the projection and SMiN representations in Lemma \ref{alpha_representation} to obtain a parametric closed form representation which can be evaluated.
\end{proof}

%
%%\section{Proof of Capacity bound}
%%\label{theorem_7}
%%\input{Annex_III}
%

\bibliographystyle{IEEEtran}
\bibliography{IEEEabrv,InterferenceAlphaStableb}

\end{document}